\documentclass[12pt,preprint]{aastex}
\usepackage[usenames]{color}

\def\l{$\lambda$}
\def\mbh{$M_{\rm BH}$\/}
\def\nh{$n_{\mathrm{H}}$\/}

\def\ne{$n_{\rm e}$\/}
\def\nc{$N_{\rm c}$\/}
\def\rfe{$R_{\rm FeII}$}

\def\feiiq{\rm Fe{\sc ii}$\lambda$4570\/}

\def\ltsima{$\; \buildrel < \over \sim \;$}
\def\ltsim{\lower.5ex\hbox{\ltsima}}  
\def\simlt{\lower.5ex\hbox{\ltsima}}  
\def\gtsima{$\; \buildrel > \over \sim \;$}

\def\gtsim{\lower.5ex\hbox{\gtsima}} 
\def\simgt{\lower.5ex\hbox{\gtsima}}

\def\lya{{ Ly}$\alpha$}
\def\civ{{\sc{Civ}}$\lambda$1549\/}

\def\cm3{cm$^{-3}$\/}
\def\hb{{\sc{H}}$\beta$\/}

\def\niv{{\sc{Niv]}}$\lambda$1486\/}
\def\ciii{{\sc{Ciii]}}$\lambda$1909\/}
\def\oiiiopt{{\sc{[Oiii]}}$\lambda\lambda$4959,5007\/}
\def\o4363{{\sc{[Oiii]}}$\lambda$4363\/}
\def\caii{{Ca{\sc ii}}}

\def\siiii{Si{\sc iii]}$\lambda$1892\/}
\def\aliii{Al{\sc iii}$\lambda$1860\/}
\def\heiiuv{He{\sc{ii}}$\lambda$1640}
\def\nv{{N\sc{v}}$\lambda$1240}

\def\feii{{Fe\sc{ii}}\/}
\def\siii{{Si\sc{ii}}$\lambda$1814\/}
\def\feiii{{Fe\sc{iii}}\/}

\def\fe{{\sc{Fe}}\/}

\def\fe76087{{\sc [Fe vii]}$\lambda$6087\/}

\def\kms{km~s$^{-1}$}

\def\rb{$r_{\rm BLR}$\/}

\def\siiv{Si{\sc iv}$\lambda$1397\/}
\def\oiv{O{\sc iv]}$\lambda$1402\/}

\def\siiiuv{Si{\sc ii}$\lambda$1533\/}
\def\feiiiuv{Fe{\sc iii}(UV34)}

\def\oi{O{\sc i}$\lambda$1304\/}
\def\R{$r_\mathrm{BLR}$\/}
\def\nhu{$n_\mathrm{H}U\/$}
\def\rbp{$r_\mathrm{BLR,\Phi}$\/}
\slugcomment{}

\shorttitle{A new method to obtain the Broad Line Region size of high redshift quasars}
\shortauthors{Negrete et al.}

\begin{document}
\title{A new method to obtain the Broad Line Region size of high redshift quasars
 \altaffilmark{\clubsuit}}

\author{C. Alenka Negrete}
\affil{Instituto Nacional de Astrof\'isica, \'Optica y Electr\'onica, Mexico}
\email{cnegrete@inaoep.mx}

\author{Deborah Dultzin}
\affil{Instituto de Astronom\'ia, Universidad Nacional Aut\'onoma de M\'exico, Mexico}
\email{deborah@astro.unam.mx}

\author{Paola Marziani}
\affil{INAF, Astronomical Observatory of Padova, Italy}
\email{paola.marziani@oapd.inaf.it}

\and

\author{Jack W. Sulentic}
\affil{Instituto de Astrof\'isica de Andaluc\'ia, Spain}
\email{sulentic@iaa.es}

\altaffiltext{0}{Based on observations made with ESO Telescopes at  Paranal Observatory under programme ID 078.B-0109(A) }

\begin{abstract}
 We present high S/N UV spectra for eight quasars at $z\sim3$\ obtained with VLT/FORS. The spectra  enable us  to analyze in detail the strong and weak emission features in the rest-frame range 1300-2000 \AA\ of each source (\ciii, \siiii, \aliii, \siii, \civ\ and blended \siiv+\oiv). Flux ratios \aliii/\siiii, \civ/\aliii, \siiv+\oiv/\siiii\ and \siiv+\oiv/\civ\  strongly constrain ionizing photon flux and metallicity through the use of  diagnostic maps built from  {\sc cloudy} simulations. The  radius of the  broad line region is then derived from the ionizing photon flux applying the definition of the ionization parameter. The \rb\ estimate and the width of a virial component isolated in  prominent UV lines yields an estimate of black hole mass. We compare our results with previous estimates obtained from the \rb\ -- luminosity correlation customarily employed to estimate black hole masses of high redshift quasars.  
\end{abstract}


\section{Introduction}

\defcitealias{negreteetal12}{Paper I}
\defcitealias{negreteetal13}{Paper II}

Measuring relevant physical parameters from the observed broad-line spectra of quasars is still an open challenge. Identification and intensity measurements of the strongest emission lines has made possible a rough inference about typical conditions in the emitting gas from the earliest days of quasar spectroscopy. The first intermediate redshift quasars discovered in the 1960s showed a fairly high ionization spectrum with prominent lines of \civ, and \heiiuv\ in addition to strong Balmer lines seen in the lower redshift sources. Photoionization by a central continuum source was considered the principal heating mechanism of the emitting gas. Significant \ciii\ emission suggested electron densities (\ne) in the range $10^9 - 10^{10}$ \cm3. The observed intensity ratio \ciii/\civ\ indicated ionization parameter ($U$; defined by Eq. \ref{eq:u} later in this paper) values of the order of $10^{-1}$. This photoionization scenario was successful in explaining at least some quasar optical and UV spectra (see the review by \citealt{davidsonnetzer79} for a synopsis).

More recent work recognized the existence of  several problems with the original scenario. Low ionization lines (LILs), and especially \feii\ are too strong to be explained by a photoionized region of moderate density and column density (see for example \citealt{dumontmathez81, joly87, collinsouffrinetal88, dumontcollinsouffrin90b}). These authors stressed that the LILs required a  denser, low-temperature environment.  Even more recently, \citet{baldwinetal96,laoretal97b,baldwinetal04} point towards high density at least for the  LIL emitting zone. This low ionization broad line region (LIL BLR) has very similar properties to the O{\sc i}  and Ca{\sc ii} emitting region identified by \citet{matsuokaetal08}. The region where these LILs  are produced cannot emit much \ciii\ if the electron density exceeds 10$^{11}$ \cm3.   BLR conditions are certainly complex and the assumption of a single emitting region cannot explain both LILs and high ionization lines (HILs) in all quasars \citep[][and references therein]{marzianietal10,wangetal11}. 

In \citet{negreteetal12}, hereafter \citetalias{negreteetal12}, we report  an analysis based on several diagnostic ratios used to constrain density, ionization parameter and metallicity in the BLR of two sources that are representative of extreme Population A (narrow Line Seyfert 1 - NLSy1) sources. These sources showed weak \ciii\ emission (relative to \siiii) which simplified our interpretation of the emission line spectra. Diagnostic ratios indicate a very dense (\ne$\sim 10^{12}$ \cm3), low ionization ($U\sim10^{-2.7}$) region that, in a photoionization scenario, is expected to also emit \feii\ and \caii\ lines \citep[e.g.,][]{bruhweilerverner08,matsuokaetal08}. In \citet{negreteetal13}, hereafter \citetalias{negreteetal13}, we show that \ciii\ is not associated with the high density region. However \ciii\ is strong in most sources,  dominating the emission of the 1900\AA\ blend. The presence of strong \ciii\ indicates that the BLR cannot be characterized anymore by a narrow range of density and ionization parameter: a gradient of ionization, of density, or both may be present. Since the BLR is not spatially resolved, the meaning of emission line ratios becomes much more ambiguous, and we cannot obtain meaningful single-value measures of \nh\ and ionization. Nonetheless, the ionizing photon flux (i.e., the product \nhu) can be retrieved with an accuracy comparable to estimates obtained from reverberation mapping, and then used to estimate \R\  \citepalias{negreteetal13}.
 
In this paper we apply the method discussed in \citetalias{negreteetal13} to a pilot sample of high redshift ($z \sim$ 3) sources. The technique we present should allow one to easily compute \R\ for large samples of high $z$\ sources. Moderate resolution and high S/N dedicated observations allowed us to detect and measure faint  and blended lines in order to analyze all physical information that can be  retrieved from rest frame UV spectra of high-$z$ quasars.  The pilot sample helps us to explore  challenges that exist for studying high-$z$\ sources using only rest-frame UV spectra. Estimation of redshift cannot rely on measures of low-ionization narrow lines that are  the most credible diagnostics at low z. \citep{eracleoushalpern03,huetal08}. Narrow lines are weak and often undetectable in very high luminosity sources \citep{marzianietal09}, a phenomenon sometimes called the \oiiiopt\ Baldwin effect; Zhang et al. 2011). On the other hand, the criteria for population and spectral type identification were set from properties of the \hb\ spectral spectral range \citep{sulenticetal00a} that is customarily not available for high $z$\ quasars.  

We present the spectra of 8 pilot sources obtained with the VLT/FORS1 in Section \ref{sec:observations};   
Section \ref{sec:reduction} summarizes data reduction including details of redshift estimation (Section \ref{z}). Before discussing the analysis of the data, we present a synopsis of quasar systematics along the so-called ``eigenvector 1'' (E1), with special attention on the interpretation of the line profiles.  
In Section \ref{sec:data_analysis} we apply the interpretation of the line profiles derived from low-$z$ sources to fit the line profiles in this high z sample.
In Section \ref{sec:component_analysis} we discuss the population assignment and the properties of each source, 
while Section \ref{sec:physical_conditions} explores  BLR physical conditions, with special reference to metallicity issues (Section \ref{metals}).
Section \ref{sec:results} presents results from application of the photoionization method to our pilot sample.
Section \ref{rblr} presents derivations of  the BLR radius (\rb; its distance from the ionizing source) and and black hole mass (\mbh). 
Section \ref{sec:discussion} compares our results with previous work and 
Section \ref{sec:conclusions} considers the the prospects for  application of our technique for single epoch \mbh\ estimates in high redshift samples.
All the computations were made considering $H_0$ =70 km s$^{-1}$ Mpc$^{-1}$  and  relative energy densities $\Omega_\Lambda=0.7$ and $\Omega_\mathrm{M}=0.3$.

\section{Observations}
\label{sec:observations}

Data were obtained between Nov. 2006 and Jan. 2007 using the VLT2/FORS1 telescope operated in service mode. FORS1 is the visual and near UV focal reducer and low dispersion spectrograph of the Very Large Telescope (VLT) operated by European Southern Observatory \citep[ESO;][]{appenzelleretal98}. A pilot sample  of 8 quasars at $z\sim3$ was observed with long exposure times to ensure high S/N in the continuum, in all cases above $\approx$ 40 , and otherwise in the range 50 -- 100. Table \ref{tab:obs} provides a log of observations that is organized as follows: 
Column (1) object name, 
Col. (2) apparent B magnitude, 
Col. (3) redshift, 
Col. (4) redshift uncertainty, 
Col. (5) line used for redshift estimation: 1)  \oi , 2) \ciii; 
Col. (6) absolute B magnitude, 
Col. (7) flux at 6 cm taken from FIRST \citep{beckeretal95}, 
Col. (8) date (refers to time at start of exposure), 
Col. (9) Digital Integration Time, 
Col. (10) number of exposures with integration time equal to DIT, 
Col. (11) average of the airmass,  
Col. (12) $S/N$ in the continuum around 1700\AA.

The observation of one of our 8 quasars, J00521-1108,  yielded the lowest S/N spectrum which we retain  because  observed features in the blend at $\sim 1900$\AA\ are clear enough to fit the individual lines. Two sources, J01225+1339 and J02287+0002, are high-ionization broad absorption lines (BAL) quasars with deep absorption throughs deeply affecting the \civ\ emission profile. 

\section{Data Reduction} 
\label{sec:reduction}

Data were reduced using  standard {\sc iraf} tasks. All spectra were wavelength and flux calibrated in the observed frame and then corrected for Galactic extinction. Flux correction was applied using meteorological data provided by ESO. The observed flux was multiplied by the inverse of the light lost computed from the ratio seeing over slit width in arcsec. Correction to rest frame requires estimating the redshift $z$\ which is not a trivial task as outlined below. Rest frame correction also involved scaling the specific flux in flux per unit wavelength interval by a factor $(1+z)^{3}$. Measurements were carried out on the rest-frame spectra. It is necessity to describe below two important aspects of the data reduction.

\subsection{A \& B Atmospheric Bands Correction}
\label{ab}

The A or B atmospheric band falls on top of the 1900\AA\ blend in many of the spectra. This is an important region for this study especially because it involves  \siiii,  \aliii, and  \siii. In order to remove these absorption features we created an  A+ B band template from standard star spectra used as specific flux calibrators. We scaled this template to find a best fit. Fig. \ref{fig:sample} shows the A and B absorption correction where we overtrace the uncorrected spectrum   to illustrate which lines are affected.  In cases where the A or B bands overlap a weak line like \siii\ the effect is considerable and measures of \siii\ should not be considered at all or with extreme care.  This happens for sources J00103-0037, J03036-0023, and J20497-0554. In  cases where one of the bands overlaps a stronger line like  \siiii\ or \aliii, the correction was good enough to permit accurate measures.

\subsection{Redshift Estimate}
\label{z}

Normally one uses strong narrow emission lines, preferentially from low ionization ionic species, to set the rest frame in  quasars. In our case no strong narrow lines are available so we consider the peaks of \lya, \civ\ and \ciii. The \lya\ peak is affected by absorption and  \civ\ is a HIL feature often showing blueshifts and/or asymmetries \citep{gaskell82, espeyetal89, corbin90, tytlerfan92, marzianietal96, richardsetal02, baskinlaor05b, sulenticetal07}. \ciii\ is  blended  with \siiii\ and \feiii\ that is especially prominent in this region and could well affect the peak wavelength of the blend. This is especially true in Pop. A sources.  Pop. B sources show a rather weak Fe spectrum making the \ciii\ peak a more reliable $z$\ estimator. 

Another option is to use the low ionization line \oi\ whenever  it is strong. It is unfortunately blended with low ionization Si{\sc ii}\l\l1304,1309 ($^2 P^0_{3/2,1/2} - ^2S_{1/2}$). Both \oi\ and Si{\sc ii}\l\l1304,1309 are broad lines; however they are of low ionization and in both Pop. A and B their peak shifts with respect to rest frame should be consistent with \hb\ or \feii, and therefore be rather modest, within a few hundred \kms\ \citep{sulenticetal12,marzianietal13a}. There is no hint of large systematic \hb\ broad component (where the broad component represents the line core and will be defined in Section \ref{compo}) peak shifts for all Pop. A sources (the majority in our sample) and $\approx 60$\%\ of Pop. B sources \citep{marzianietal03a}.

\begin{figure}
\epsscale{1.1}
\plotone{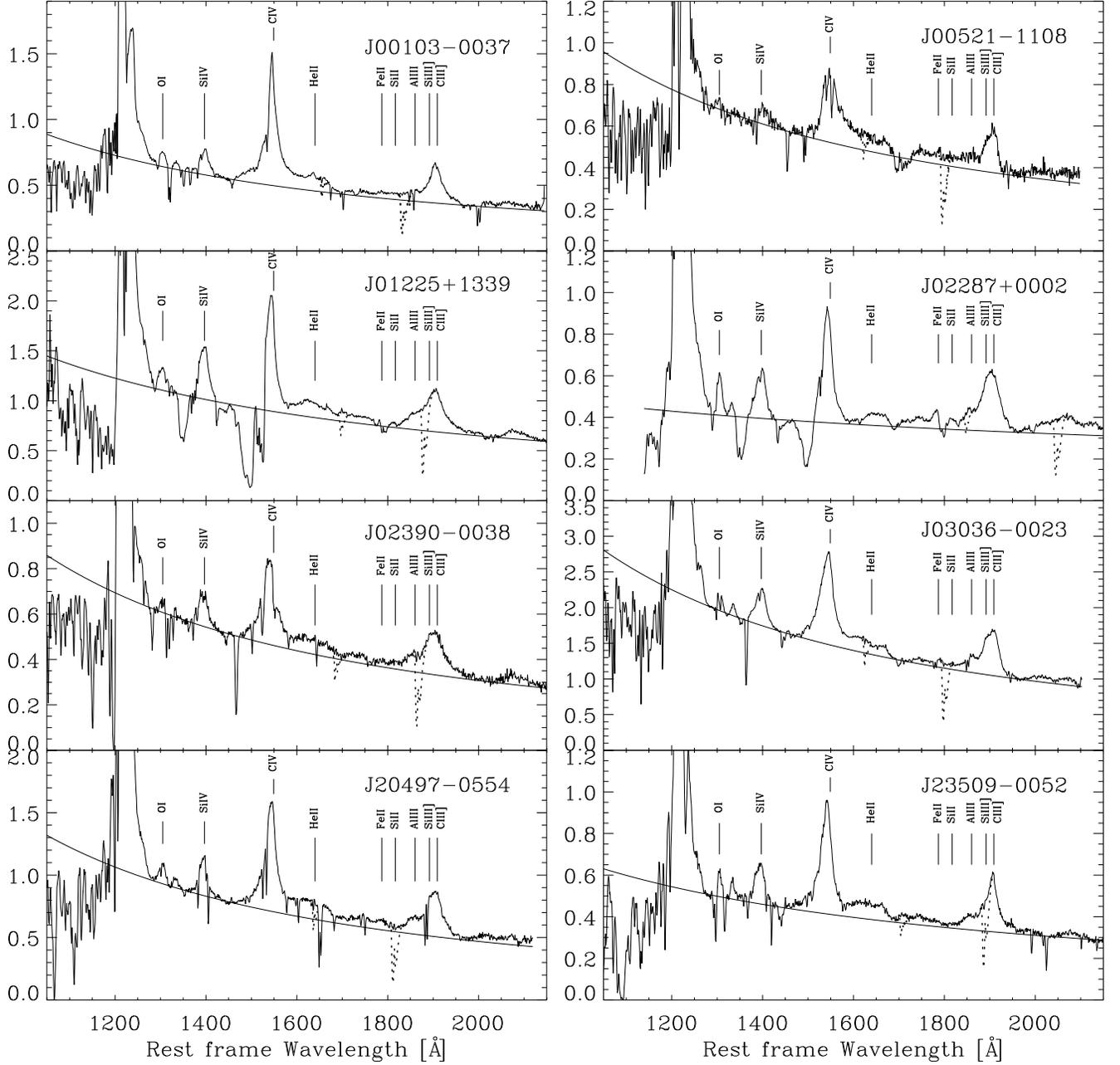}
\caption{Sample of 8 VLT spectra in rest frame wavelength. Abscissa is observed wavelength in \AA, ordinate is specific flux in units 10$^{-13}$ ergs s$^{-1}$ cm$^{-2}$ \AA$^{-1}$ corrected for Milky Way Galactic extinction. The superimposed dotted line is before atmospheric bands subtraction. We show the fitted continuum and the positions of the lines of our interest \ciii, \siiii, \aliii, \siii, Fe{\sc ii}$\lambda$1787,  \civ\ and \siiv. J01225+1339 and J02287 are BAL quasars. 
\label{fig:sample}}
\end{figure}

Photoionization simulations in the (\nh, $U$) region of interest show \oi\ $\approx$\ 2 Si{\sc ii}\l 1304, 1309 and this is confirmed in the spectrum I Zw 1 where \oi\ and Si{\sc ii}\l\l1304,1309 are resolved. The two components of the Si{\sc ii}\l\l1304,1309 doublet are set to the same intensity (i.e., we assume an optically thick case). We model the blend  \oi\ + Si{\sc ii}\l\l1304,1309 with 5 Gaussians; the three components of the O{\sc i}  feature are produced by Bowen florescence mechanism, and should show ratios consistent with their transition probabilities. Generating a model spectrum in {\sc iraf} (lines broadened to 4000 \kms) yields a rest frame peak wavelength of 1304.8 $\pm$\ 0.2 \AA\ (in vacuum) which we use as a reference for our VLT spectra. 
We also used \heiiuv\ to check the restframe assignment, in the objects when it is clearly visible (J00103-0037, J00521-1108, J02287+0002, J02390-0038 and J23509-0052).
Examination of Fig. \ref{fig:sample} reveals that the peak of \oi\ in source J00521-1108 is not observed clearly. We use \ciii\  to set the rest frame in this case.  Redshifts obtained for  three quasars, J01225+1339, J03036-0023 and J23509-0052, were obtained from \oi, and  are consistent with the redshift obtained with \ciii.  There are other sources J00103-0037, J02287+0002, J02390-0038 and J20497-0554 where the redshift estimation using both \oi\ and \ciii\ are not in good agreement (see Cols. 3 and 4 of Table \ref{tab:obs}). 
 The largest  disagreement  was found for J02287+0002, with $\Delta z \approx$ 0.0097. In this case, the uncertainty is so large to affect the interpretation of line profiles, and we are forced to repeat the line fitting with the assumption of two different redshifts. 

Fig. \ref{fig:sample} shows the deredshifted VLT-FORS1 spectra for our sample of 8 quasars.

\section{Data Analysis}
\label{sec:data_analysis}

\subsection{Quasar Systematics}

Quasar spectra are not all alike. There are significant differences in line intensity ratios and broad line profiles from object to object \citep{bachevetal04,marzianietal10}. More importantly, these differences can be organized in a systematic way  \citep{borosongreen92}. Since the early 1990s several authors have stressed the importance of the E1 of quasars \citep[e.g., ][]{gaskelletal99}. \citeauthor{sulenticetal00a} (\citeyear{sulenticetal00a,sulenticetal07}) expanded the E1 trends into a 4-dimensional space involving optical, UV and X-ray measures. They also defined spectral types along a sequence occupied by AGN in an optical plane involving \feii\  and FWHM H$\beta$ parameters \citep{sulenticetal02}. Objects at extreme ends of the E1 sequence are very different at almost all wavelengths and median spectra computed in spectral bins within this plane emphasize systematic changes in broad line properties \citep{sulenticetal02,sulenticetal07}. The most effective divider of the two quasar types appears to be  at FWHM of the H$\beta$\ broad component (BC) H$\beta _\mathrm{BC}  \approx$ 4000 \kms\ for low-to-moderate luminosity  sources \citep{marzianietal09}. The limit on FWHM corresponds to  Eddington ratio $L/L_\mathrm{Edd} \sim 0.2 \pm$0.1 \citep{marzianietal03b}.

\citet{sulenticetal02} gridded the BC of FWHM(\hb$_{BC}$) versus \rfe = W(\feiiq)/ W(\hb$_\mathrm{BC}$) parameter plane into bins of fixed $\Delta$ FWHM = 4000 \kms\ and $\Delta$ \rfe = 0.5. Quasar spectra  in different bins are different in many measures. As mentioned earlier, the largest differences are found between NLSy1-like objects, Pop. A, and  broader  sources of Pop. B  with FWHM(\hb$_{BC}$) $\ga$ 4000 \kms. 

Bins A1, A2, A3 are defined in terms of increasing FeII$\lambda$4570 (see Fig. 1 of \citealt{sulenticetal02}). Median composite UV spectra  of low-$z$\ quasars were computed by \citet{bachevetal04}.  However, assignment to population and spectral type was originally done on the \hb\ spectral range that was available for every source included in the composites. In the case of high redshift quasars, the \hb\ spectral range is customarily not available. A proper analysis of the spectrum then requires that sources are assigned to either Pop. A or B, if not to a spectral type, from UV data alone. The two following criteria comes from the analysis of optical lines profiles; previous work has shown that they can be applied to the UV lines as well.

\begin{enumerate}
\item Broad line width. The intermediate ionization lines \aliii\ and \siiii\ are found to be equivalent to \hb\ broad component, with their FWHM in close agreement \citepalias{negreteetal13}.  At high luminosity the limit  FWHM = 4000 \kms\ must be increased to consider its dependence on luminosity.  Following \citet{marzianietal09}, we can set FWHM $\approx$ FWHM$_{0} \left(\frac{L}{10^{45}}\right)^{\frac{1-\alpha}{2}}$\ with FWHM$_{0} \approx$    4000 \kms.  The limit should be considered as indicative since the value of $a$\ is not known with a high precision. However, high luminosity sources with FWHM $\ltsim$ 4000 \kms\ are found to be of Pop. A;
\item Evidence of a prominent red wing indicative of a possible very broad component (VBC); Pop. B sources show \civ\ profiles that resemble the \hb\ ones \citep{marzianietal96}. A prominent redward asymmetry described by several authors \citep[w.g.,][]{willsetal93,corbin95,punsly10,marzianietal10} is present in both \civ\ and \hb. This feature, if prominent, easily and uniquely identify Pop. B sources.
\end{enumerate}

Two additional  discriminating UV spectral properties are the following:

\begin{enumerate}
\item \civ\ equivalent width. According to \citet{bachevetal04} there  is a rather abrupt discontinuity in W(\civ) between spectral types A1 and A2, from $\approx$ 80 \AA\ (A1) to  $\approx$ 40 \AA\ (A2). W(\civ) $\simlt$ 40$+ 2 \sigma$ \AA\ is a sufficient condition to identify a Pop. A source albeit not a necessary one, since A1 sources do not satisfy this condition. 
\item  Prominence of \aliii\ and \siiii\ with respect to \ciii. The A3 and A4 spectral types show intensity ratios \aliii/\siiii $\simgt$ 0.5 and \siiii / \ciii $\simgt 1$. 
\end{enumerate}

In addition, Pop. A sources show \civ\  blueshift at half maximum intensity that are of larger amplitude than in Pop. B. However, not all Pop. A sources show blueshifts like most Pop. B sources \citep{sulenticetal07}. \civ\ blueshifts are expected to be orientation dependent if the \civ\  emitting gas is due to an outflow or wind \citep[e.g.,][and references therein]{flohicetal12,richards12}. A criterion based on \civ\ blueshift will be useful only to identify and confirm the most extreme Pop. A sources, and will not be of general validity to distinguish Pop. A and B. 

Previous and current work indicates that assignment to population  is relatively straightforward for most sources \citep[][Sulentic et al., in preparation]{bachevetal04,marzianietal10,sulenticetal13}. Some ambiguity may be present for sources  right at the boundary between Pop. A and B i.e., between spectral types A1 and B1.

\subsection{Methodological Considerations on Multicomponent Fits}

The {\sc specfit}  {\sc IRAF} task \citep{kriss94}  allows us to fit the continuum, emission and absorption line components, \feii\  and \feiii\  templates, etc. We fit the emission lines in three spectral ranges: (1) 1340-1450 \AA\ to model the 1400\AA\ blend most likely due to \siiv\ and \oiv\ \citep{willsnetzer79}; (2) 1450--1680 \AA\ for analysis of \civ\ and (3) 1750--2050 \AA\ for analysis of the 1900\AA\ blend.

\subsection{Line Components}
\label{compo}

We base our {\sc specfit} analysis on several previous observational results described in the previous Paper I and II that  point toward three different components in broad line profiles \citep[see][]{marzianietal10}.

\begin{enumerate}

\item A BC showing a roughly symmetric profile with FWHM in the range 1000-5000 \kms. It is consistent with the component identified by \citet{matsuokaetal08}.  This BC dominates LILs in Balmer lines of Pop. A sources while it becomes less prominent in Pop. B.   The most relevant one is that  \hb\ can be described by a Lorentz function in Pop. A sources \citep{veroncettyetal01,sulenticetal02} and by the sum of 2 Gaussians in Pop. B sources (a BC unshifted + a broader redshifted component, the very broad component VBC \citep[e.g.,][]{zamfiretal10}. 

\item A VBC, as seen in LILs and HILs of most Pop. B sources but is absent from Pop. A profiles.  The VBC can be modeled as a Gaussian (FWHM $\sim$ 10000 \kms) often with a significant shift to the red. It can be called a defining property of Pop. B sources.  This component is clearly identified  in the \civ\ line of Pop. B objects, and is also appreciable on the red side of  \ciii\ of Pop. B  objects J00103-0037 and J02390-0038 that are discussed in this paper.

\item A blueshifted broad component (BLUE), defined as the residual emission in the \civ\ line after subtracting a scaled \hb\ profile \citep{marzianietal10}.  This BLUE component is often prominent in \civ\ and \lya\ of Pop. A sources. It is much less intense in radio-loud Pop. B sources \citep{marzianietal96, punsly10, richardsetal11}. We model this profile as a blueshifted Gaussian. The  Gaussian approximation is probably inappropriate especially if the BLUE component is strong: this component is believed to be produced in a partially-obscured radial flow, not in a virialized emitting system. 
\end{enumerate}

{ Our multicomponent description of the broad line profile is rather crude as it involves mostly symmetric functions. While an unshifted symmetric function is expected for virial broadening, BLUE and the VBC are not necessarily Gaussian. A more accurate representation of BLUE may be obtained through a skew Gaussian. However, in the fitting of \hb\ for Pop. B sources, two Gaussians provide a remarkably good fit in most objects and in median composites \citep{zamfiretal10, marzianietal13b}.

BLUE has been introduced because of the large blueshifted emission observed in \civ\ of A3 and A4 sources: the large \civ/\hb\ ratio on the blue side of the line profile clearly indicates different physical conditions than in the line core. The origin of BLUE may be ascribed to a predominantly radial outflow in a radiation driven wind context \citep[e.g.][]{elvis00}. ÊA physically motivated definition of the VBC involves Êthe ionization stratification that is likely found in Pop. B sources (as per echo mapping studies, e.g. \citealt{petersonwandel99}): since no significant VBC was detected in the FeII blends even at the highest S/N, the VBCÊcan be properly associated with the innermost BLR part where ionization is too high to allow for significant singly-ionized iron emission. The VBC can account for the difference in full profile line width between Pop. B \civ\ and \hb. At the same time, if motion is predominantly virial, the VBC gas will be responding first to continuum changes,  as found in some reverberation studies \citep[e.g.][who consider Pop. B sources]{petersonwandel99}.  In \citepalias[][Table 1]{negreteetal13} we tested that our method yields results consistent with \hb\ RM, and not with \civ\ RM.
}

We then use the results of  \citet{marzianietal10} and \citetalias{negreteetal13}: the BC of \siiii, \aliii\ and \civ\ lines is similar to the one of \hb, including the FWHM and profile shape, either Gaussian or Lorentzian. The similarity helps us to define whether an object is Population A or B in this paper. { We use the BC (and not the full line) intensity to compute the line ratios considered in our method Ê(Section \ref{sec:ionizingphf}).}

\citet{baldwinetal96} presented a similar analysis. Their Fig. 2 organizes spectra in a sequence that is roughly corresponding to E1, going from \aliii-strong sources to objects whose spectra show prominent \ciii\ along with weak \aliii\ \citep{bachevetal04}.  Two of the three line components they isolated correspond to the ones we consider in this paper: a blue-shifted feature, and a more symmetric, unshifted and relatively narrow component that we call LIL-BC. Less obvious is the correspondence of a third feature, although it appears to be the redshifted part of what we call the VBC.

\subsection{\feii\ and \feiii\ Emission}
\label{sec:fe2_fe3}

 \feii\ emission is not strong in the spectral range we studied.  Significant \feiii\  emission is however expected close to and underlying the 1900\AA\ blend, especially for Population A sources \citep{francisetal91,vandenberketal01}. Our approach is completely empirical and employs an  \feii\  + \feiii\  template taken from templates successfully used in \citetalias{negreteetal12} (Sec. 3.2 provides more details) and \citetalias{negreteetal13}. We adopt an \feii\  template  based on a {\sc cloudy} simulation and  is not very far from the preferred model of \citet{bruhweilerverner08}, and  the \feiii\ template of \citet{vestergaardwilkes01}. The \feii\ intensity scale of the template is anchored to the  \feii\ UV 191 multiplet intensity, although we repeat that \feii\ is in general weak. This allows to reliably constrain  \feii\ emission underlying \civ\ and the 1900\AA\ blend. Similarly the \feiii\ intensity is set by a feature external to the 1900\AA\ blend (2080\AA).

\subsection{Expected emission from the three components} 
\label{Expected}

We looked for evidence of three possible components as described above: BC and BLUE component for Pop. A and BC, BLUE, VBC for Pop. B sources. The three components  show different strength in different lines, and are negligible in some lines, simplifying the interpretation of blended features. 

\subsubsection{Pop. A}

The \civ\ and \heiiuv\ are modeled assuming that a BC and a BLUE component is present in both lines. The BC and BLUE component of \heiiuv\ account for the flat shape of broad \heiiuv\ emission on the red side of \civ.

The 1400\AA\ blend is modeled assuming a single BC and a BLUE component similar to the ones of \civ. Since \oiv\ and \siiv\ are inextricably blended together, we  consider the total blend flux  for metallicity analysis (Section \ref{metals}).  However, high-density gas should emit negligible intercombination \oiv.  In the extreme Pop. A sources the BC could be ascribed to mainly \siiv, while the BLUE  component only is a blend of both \siiv\ and \oiv. In those cases the \siiv\ + \oiv\ blend closely resembles the shape of \civ. 

We expect that the \aliii\ doublet is emitted exclusively in the BC, the region where \feii\ is also emitted. This is empirically confirmed  by the aspect of the 1900\AA\ blend in many sources, where we do not see any evidence of BLUE nor VBC in \aliii. We remark that the \aliii\ doublet is relatively unblended, and that a BLUE  feature as strong as in the \civ\ profile of Pop. A sources would not easily escape visual detection. The same is also true for \siiii. Several fits that included a BLUE component in \ciii\ yielded 0 intensity, implying a large \civ/\ciii\ \citep{marzianietal10}.   

The BLUE component is very weak or undetectable in the vast majority of the \hb\ profiles analyzed in \citealt{marzianietal03a} (but see \citealt{zamfiretal10} for several cases of \hb\ BLUE), while prominent in \lya; the \lya/\hb\ ratio in this component is high.  In summary, the BLUE  component is visually strong in \lya\ and  \civ. A \heiiuv\  BLUE component is needed for a self-consistent fit of the \civ+\heiiuv\ blend. 

Since the shift and FWHM are assumed the same for all lines (and templates) in the 1900\AA\ blend, the only  free parameters in addition to shift and FWHM are the intensities of  6 components: two  from the templates, \siii\ and \aliii\ that are not heavily blended, and \siiii\ and \ciii.  The {\sc specfit} analysis is especially helpful to measure in a non-subjective way, taking all constraints into account, the  two parameters that are most affected by  blending: the intensity of \siiii\ and \ciii\ (any \feiii\ $\lambda$1914 contribution in excess to the one of the adopted template is included in the estimated \ciii\ intensity). 

\subsubsection{Pop. B} 

As pointed out by \citet{marzianietal10} the ``plateau'' appearance of the far red wing of \civ\ is accounted for a BLUE component and a redshifted VBC of \heiiuv, whose shifts and widths match the ones of \civ. 

\begin{figure}
\epsscale{0.6}
\includegraphics[scale=0.35]{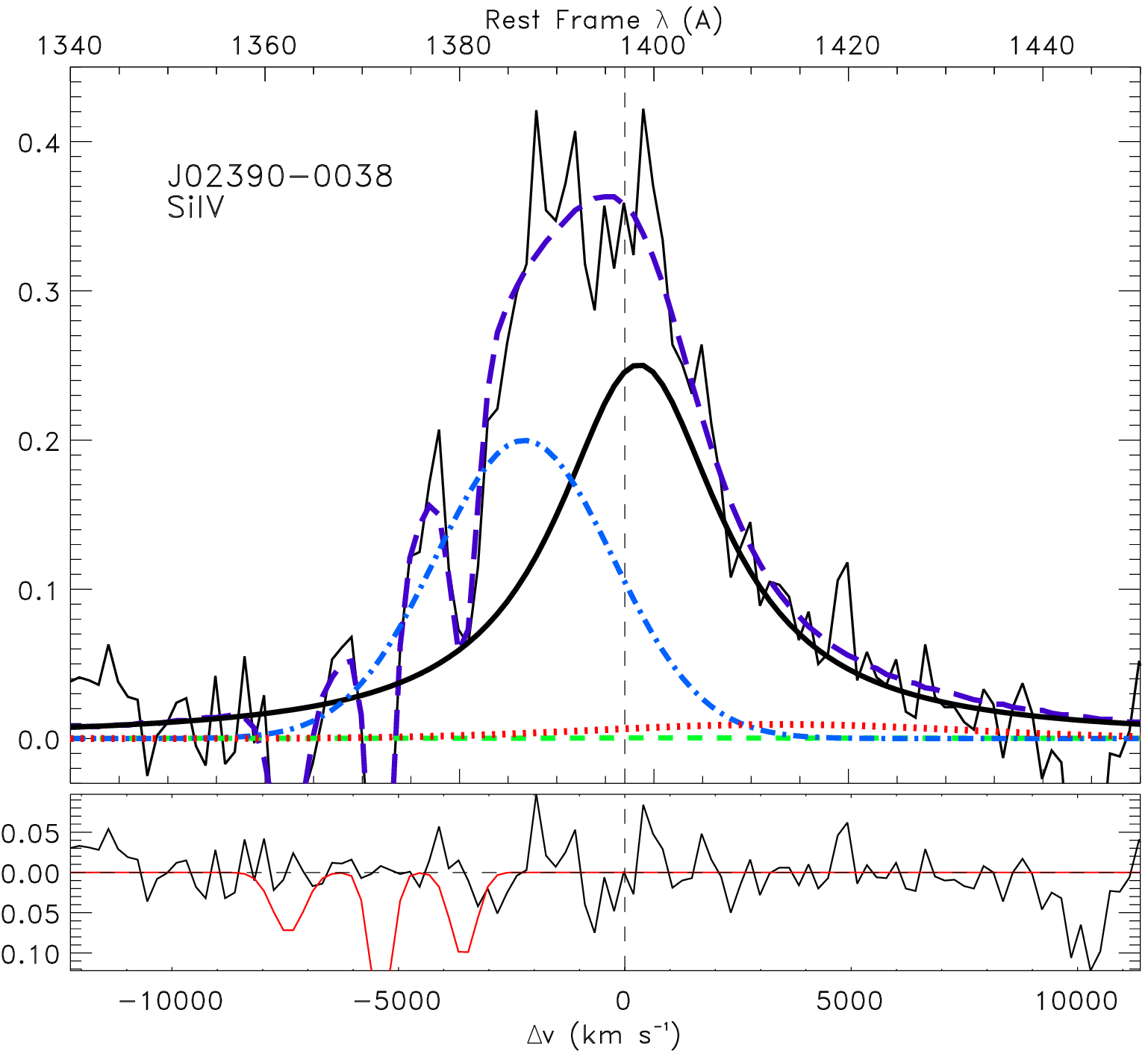}\includegraphics[scale=0.35]{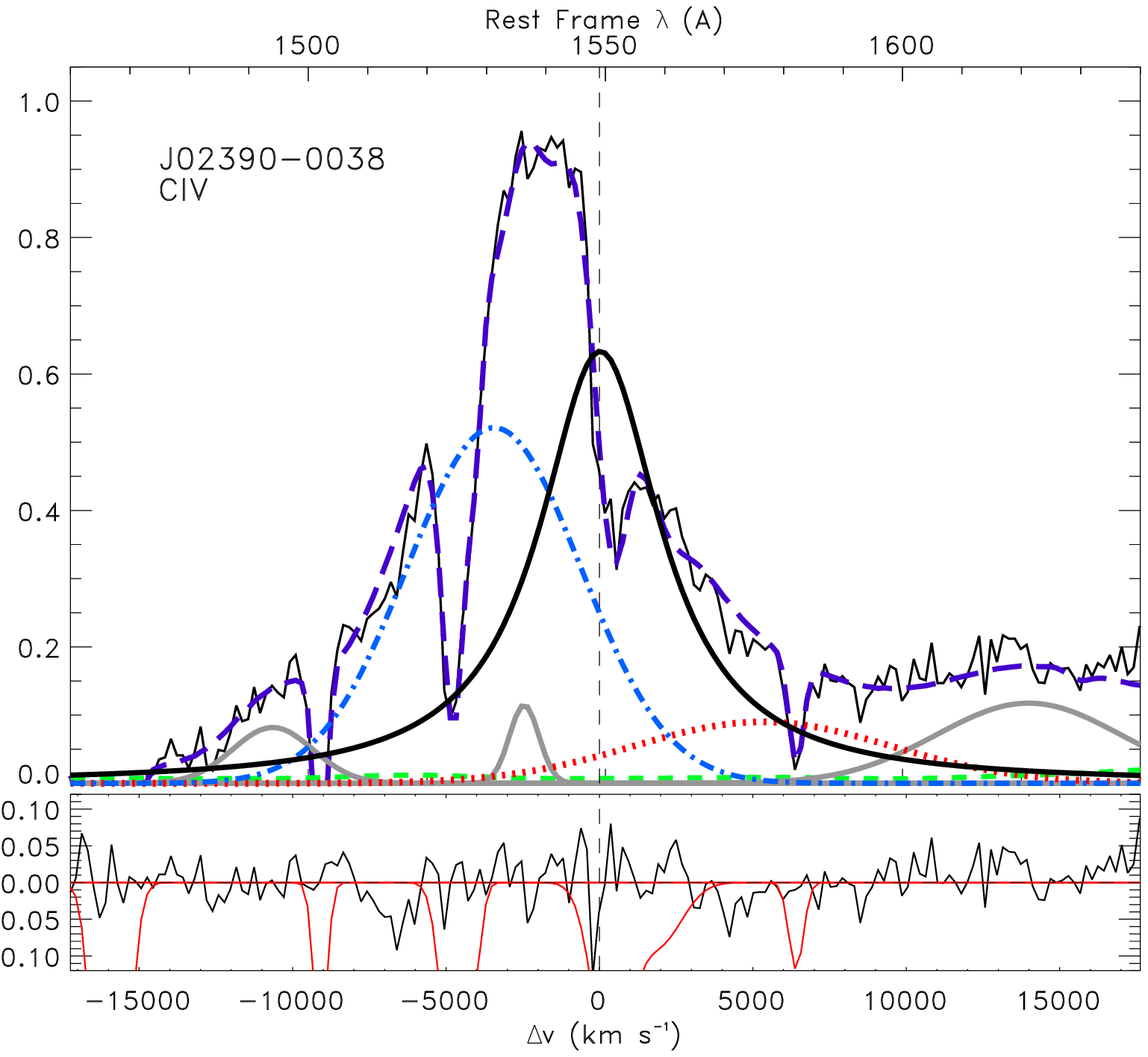}\includegraphics[scale=0.35]{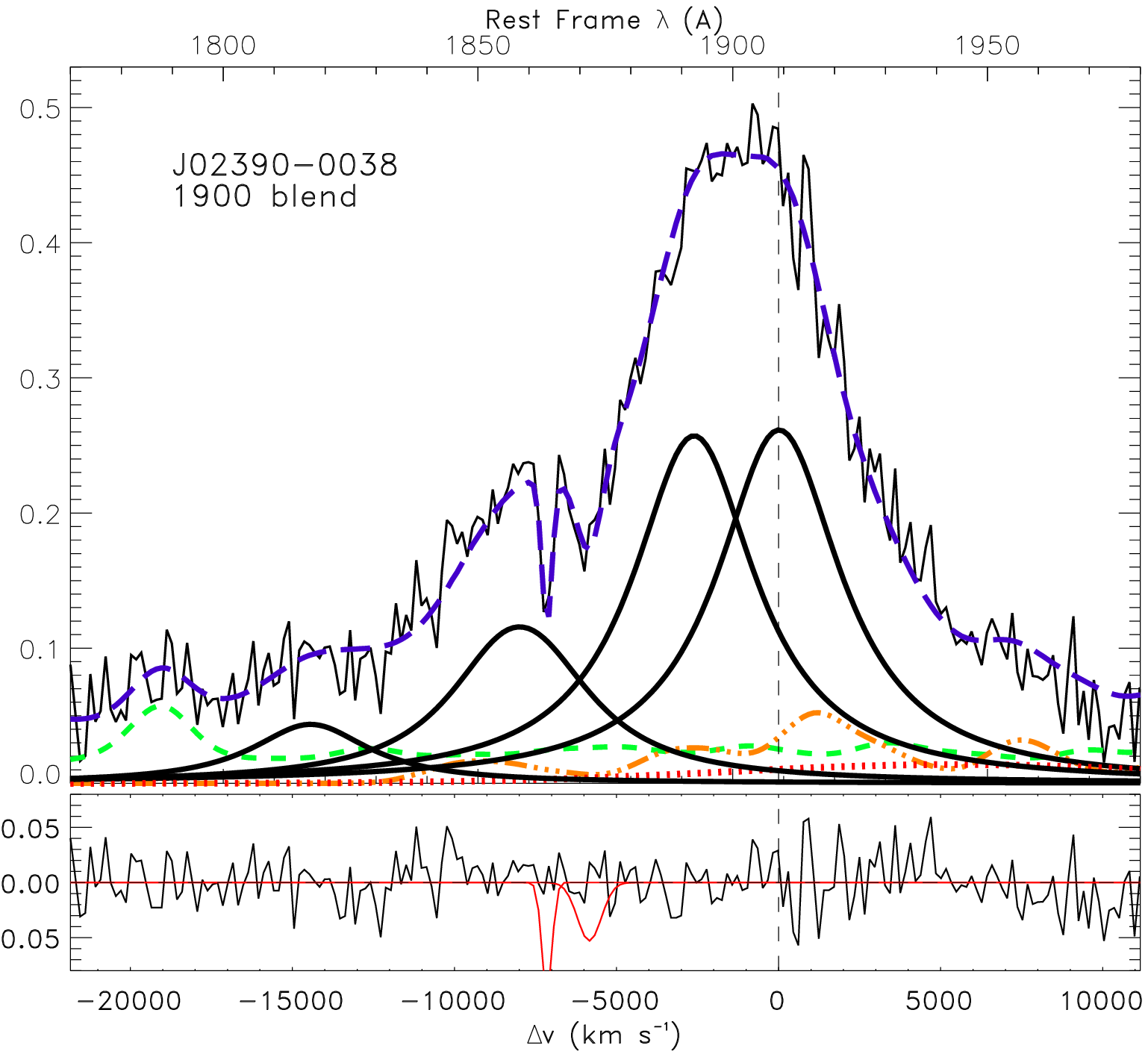}\\
\includegraphics[scale=0.35]{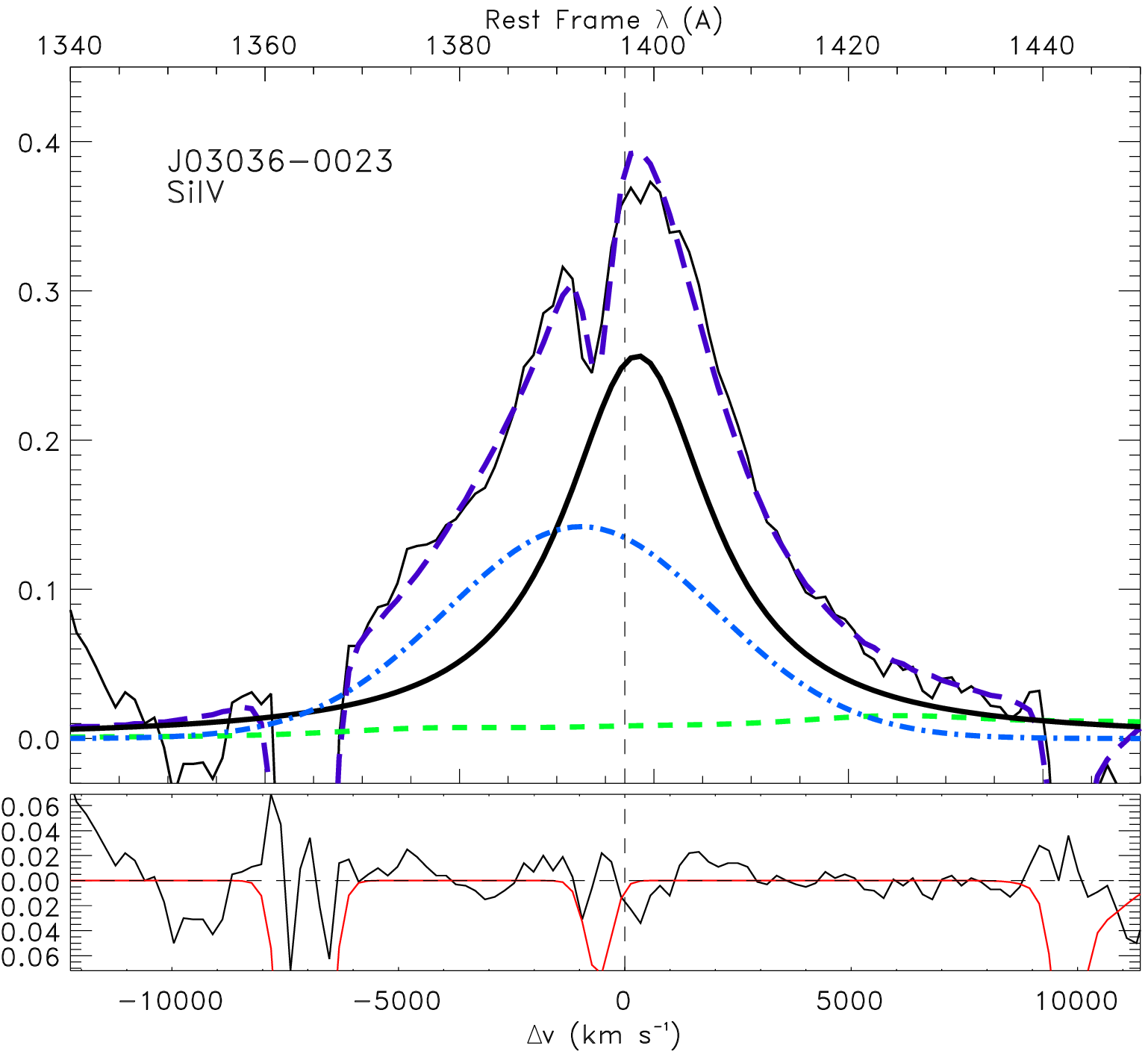}\includegraphics[scale=0.35]{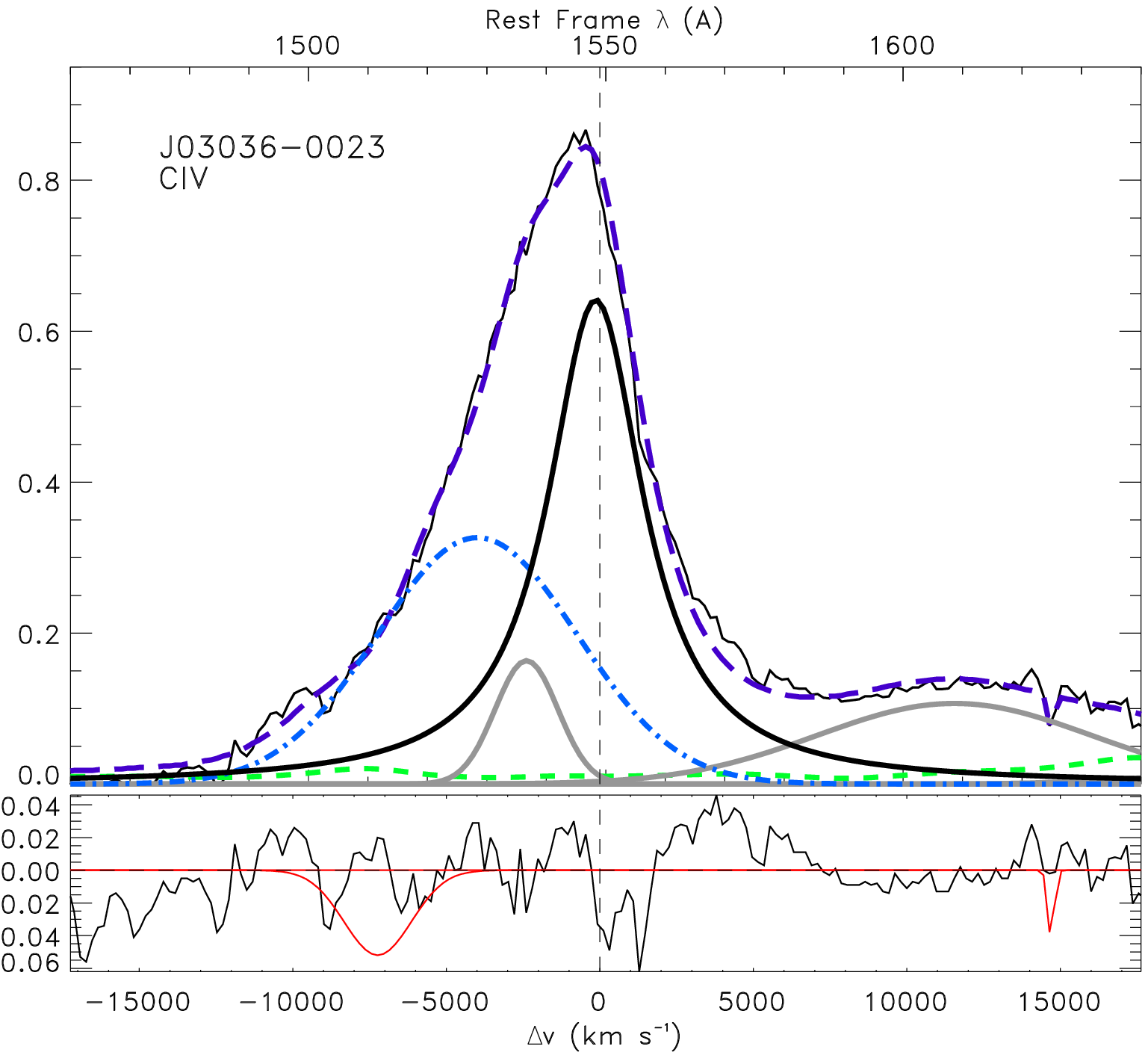}\includegraphics[scale=0.35]{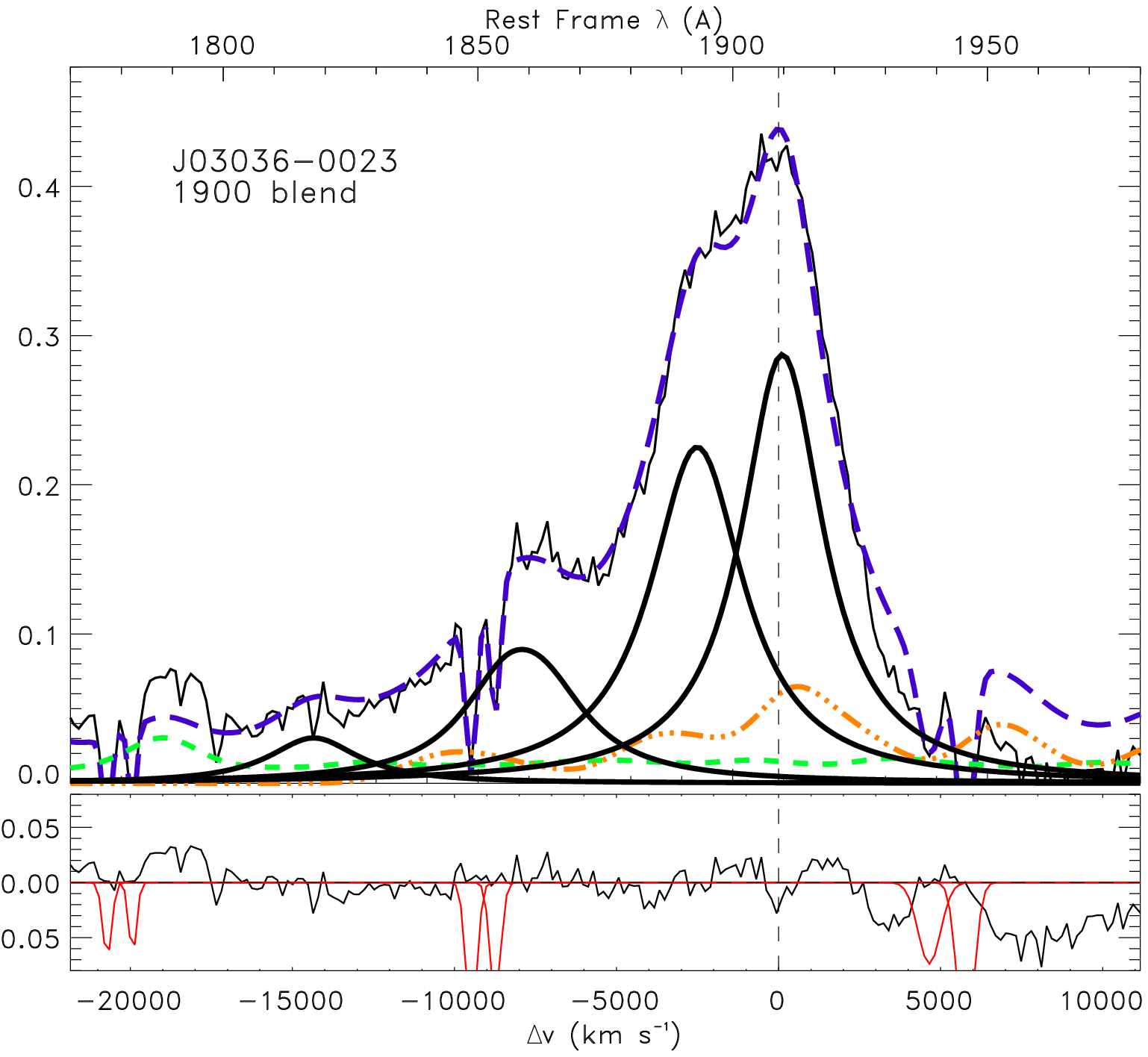}\\
\caption{Fits for Pop. A objects: J02390-0038, J03036-0023, J20497-0554 and J23509-0052. Upper panels show the fits and the lower panels under the fits show the residuals and also the fitted absorptions lines { shown in red}. 
Upper abscissa is rest frame wavelength in \AA, lower abscissa is in velocity units, ordinate is specific flux in arbitrary units. 
Vertical dashed line is the restframe for \civ\ and \ciii. 
Purple long dashed line is the fit, solid black lines are the BCs: \siiv\ in left panels, \civ\ in center panels and \ciii, \siiii, \aliii, \siii\ in right panels. For \aliii\ we show the sum of the doublet. Short green dashed line is \feii. \feiii\ is shown in orange dash-triple-dot line in the right panels. Blue dash-dot line in the left and center panels is the blue-shifted component of \siiv\ and \civ\ respectively. Dotted red line is the VBC, present also in \ciii\ for Pop. B objects. In the center panels we show with grey lines the contribution of {N\sc{iv}}$\lambda$1486, {Si\sc{ii}}$\lambda$1533 and \heiiuv\ core and blue-shifted components. For colors see online figures. \label{fig:fitsA}}
\end{figure}

The 1400\AA\ blend is fit with a single BLUE, BC and VBC. The VBC is rather faint (left panels of Figs. \ref{fig:fitsA}, \ref{fig:fitsB} and \ref{fig:fits_bal}) allowing for an estimate of the BC. However, considering that the blend is due to five components of \oiv\ and two of \siiv, whose relative intensities are unknown, we again consider the total blend flux for metallicity analysis (Section \ref{metals}). 

No VBC emission is observed (or expected) in \aliii, \feiii\ and  \feii.  These constraints help also to make the fits less ambiguous. However, the presence of a VBC  in \ciii\ and \siiii\ complicates the fit  of the 1900\AA\ blend. In any case, considering that we can expect the  VBC to be assimilable to a shifted Gaussian with FWHM $\sim 10000$\kms, the unblended part of the \ciii\ VBC provides a strong constraint.  Considering that the wavelength separation between \siiii\ and \ciii is $\approx$ 3000 \kms, $\ll$ FWHM VBC, what we model is most likely blended \siiii\ and \ciii\ VBC emission. 

{ We also fit absorption lines when they were needed in the {\sc specfit} routine along with the emission line components described above. They are shown in red in the lower panels of the Figures \ref{fig:fitsA} to \ref{fig:fits_bal}. In the case of the BAL quasars, we modeled the BAL profiles using several absorption components.}

\subsection{Errors} 
\label{errors}
Apart from the effect of noise that is treated as statistical source of error, we identify five main sources of error  that may significantly affect our measures.

\begin{enumerate}

\item A \& B atmospheric bands correction (already described in Section \ref{ab}). The most serious effect is  when the A/B band overlaps \siii.  

\item Line profile shape, Gaussian or Lorentzian (Pop. A or B).  The distinction between Pop. A and B is based on line width with the boundary at  FWHM H$\beta \approx$ 4000 \kms\ in low luminosity quasars and around $\approx$ 5000 \kms\  at higher luminosity such as the eight sources presented here.  Most of our quasars are unambiguously Pop. A or B because of line width and because Pop. B sources show an H$\beta$\ VBC, while Pop. A sources a prominent \civ\ BLUE component. In these cases only one profile shape (Gaussian or Lorentzian) was fitted.  A posteriori, we can say that the estimated line ratios are rather insensitive to the emission component profile shape: assuming a Gaussian or Lorentzian profile yields the same ratio for the strongest lines (i.e.,  \civ, \siiii, \aliii) upon which our analysis is based (with an uncertainty of $\sim$10\%). 

\item Rest-frame determination using \oi\ or \ciii.  
In some cases the redshift estimates derived from the two lines do not agree, most likely because of absorptions present in \oi\ and because this is not a very intense line. The principal impact  of uncertainty in the rest frame placement is estimation of the peak wavelength of \civ. If the line peak  differs from \l1549\AA, the BC intensity is diminished and we infer a greater contribution from the BLUE  component. Similarly, for the blend 1900\AA, the rest  frame shift may increase or decrease our estimate for the strength of  \ciii\  with consequent decrease or increase of the \siiii\ contribution. This additional source of uncertainty affects   J02287+0002, J02390-0038 and J20497-0554.  However, only in the case of  J02287+0002 the redshift difference produces a significant effect due to a $\Delta z \approx 0.0097$. 

\item \feii\  intensity (continuum placement). Broad \feii\  emission can produce a pseudo-continuum affecting our estimates of emission line intensities.  \siii\ is especially affected in our spectra because it is weak.  \aliii\ is similarly affected when it is weak. The effect is less noticeable for \civ\ since expected \feii\  emission  underlying the \civ\ line  is weak also for strong \feii\ emitters. The \feiiiuv\ and \feii\ emission is not very strong and a posteriori we find that there is no significant effect with the exception of \siii\ measures but the uncertainty can be included in the one associated with continuum placement, that contributes to the statical error.

\item BAL quasars principally affect the blue side of  \civ. We also find an absorption feature between FeII\l1787 and \siii\ (eg. Fig.  \ref{fig:fits_bal}). In  sources J01225+1339 and J02287+0002 Êderived  line intensities should be viewed with care since we fit unabsorbed components where the total flux eaten by absorptions is unknown. 

\end{enumerate}

In the case of \ciii\ we need to consider the possibility that the profile is narrower because there might be a contribution from different regions: indeed, the {\sc specfit} routine usually converges toward a narrower profile if the \ciii\ width is not constrained. The effect depends on the strength of the \feiii\ $\lambda$ 1914 feature which is expected to be prominent only in extreme Pop. A sources \citepalias{negreteetal12}. These sources of uncertainty are included, when appropriate, in the errors reported in Table \ref{tab:fluxes}. From the above considerations it is however clear that a faint line like \siii\ can be used mainly for confirmatory purposes because of the large errors plaguing its intensity estimates.

{ Finally, we note that the errors derived from the {\sc specfit} task are much smaller than the errors reported Êhere. The reason is that the errors reported by the tasks are formal errors associated with the multicomponent fit only.}

\section{Results of Line Component Analysis on Individual Objects}
\label{sec:component_analysis}

In Figures from \ref{fig:fitsA} to \ref{fig:fits_bal} we show our best fits for the VLT sample taking into account the considerations reported in Section \ref{sec:data_analysis}. In the left panels we present the fits for the \siiv\ line, the center panels are for the \civ\ line and the right panels are for the the 1900\AA\ blend. The fluxes and equivalent widths are in Tables \ref{tab:fluxes} and \ref{tab:ew}. Table \ref{tab:weak} shows the weak lines around \civ. For \civ\ and \siiv\ lines we show the BC, BLUE and the VBC. Errors are at a $2 \sigma$ confidence level, and include the sources of uncertainty described in Section \ref{errors}.  Errors are then quadratically propagated according to standard practice to compute intensity ratios and their logarithm.

We present in the following a phenomenological description of the fits. 

\subsection{Pop. A Objects}

We show the fits for our four Pop. A objects in the Figure  \ref{fig:fitsA}.  

\begin{itemize}

\item J02390-0038 -- This object is  borderline since  FWHM(BC) $\sim$ 4600 \kms\  places it close to the boundary between Pop. A and  Pop. B, at high $L$.  In support of Pop. B assignment,  we note that the \civ\ profile is best fitted assuming a VBC emitting region. Also, we find  {Fe\sc{ii}}\l1787 to be weaker than \siii.   On the other hand, \ciii\  is flat topped and has a very similar intensity as \siiii\  and the blend is best fitted using Lorentzian profiles. This object shows a very strong BLUE component in \civ\ that is also indicative of a Pop. A source. \siiv\ has a similar flat topped profile as in \ciii\ and probably as in \civ. We fit \siiv\ + \oiv\ blend following the \civ\ fit. Since the strongest and less ambiguous features indicate Pop. A, we assign this object to Pop. A and use Lorentzian profiles to fit all the BCs.

\item  J03036-0023 -- We estimate for this source a FWHM(BC) $\sim$ 3700 km s$^{-1}$\ and we use a Lorentzian function to fit the BCs.  The peak of \civ\ is blueshifted and requires a strong BLUE component. The bump on the red side of \civ\ can be accounted for by \heiiuv\ BC and BLUE components. There is no evidence for a red shifted component in \ciii. \aliii\ is prominent. Unfortunately the blue wing of \siii\ and the red wing of {Fe\sc{ii}}\l1787 are affected by A band absorption. 

\item J20497-0554 --  This source shows FWHM(BC) $\sim$ 3800 \kms. As for J03036-0023, the \civ\ line can be accounted for by an unshifted BC (assumed Lorentzian) and a considerable contribution of a BLUE component. We see a prominent \aliii\ line and FeII\l1787. \siiii\ is affected by several narrow absorption lines; however, it is obviously strong. The lack of a red wing on  \ciii\  suggests that no VBC is present. There are several absorption lines in the wings of \siiv. However, the line profile can be extrapolated over them.

\item  J23509-0052 -- {This source has  FWHM(BC) $\sim$ 3600  \kms. \civ\ shows a slight blue asymmetry with a BLUE component required to model it. The contribution of \feii\  is small and {Fe\sc{ii}}\l1787 is weak. \ciii\  is very strong.  \aliii\  is affected by A band absorption;   the profile  we fit is probably an upper limit. This object could well belong to spectral type A1 that includes Pop. A sources with the lowest \rfe ($\simlt 0.5$). {  Five absorption lines affect the \siiv\ + \oiv\ blend; luckily, they do not affect the  core of the blend.}}

\end{itemize}

\subsection{Pop. B Objects}.

\begin{itemize}

\item J00103-0037 --  This source has a FWHM(BC) $\sim$ 4500 kms$^{-1}$. The red side of \civ\ is blended with \heiiuv.  Fitting a BC with no shift plus a BLUE component to \civ\ leaves a very large residual on the red side. A redshifted VBC is needed to model the spectrum (Fig.\ref{fig:fitsB} upper center).  The faint narrow line under \civ\ can be explained as the narrow component (NC) of \civ\ \citep[see][]{sulenticetal07}. The presence of a similar NC in \ciii\ could possible explain the large residual seen $\sim$1900\AA.  We specifically note the prominent  \ciii\  emission and weak (but detected) \aliii\ (Fig.  \ref{fig:fitsB} upper right). The \feii\ ``bump'' at 1787\AA\ (UV 191) is appreciable. Fainter \feii\ emission is relatively unimportant because \feii\ creates a pseudo-continuum. \siii\ is compromised by A-band absorption. The blend at 1900\AA\ includes a  \ciii\  VBC and the fit indicates \civ/  \ciii\  (VBC) $\approx$ 7 which is reasonable.  We found several absorption lines on the blue side of the \siiv\ + \oiv\ blend (Fig. \ref{fig:fitsB} upper left), significantly affecting  its BLUE component that seems at any rate to be fainter than the  one of \civ.

\item J00521-1108 -- This source shows the noisiest spectrum in the sample.  We fit a FWHM(BC) $\sim$ 5300 \kms\ with \civ\ requiring a large VBC to account for the red wing. (Fig. \ref{fig:fitsB} lower center). Absorption features seriously affect the \civ\ profile. The profile of \ciii\ is strongly asymmetric due to some sort of absorption on the red side. \aliii\ is weak and consistent with pop. B (Fig. \ref{fig:fitsB} lower right).   We decide not to carry out a fit of the \siiv\ + \oiv\ blend because it is too noisy.

\end{itemize}

\subsection{BAL QSOs}

  We fit our two BAL quasars using Lorentzian profiles following \citet{sulenticetal06a}. We remind that the identification of \ciii\ in the BAL quasars and  in sources with strong \aliii\ is debatable \citep{hartigbaldwin86}: strong \feiii\l 1914 could take the place of most \ciii\ emission. 

\begin{itemize}

\item J01225+1339 --  \civ\ is highly affected by two broad absorption lines (Fig.  \ref{fig:fits_bal} upper center) with blueshifts of 5200 and 10800 \kms\ at peak absorption with equivalent widths/ FWHM  -12\AA\ / 3900 \kms\ and -25\AA/5200 \kms, respectively.  The blueshift of the \civ\ peak  leads us to suspect a large BLUE emission component. The 1900\AA\ blend shows absorptions coincident with  {Fe\sc{ii}}\l1787  and the blue side of \siii\ which is however unambiguously detected (Fig. \ref{fig:fits_bal} upper right).  \aliii\ is prominent which implies that this a Pop. A source.  The FWHM(BC) $\sim$ 4400 \kms\ is consistent with a high-luminosity Pop. A source.  It is also possible  that this BAL quasar is an outlier like Mark 231 at low-$z$\ \citep{sulenticetal06}, in other words an extreme Pop. A object. The  \ciii\  is well fitted with a Lorentzian profile. Broad A band atmospheric absorption lies over \siiii. {The 1400\AA\ blend  is also affected by two broad absorption line on both sides of the line (Fig. \ref{fig:fits_bal} upper left). As in previous cases, we made a fit following the \civ\ fit.}

\item J02287+0002 -- This object has a very complex spectrum. On the one hand, it has a FWHM(BC) $\approx$ 4700 \kms.  Considering that the FWHM limit  between Pop. A and B is increasing with luminosity, the FWHM(BC) is within the limit of Pop. A but close to the boundary with Pop. B. The lines profiles are better fit with Lorentzians. On the other hand, however, it shows features that are typical of extreme Pop. A sources:  prominent {Fe\sc{ii}}\l1787, strong  \aliii, no  \ciii\  VBC  (Fig. \ref{fig:fits_bal} middle right).  The \ciii\ line is not very flat topped but the similar intensities of \ciii\ and \siiii\ remind the case of  J02390-0038.  We assign J02287+0002 to Pop. A also because it has a strong blue-shifted component in \civ\ atypical to Pop. B objects  (Fig. \ref{fig:fits_bal} middle center). The \civ\  BAL shows a blueshift of 9100 \kms at deepest absorption, a EW of --14 \AA\ and a FWHM of 4600 \kms.   As in the case of J01225+1339, the \siiv\ + \oiv\ blend profile is severely affected by two broad absorption lines on both sides  (Fig. \ref{fig:fits_bal} middle left). 

The estimated rest frame of this quasar differs by $\sim$ 1300 \kms\ using  \oi\ and \ciii.  This is the  largest discrepancy in our sample. In order to evaluate the effect of the $z$\ discrepancy we  performed two fits  using both rest frames. The middle panels of Figure \ref{fig:fits_bal}  use the  \oi\  restframe.  In  the 1900\AA\ blend, we found a contribution of \siiii\ similar to \ciii.   If we use the \ciii\ inferred rest frame that we show in the lower panels of Figure \ref{fig:fits_bal},  \ciii\  becomes stronger with a resultant decrease of \siiii. A similar effect occurs for \civ\ broad and blue-shifted components.

\end{itemize}

Summing up, we are able to assign a Pop. A/B identification to all sources in our sample. The two BAL quasars appear as objects of extreme Pop. A; for  J02287+0002 the assignment depends on the assumed redshift. However, the most likely estimate  is based on the \oi\ line and is $z \approx 2.726$, and implies Pop. A features. Fig. \ref{fig:fits_bal} stands as a neat example of the importance of accurate redshift determination for quasars, as the interpretation of the source spectrum can be made  different by a $\Delta z \approx$ 0.0097.

\section{Diagnostics of Ionizing Photon Flux}
\label{sec:physical_conditions}

The physical conditions of photoionized gas can be described by electron density \ne, hydrogen column density \nc, metallicity  ($Z$; normalized to solar), shape of the ionizing continuum, and the ionization parameter $U$. The latter represents the dimensionless ratio of the number of ionizing photons and the electron density \ne\ or, equivalently, the total number density of hydrogen \nh, ionized and neutral.\footnote{In a fully ionized medium \ne $\approx 1.2$ \nh. We prefer to adopt the definition based on \nh\ because it is the one employed in the  {\sc cloudy} computations.} Both $U$\ and \nh\ are related through the equation
\begin{equation}
\label{eq:u}
U = \frac {\int_{\nu_0}^{+\infty}  \frac{L_\nu} {h\nu} d\nu} {4\pi n_\mathrm{H} c r^2} = \frac {Q(H)} {4\pi n_\mathrm{H} c r^2} 
\end{equation}
where $L_{\nu}$\ is the specific luminosity per unit frequency, $h$\ is 
the Planck constant, $\nu_{0}$\ the Rydberg frequency, $c$ the speed of light, and $r$\ can be interpreted as the distance between the central source of ionizing radiation and the line emitting region. $Q(H)$\ is the number of H-ionizing photons. Note that \nhu\ is, apart from the constant $c$, the ionizing photon flux
\begin{equation}
\label{eq:phi}
c\, n_\mathrm{H} U = \Phi(H) =  \frac{Q(H)} {4\pi r^2}.
\end{equation}
If we know {\em the product of } \nh\ and $U$, we can estimate the radius $r$\ of the BLR from Eq. \ref{eq:u}:
\begin{equation}
\label{eq:erre}
r_\mathrm{BLR,\Phi} =  \sqrt{ \frac{Q(H)} {4\pi  c (n_\mathrm{H}U)}.}
\end{equation}

\rbp\ is the ``photoionization'' \rb\ as defined in \citetalias{negreteetal12}. The dependence of $U$\ on \rb\ was used by \citet{padovanirafanelli88}  to derive central black hole masses assuming a plausible  average value of the product \nhu. The typical value of \ne\ was derived at that time from semiforbidden line \ciii\ which implied that  the density could not be much higher than  \ne $\approx 10^{9.5}$ cm$^{-3}$ \citep{osterbrockferland06}.  \citet{padovanirafanelli88} derived an average value $<U \cdot$ \ne $ >  \approx 10^{9.8}$\ from several sources where \rb\ had been determined from reverberation mapping, and for which the number of ionizing photons could be measured from multiwavelength observations. The average value was then used to compute black hole masses for a much larger sample of Seyfert 1 galaxies and  low-$z$\ quasars \citep{padovanirafanelli88, padovanietal90}. \citet{wandeletal99} compared the results of the photoionization method with the ones obtained through reverberation mapping, found a very good correlation for the masses computed with the two methods, and concluded that ``both methods measure the mass of the central black hole.'' A similar method based on the \hb\ luminosity was proposed by \citet{dibai84}, who retrieved the ionizing luminosity from the luminosity of \hb.  \citet{bochkarevgaskell09} verified that Dibai mass estimates agree with reverberation-mapping mass estimates.

 In  \citetalias{negreteetal13}  we showed that the ionizing photon flux estimated from  UV lines is also in close agreement with reverberation mapping BLR distance from continuum source. Photoionization methods can at least provide an alternative method the deduce \rb\ with an accuracy comparable to reverberation mapping.   The approach of this paper follows from  the results of \citetalias{negreteetal13}, in which intermediate and high-ionization lines are used as a diagnostic of the ionizing photon flux.  There are three main issues that are of relevance for a general population of quasars: (1) the actual physical conditions within the BLR, and specifically (2) the contribution of relatively low density gas (\nh $\sim$ 10$^{10}$ \cm3). A third special issue  is the estimate of the quasar metallicity since previous studies agree on super solar metallicity for high $z$\ quasars \citep[e.g.][]{hamannferland93,ferlandetal96,nagaoetal06,simonhamann10}. 

\subsection{Physical Conditions in the Emitting Regions} 

The method used to estimate the physical conditions of the emitting region is described in \citetalias{negreteetal12}. Here we summarize the basic aspects as well as the complications arising from dealing with sources that are different from  extreme sources Pop. A. To this aim, we apply and further develop the empirical technique that has been discussed in \citetalias{negreteetal13}. 

We base our interpretation of line ratios on a multidimensional grid of {\sc cloudy} \citep{ferlandetal98,ferlandetal13} simulations, (see also \citealt{koristaetal97}) to derive $U$\ and \nh\ from our spectral measurements. Simulations span the density range $7.00 \leq \log$ \nh$ \leq 14.00$, and $-4.50 \leq \log U \leq 00.00$, in intervals of  0.25. Each simulation was computed for a fixed ionization parameter and density assuming plane parallel geometry. The 2D grid of simulations was repeated twice assuming \nc$ = 10^{23}$\ and $10^{24}$ cm$^{-2}$. Several cases were computed also for \nc =  $10^{25} $cm$^{-2}$.  Metallicity was assumed to be either solar or five times solar. Two alternative input continua were used: 1) the standard AGN continuum of {\sc cloudy} which is equivalent to the continuum described by \citet{mathewsferland87} and 2) the low-$z$\ quasar continuum of \citet{laoretal97a}. Computed line ratios are almost identical for fixed ($U$, \nh). 
 { For this paper we use the ($U$, \nh) maps computed for  the \citet{laoretal97a} continuum (Figure 5 of \citetalias{negreteetal12}).  It is the ionizing luminosity that differs by more than a factor of 2 for a fixed specific continuum luminosity (as considered in Section \ref{rblr}) since the same $U$ is reached at smaller distance for the continuum with fewer ionizing photons.  This is taken into account when estimating \R\ (Section \ref{rblr}).}

\citet{feldmanetal92} give a critical density value for \siiii\ $n_e \sim 2 \cdot 10^{11} $ \cm3. \aliii\ is a permitted transition with large transition probability ($A \sim 5 \cdot 10^8$ s$^{-1}$) and has very high  and ill-defined critical density (i.e., its equivalent width goes to zero toward thermodynamic equilibrium, which occurs at very high density, when all emergent line emission is zeroed by equilibrium between collisional excitation and deexcitation). Our 2D array of {\sc cloudy} simulations shows that the ratio \aliii/\siiii\ is well suited to sample the density range $10^{10} - 10^{12.5} $ \cm3.  Within this range the \siiii\ intensity decreases smoothly by a factor 10;  above the upper limit in density, the predicted  intensity of  \siiii\ decreases \citepalias{negreteetal12}.   The ratio \aliii/ \siiii\ alone is, generally speaking, insufficient to constrain \nh.  A second diagnostic ratio is needed to constrain $U$\ and to unambiguously derive \nh. We consider  \civ/ \siiii\ and \siiv+\oiv/ \siiii\  as two diagnostic ratios suitable for constraining $U$.

\subsection{The Contribution of Lower Density Gas}
\label{sec:ciii_contr}

The presence of significant \ciii\ emission complicates  the analysis. As pointed out, the photoionization solution for the BC suggests very high density, and in this region no \ciii\ emission is expected.  The ratio \aliii/ \siiii\ is   diagnosing high density gas, while the \ciii/ \siiii\ ratio covers the domain of  \nh $ \sim 10^{10}$\cm3.    The spatially unresolved line emission is probably a mixture of gas in different density ionization condition, possibly following a smooth gradient.  The much lower \ciii\  critical density  implies that the \ciii\ line should be  formed farther out than \siiii\ and \aliii\   if all these lines are produced  under similar ionization conditions. 

For the BAL quasars in our sample and \aliii-strong sources most of what we ascribe to \ciii\ could be actually \feiii, as suggested by \citet{hartigbaldwin86}.  For Pop. A sources of spectral types A3, A4, as well as for BAL quasars belonging to these A spectral types, where \ciii\ is weak with respect to \siiii, any contribution due to lower density gas could be  neglected.

Among Pop. A2 and some A3 objects it is not so obvious that the profile of \ciii\ and \siiii\ is the same. It could be well that the \ciii\ profile is narrower than the ones of \siiii\ and \aliii\ (as found for SDSS J12014+0116, see \citetalias{negreteetal12}), justifying the idea of \ciii\ emission from a disjoint region or perhaps from the lower-density tail of comet-shaped clouds \citep{maiolinoetal10}. Reverberation mapping studies of the 1900\AA\ blend are scant, but they indicate that time delays in \ciii\ are a factor 2--3 larger than in \civ\ \citep{onkenpeterson02,metzrothetal06}.  

Whenever strong \ciii\ is observed, as in the case of spectral type A1 and A2 and even more of Pop. B, we could reverse the question: how much does any \ciii\ emitting gas contribute to the lines used for diagnostic ratios? Negligible contribution is expected to \aliii. However, this is not true for \civ\ and \siiii.   A realistic estimate of the low density contribution  to \civ\ and \siiii\ flux will depend on the assumptions concerning  ionization and density gradient, and will be therefore model dependent.

In \citetalias{negreteetal13} we showed that diagnostic ratios involving \ciii\  are not representative of \hb\ emitting gas that is responding to continuum variations. Diagnostic ratios with \ciii\ will be avoided. Here we will follow the same approach applying an empirical correction dependent on the ratio \aliii/\ciii. Generally speaking the \nhu\ value derived from the ratios \aliii/\siiii\ and \siiii/\civ\ is associated to a single point in the parameter plane (\nh, $U$). This corresponds to a well defined solution that  is biased toward high density (and therefore low \R).  As a consequence, the BLR physical conditions cannot be described by a single value of density and ionization parameter.  Even  the time lag from reverberation should be interpreted with some care, since it is a single number that is a rather abstract representation of the BLR distance from the central continuum source and may not have a well-defined structural counterpart \citep{devereux13}. Nonetheless, a correction based on the \aliii/\ciii\ ratio can lower the product \nhu\  increasing the derived \R\ and minimize the bias in the photoionization estimates with respect to reverberation lag { (i.e. the product of the speed of light times the time lag, $c\tau$)}, as shown in \citetalias{negreteetal13}.  We apply the following correction to the derived \R: 
 
\begin{equation}
\label{eq:low_dens}
\log r_\mathrm{BLR,\phi} - \log c\tau \approx 0.69 \log {\mathrm W}(\mathrm{\mathrm AlIII}\lambda1860)/\mathrm{\mathrm  W(CIII]}\lambda1909) + 0.49
\end{equation}

This is the same correlation found by \citetalias{negreteetal13} updated including an improved $c\tau$ \ value \citep{negreteetal13a}.

\subsection{Metallicity}
\label{metals}

In  high $z$ quasars, the strength of \nv\ relative to \civ\ and \heiiuv\ suggests supersolar chemical abundances \citep{hamannferland93, hamannferland99}.  Chemical abundances may be well 5 to 10 times solar \citep{dhandaetal07}, with Z $\approx$ 5$Z_{\odot}$ reputed  typical of high $z$\ quasars \citep{ferlandetal96}. The E1 sequence seems to be mainly a sequence of ionization in the sense of a steady decrease in prominence of the low-ionization BC toward Population B (\citealt{marzianietal01}, \citeyear{marzianietal10}). However, this is not to neglect that metal-enrichment also plays a role, especially for the most extreme Pop. A sources i.e., those in bin A3\ and higher \citep{sulenticetal01,sulenticetal13}. 

The lines employed in the present study come from carbon, silicon and aluminium; all these element can be significantly depleted from gas if dust grains are formed \citep[e.g., ][]{mathis90}. However, the emitting regions where our lines are produced are thought too hot to contain  significant amount of dust (a definition of BLR is right the central engine region below the dust sublimation region: e.g., \citealt{elitzur09}). In addition Si and Al are expected  to be produced under similar circumstances in the late stage of evolution of massive stars \citep[][Ch. 7]{clayton83}.  We considered three metallicity cases  (1) solar; (2) constant  solar abundance ratio Al:Si:C with $Z=5 Z_{\odot}$\ (5Z); (3) an overabundance of Si and Al with respect to carbon by a factor 3, again with $Z= 5 Z_{\odot}$\ (5ZSiAl) following \citetalias{negreteetal12}. This condition comes from the yields listed by  \citet{woosleyweaver95} from type II Supernov\ae, and is meant to represent extreme cases associated with a circumnuclear burst of star formation significantly affecting the BLR gas chemical composition \citep[e.g.,][and \citetalias{negreteetal12}]{sanietal10}.

Two additional arrays of simulations as a function of ionization parameter and density were computed assuming the metallicity  conditions (2) and (3) listed in the previous paragraph.  If   solar metallicity is simply scaled by a factor (5Z), the ratio \aliii/\siiii\ is not strongly dependent on $Z$\ since it  increases by about 40\%\ passing from $Z = 1 Z_{\odot}$\ to $Z = 5 Z_{\odot}$, for $\log $\nh $\approx 12$ and  $\log U \approx$--2. The same is true for the \siii/\siiii\  and \siiv+\oiv/\siiii\ ratios, and for the ratios involving \civ\ in case 5Z (and it should be even more so if a metallicity increase is $1 \la Z/Z_{\odot} \la 5 $).  Therefore, we do not expect that the ratios employed are  sensitive to $Z$; this is a major advantage of the method and will be verified a posteriori (Section \ref{zeff}). To gain some information on $Z$, we consider the ratio \siiv+\oiv/\civ\ that has been extensively used in investigation of metallicity for low and high-$z$\ quasars \citep[e.g.,][]{nagaoetal06} and that is fairly well correlated with \nv/\lya\ \citep{shinetal13}. {We avoid considering the ratio involving \nv\ since (1) the deblending is extremely difficult without a very careful model of \lya\ that is customarily  heavily affected by narrow absorptions; (2) using \lya\ will restrict the redshift range for which this method can be employed with our resorting to far UV space-based observations,  from  a minimum  $z \approx$ 1.4 to $z \approx 2.0$. In addition, the goal of the present method is to estimate \R, not $Z$; we only try to analyze how $Z$\  can affect \R. }  

\section{Results on the  $z \approx$ 3 Quasars}
\label{sec:results}

\subsection{The ionizing photon flux}
\label{sec:ionizingphf}

 To estimate $\log$   \nhu\  we use the {\sc cloudy} contour plots of the ratios \aliii/ \siiii,   \civ/ \siiii,  \siiv+\oiv/ \siiii, \civ/ \aliii, \aliii/ \siiv+\oiv\ and \siiv+\oiv/ \civ\   showed in Fig.  5 of \citepalias{negreteetal12}.\footnote{Note that there are regions  where the ratio values are actually undefined:  close to the high $U$\ limit ($\log U \ga -0.3$),  ratio \aliii/ \siiii\ (with \nh $\la 10^9$\cm3) should not be considered. } The data points of our objects are in regions where the ratios are well-defined. The ratio \siiv+\oiv/ \civ\ is used only to give a hint of the metallicity (see Section \ref{zeff}). The ratios \civ/ \siiii, \siiv+\oiv/ \siiii, \civ/ \aliii\ and \aliii/ \siiv+\oiv\ are mainly sensitive to the ionization parameter $U$, while \aliii/ \siiii\ and \ciii/ \siiii\ (the latter is not considered in the following as discussed in Section \ref{sec:ciii_contr})\ are mainly sensitive to density.

The diagnostic ratios were computed from the intensity of the BC of  \siiii, \aliii, \siiv\ and \civ\ reported in   Table \ref{tab:fluxes}.   We  display  on a graph  a line representing the behavior of  each ratio under the assumption of solar metallicity; the ideal point where the lines representing different diagnostic ratios cross  determines the values of $\log$\nhu. Figure \ref{fig:neu} shows the contour plots where we can see that the diagnostic ratios (except \siiv+\oiv/ \civ) converge to rather well defined values. The crossing point is very precise for the objects J00521-1108, J01225+1339, J02287+0002 (using $z_{CIII}$), J02390-0038, J03036-0023, J20497-0554 and J23509-0052; for the remaining objects  J00103-0037 and J02287+0002 (using $z_{OI}$),  the crossing point is slightly different.  

Table \ref{tab:neu} summarizes the $\log n_\mathrm{H}$ and $\log U$ values including their uncertainty. Since $U$\ and \nh\ are not independent quantities (their correlation coefficient is found to be 0.55),  we adopt the appropriate formula for the errors on the product \nhu\ \cite[following ][]{bevington69}. As discussed in \citetalias{negreteetal13}, \nh\ and $U$ \ cannot be separately estimated  unless a source is of extreme Pop. A.

\subsection{Effects of Metallicity }
\label{zeff}

The crossing point of the ratios \siiv+\oiv/ \siiii\ and \siii/ \siiii\ is in principle independent on metallicity. Therefore, any significant disagreement between this crossing point and the ratios based on \civ\  may indicate  chemical composition different from the assumed solar one (Section \ref{metals}). We note also that the \civ/\siiii\ and \civ/\aliii\ usually give results that are in perfect agreement in the plane (\nh,$U$). These findings support our assumption that, if metallicity variations are present, the relative  abundance Al to Si remains constant. 

We attempted  to isolate a \civ\ and a \siiv\ component that corresponds to the \aliii\ and \siiii\ lines. A large part of the emission  in these lines is due to a BLUE component that is emitted in physical and dynamical conditions different  from the ones of the BC emitting gas. \civ\ shows a large blueshift and is much broader than \hb, \siiii\ and \aliii\   (Fig. 2 of \citealt{marzianietal10}). In spite of this consideration, we measured the intensity of the full profile. In the case of the 1400\AA\ blend we also considered the full intensity, regardless of the line components decomposition. The 1400\AA\ feature is too complex, a blend for the \siiv\ and \oiv\ contribution to be reliably singled out, for the wide majority of sources that are not extreme Pop. A\ \citepalias{negreteetal12}. Even if two components emitted under different physical and/or dynamical conditions are responsible for the 1400\AA\ blend, the chemical composition of both components is expected to be the same  (and in this case the full \civ\ intensity was used for normalization).   In addition, the 1400\AA\ full blend intensity normalized by full \civ\ intensity has been used as a metallicity indicator in recent works \citep[e.g.,][]{nagaoetal06,shinetal13}. 

Generally speaking,  appreciable discrepancies in the crossing point of the 1400\AA/\civ\ ratio  may signal a non-solar metallicity yielding an higher and a lower 1400\AA/\civ\ ratio in the case of super solar  and sub solar metallicity respectively,  for a given ionization and density solution. However, the 1400/\civ\ ratio does not provide a good constrain of metallicity because  the uncertainty bands in the (\nh, $U$) plane are very large. 
In the case of the BAL quasars J01225+1339, and J02287+0002 if $z$ \oi\ is assumed, this ratio gives some indication of super solar metallicity. For this reason we calculated contour plots with higher metallicity values. These new plots are shown for J01225+1339, and J02287+0002 in Fig. \ref{fig:z5} in the case of $5Z_{\odot}$ and $5Z_{\odot}$SiAl metallicity.  The agreement in the intersection points for J01225+1339, remains well defined in the case of $Z = 5 Z_{\odot}$. For J02287+0002 the agreement in the intersection point becomes better for  $Z = 5 Z_{\odot}$SiAl. The $Z = 5 Z_{\odot}$SiAl case yields higher $U$\ and smaller \nh\ if emission line ratios involving \civ\ are considered. This reflects the increase in abundance of Si and Al with respect to C, and the fact  the \siii, \siiii, \aliii\  lines are emitted at  lower ionization than \civ.

 In all other cases the metallicity is not well constrained, and could be well within 1 and 5 times solar. In order to be sure however, we repeated the \nhu\ derivation for all sources in the $5Z_{\odot}$ case: there is no systematic difference  and the dispersion is also less than the estimated \nhu\ uncertainties:  $<$ \nhu ($1Z_{\odot}$) --  \nhu ($5Z_{\odot}$) $>$  $\approx$ --0.02, with rms $\approx$ 0.09. If a solar metallicity had been assumed for J02287+0002,   \nhu ($1Z_{\odot}$)  --  \nhu ($5Z_{\odot}$SiAl )    $\approx$ 0.16. Significant changes are expected only in the cases there is a selective enhancement of some elements over others involved in the computation of the diagnostic intensity ratios. The average effect would be  \nhu ($1Z_{\odot}$)  --  \nhu ($5Z_{\odot}$SiAl )    $\approx$ 0.11 with an rms of $\approx$ 0.2.  We do not report all super solar cases in Table \ref{tab:neu} since they are believed to be not correct save for J02287+0002 and J01225+1339. Numbers in bold identify the preferred $Z$ (\nh, $U$) solutions.
   
We conclude that scaling the metallicity up to $Z=5 Z_{\odot}$\  from  $Z=1 Z_{\odot}$\ yields an effect that is well below  the uncertainty associated with  the method, and particularly small if the ratios  \aliii/\siiii, \siiv+\oiv/\siiii\ and \civ/\aliii\ are considered to compute \nh\ and $U$. It is significant if strong enrichment of Al and Si over C occurs, so that the efforts should be focused on the identification of such cases. These cases are however believed   to be relatively rare ($\approx$ 10 \% at low $z$), and are easily identified since they generally satisfy the condition \aliii\ $\ga$ 0.5 \siiii, and \siiii\ $\ga$ \ciii\ \citep{marzianisulentic13}.

\section{Photoionization Computations of   Broad Line Region  Distance and   Black Hole Mass.}
\label{rblr}

The distance of the broad line region \rb\ and the black hole mass (\mbh) are key parameters that let us understand the dynamics of the gas in the emitting region and the quasar behavior and evolution. In this work we will use a  method based on the determination of \nhu\  to compute \rbp.  Eq. \ref{eq:erre} can be rewritten as

\begin{equation}
r_{\rm BLR,\Phi} = \frac 1{h^{1/2} c} (n_\mathrm{H} U)^{-1/2} \left( \int_{0}^{\lambda_{Ly}} f_\lambda \lambda d\lambda \right) ^{1/2} d_c \label{eq:rblr1}
\end{equation}
where  $d_\mathrm{c}$ is the radial comoving distance. The integral is carried out from the Lyman limit { ($\lambda_{Ly}$)} to the shortest wavelengths on the {\em rest frame} specific flux $f_{\lambda}$. For the integral we will use two Spectral Energy Distributions (SEDs): one described by \citet{mathewsferland87} and one by \citet{laoretal97a}. 

Expressing \rbp\ in units of light-days, and scaling the variables to convenient units, Eq. \ref{eq:rblr1} becomes:

\begin{equation}
r_{\rm BLR,\Phi} \approx 93 \left[ \frac{f_{\lambda_0,-15} \tilde{Q}_\mathrm{H,0.01}}{(n_{\mathrm H} U)_{10}} \right]^\frac{1}{2} 
\zeta(z, 0.3, 0.7)  ~~\mathrm{lt-day} 
\end{equation}

{ where  $f_{\lambda_0,-15}$ is the specific rest-frame flux (measured on the spectra at $\lambda_0$ =1700\AA) in units of $10^{-15} erg \, s^{-1} cm^{-2} \AA^{-1}$. The product \nhu\ is normalized to $10^{10} cm^{-3}$. $\zeta(z, 0.3, 0.7)$\ is an interpolation function of $d_\mathrm{c}$\ as a function of redshift.  $\tilde{Q}_{H,0.01} = \int_{0}^{\lambda_\mathrm{Ly}} \tilde{s}_\lambda \lambda d\lambda$ is normalized to $10^{-2}$ cm \AA. We use $\tilde{s}_\lambda$ to define the SED following \citet{mathewsferland87} and \citet{laoretal97a}. $\tilde{Q}_{H}$\ is { 0.00963} cm\AA\  in the case the continuum of \citet{laoretal97a} is considered; $\tilde{Q}_{H} \approx$0.02181 cm\AA\ for \citet{mathewsferland87}. }We use their average value, since the derived $U$\ and \nh\ are not sensitive to the two different shapes to a first approximation.\footnote{Since the \citet{laoretal97a} continuum produces a fewer ionizing photons, the same value of $U$ is obtained at a smaller distance. } { The two SEDs give a small difference in the estimated number of ionizing photons. 
}

Knowing \rbp\ we can calculate the $M_\mathrm{BH,\Phi}$ assuming virial motions of the gas
\begin{equation}
\label{eq:vir}
M_\mathrm{BH,\Phi} = f \frac{\Delta v^2 r_\mathrm{BLR,\Phi}}G.
\end{equation}
or,
\begin{equation}
M_\mathrm{BH,\Phi} = \frac 3{4G} f_{0.75} (FWHM)^2 r_{BLR,\Phi} 
\end{equation}
with the geometry term  $f_{0.75} \approx 1.0$ \  (\citealt{grahametal11}, see also \citealt{onkenetal04, wooetal10}). \citet{collinetal06}  suggest that $f$\ is significantly different for Pop.A and B sources; we do not consider here their important result for the sake of comparison with previous work (Section \ref{vp06}). 

Table  \ref{tab:r_m} reports the values of the \rbp\ and the \mbh$_{,\Phi}$ of our 8 objects  and the extreme objects in the last two rows.   Column 1 identifies the quasar name; Col. 2 gives the quasar comoving distance in mega parsecs [Mpc]; Cols. 3 and 4 are the continuum specific flux value at 1350\AA\, and 1700\AA\, respectively, Col. 5 reports the FWHM in km s$^{-1}$\ for the BCs, Col. 6 is the Population designation.   Cols. (7) and (8) report the logarithm of the \rb\ in cm obtained from: a) the values selected from Table \ref{tab:neu} (\rbp); and b) corrected by low density emission using Eq. \ref{eq:low_dens}. Cols. (9) and (10) list   the determinations  of  \mbh\ in solar masses in the same order as for \rb.  Finally Cols. (11) and (12) is \mbh\ computed  following \citet{vestergaardpeterson06}   (their Eq. \ref{eq:vestergaard}) and \citet{shenliu12} respectively. We will explain in Section \ref{sec:discussion} how these quantities are computed.

\section{Discussion} 
\label{sec:discussion}

\subsection{Previous work}
\label{sec:previous_work}

There have been several studies aimed at computing \rb\ and  \mbh.  A direct measure of \rb\ through reverberation mapping  requires an enormous amount of observational effort and has only been applied to a relatively small number of quasars:  slightly less than 60  objects with $z \la 0.4$\ \citep{kaspietal00, kaspietal05,petersonetal04, bentzetal10,bentzetal13}).  A second way to measure \rb\ uses a less direct method. \citeauthor{kaspietal00} (\citeyear{kaspietal00}, \citeyear{kaspietal05}) and \citet{bentzetal09} used reverberation mapping results to find, in an empirical way, a relation between \rb\ and the optical continuum luminosity at 5100\AA, 
\begin{equation}
\label{eq:r_l}
r_{\mathrm{BLR}} \propto L^\alpha
\end{equation}
with $\alpha \approx 0.52$.  \citet{vestergaardpeterson06} obtained a similar result for the optical continuum luminosity with an $\alpha \approx 0.50$ and for the UV continuum at 1350\AA, $\alpha \approx 0.53$. These relations have been used to compute the \rb\ not only for nearby objects, but also for  high redshift, high luminosity objects. There are other works that use single epoch spectra and the continuum at 3000\AA, obtaining an $\alpha \approx 0.47$ (McLure \& Jarvis 2002).

We can rewrite Eq. \ref{eq:vir} as 
\begin{equation}
M_{\mathrm{BH}} \propto f \frac{\mathrm{FWHM}^2 L^\alpha}{G}.
\end{equation}

\hb\ is a low ionization strong line  whose FWHM has been widely used to determine the \mbh\ for objects mainly up to $z\la 0.9$; above this limit IR spectrometers and large telescopes are needed to cover the redshifted line. For distant objects ($z\sim2$), an alternative is to use  \civ, a high ionization line emitted in the UV. However, this line should be used with caution  because  the line is often blueshifted. This  means that at least part of this line is likely emitted in an outflow \citep{netzeretal07,richardsetal11,marzianisulentic12}. Thus the estimation of  \mbh\ using FWHM(\civ) tend to be systematically higher than those using FWHM(\hb) for objects of Population A \citep{sulenticetal07}.

\subsection{Comparison with previous work}
\label{vp06}

 We compare \rbp\ obtained using our photoionization method with the ones estimated through the \rb\  -- $L$\   correlation in the upper panel of Fig. \ref{fig:MBHcomparison}. Our \rbp\ agree well with the \citet{bentzetal13} relation: the average \rb$_{,\Phi}$ -- \rb($L$) is $\approx$ 0.14 $\pm 0.10$ and -0.16 $\pm 0.13$ without and with low-density correction, respectively. 

\citet{vestergaardpeterson06} (VP06) used the relation $r_{\mathrm{BLR}} \propto L^{0.53}$ \ to obtain the following formula that relates \mbh\ to the FWHM(\civ$_{}$) and the continuum luminosity at 1350\AA:
\begin{equation}
\label{eq:vestergaard}
\log M_\mathrm{BH}(\mathrm{C\sc{\i v}}) = 0.66 + 0.53 \>  \log \left[ \frac{\lambda L_\lambda(1350 \mathrm{\AA})} {10^{44} \> \mathrm{ergs} \> \mathrm{s}^{-1}} \right] + 2  \> \log   \left[  \frac {\mathrm{FWHM}(\mathrm{C \sc{\i v}})} { \mathrm{km} \> \mathrm{s}^{-1}} \right]  - s_\mathrm{f}.
\end{equation}

More recently, \citet[][S12]{shenliu12} updated the \mbh\ -- Luminosity relation:

\begin{equation}
\label{eq:shenliu}
\log M_\mathrm{BH}(\mathrm{C\sc{\i v}}) = 7.295 + 0.471 \>  \log \left[ \frac{\lambda L_\lambda(1350 \mathrm{\AA})} {10^{44} \> \mathrm{ergs} \> \mathrm{s}^{-1}} \right] + 0.242  \> \log  \left[   \frac {\mathrm{FWHM}(\mathrm{C \sc{\i v}})} { \mathrm{km} \> \mathrm{s}^{-1}} \right] - s_\mathrm{f}.
\end{equation}

The scale factor $s_\mathrm{f} \approx -0.27$\ sets the masses to the $f$\ value obtained by \citet{grahametal11}.  In Cols. 9 to 12 of Table \ref{tab:r_m} we show our \mbh$_{,\Phi}$ results with those using Equations \ref{eq:vestergaard} and \ref{eq:shenliu}. We do not apply corrections for radiation-pressure effects that are likely relevant especially for objects radiating at large Eddington ratio \citep{netzer09,netzermarziani10}. The difference between this computation and the one reported in \citet{sulenticetal07} is that in the latter work the BLUE component was not separated from the BC of  \civ\ just to show how larger values of FWHM(\civ) yielded \mbh\ much larger than the ones derived from FWHM(\hb) in Pop. A objects. 

We compare the masses obtained using our photoionization method with those of VP06 and S12 in Fig. \ref{fig:MBHcomparison}, middle and lower panel. We use the FWHM of the BC reported in Col. 5 of Tab. \ref{tab:r_m} as an estimator of the virial line broadening.  The photoionization masses  before correction agree with the prediction of the the luminosity correlation  with   a systematic offset of 0.32 $\pm$ 0.10 with VP06 and 0.22 $\pm$ 0.08 with SL12. The \mbh\ values obtained after correction for low density gas are systematically lower. This happens because the correction lowers the \R\ and hence the \mbh. The agreement after correction is very good, with $<\log$\mbh$_{,\Phi}$ -- $\log$\mbh(VP06)$>$ $\approx$ 0.12 $\pm$0.14 and $<\log$\mbh$_{,\Phi}$ -- $\log$\mbh(SL12)$>$ $\approx$ 0.02 $\pm$0.07. No significant differences occur if the $z$ \ciii\ solution is used for J02287+0002.

Fig. \ref{fig:MBHcomparison} should be looked at with two cautions. First, the correlation is dominated by the luminosity dependence of \rb, used to compute \mbh\ in both cases. Second, the spread of \mbh\ values is small, less than one order of magnitude (and most objects have  indistinguishable masses within the errors).  Our estimated error bars are however smaller compared to the spread expected on the basis of the \rb-$L$\ correlation which is $\approx$0.33 for VP06 and $\approx$0.28 for S12 at 1$\sigma$ confidence level. The two shaded bands of Fig. \ref{fig:MBHcomparison} limit the region where we can expect to find data points on the basis of the  \rb-$L$\  correlation. 

The present results indicate that the photoionization relations can be extended and used for high redshift objects (or at least until z$\sim$3). In order to do this, we need:

\begin{itemize}
\item [$\bullet$] Spectra with S/N high enough to see the profile shape that allows decomposition of the \civ\ line, especially to separate the BLUE component from the BC;

\item [$\bullet$] to follow the methodological considerations explained in Section \ref{sec:data_analysis} .
\end{itemize}

\section{Conclusions} 
\label{sec:conclusions}

In this paper we presented new observations of eight high redshift quasars. The spectra were meant to provide high S/N, moderate resolution data on which  the \siiv, \civ, \siiii\ and \aliii\ emission line profiles could be accurately analyzed. Line profile fits permitted us to isolate a specific component whose  intensity ratios were used to derive values of the ionizing photon flux. 

These results allowed us to compute the  product \nhu\ and hence the size of the BLR and the central black hole mass. The method described in this paper rests on the assumption of photoionization as the mechanism of gas heating, on the assumption of isotropic luminosity, and on line ratios predicted by {\sc cloudy} simulations. 

 We found that the \mbh\ derived from the computed \rb\ and from the virial assumption are in good agreement with the ones derived from the luminosity-size relation.  The photoionization method explored in this paper offers an estimate of \rb\  for each quasar, with some advantages on the \rb\ valued derived from the luminosity-size correlation. The luminosity correlation suffers from large scatter and is simply extrapolated to very high luminosity without any support since there are, unfortunately, no conclusive results on reverberation of high luminosity quasars even if heroic efforts are underway \citep[e.g.][]{treveseetal07,kaspietal07,bottietal10}. 

We repeat that our \mbh\ and \rb\ results are based on the product \nhu\ and not on  values of \nh\ and of $U$\ taken separately. To apply the photoionization method in the most effective way, determining \nhu\  with the lowest uncertainty, spectral data should be of moderate resolution ($\lambda/\Delta \lambda \sim 1000$) as well as of high S/N. If the \siii\ line can be measured in an accurate way, it would be possible to derive reliable estimates of  $Z/Z_\odot$.  Instead, we used \siiv+\oiv/\civ\ to constrain metallicity.
Only for extreme Pop. A sources, when \siiii\ $\gtsim$ \ciii, it is possible to estimate the \nh, $U$\ and metallicity with very high S/N spectra. 

The present exploratory analysis emphasized several sources of uncertainty. However, the parameter needed for \rb\ and \mbh\ computation, the product \nhu, seems to be fairly stable and well-defined. Even with an error of a 0.3 in logarithm, the square root will be subject to a 0.15 uncertainty in logarithm, much lower than the uncertainty associated with the \rb\ -- luminosity correlation. The large intrinsic spread of the correlation at low luminosity, its uncertain extrapolation at very high luminosity   make preferable a one-by-one determination based on  physical properties of the emitting regions.

\acknowledgments
A. Negrete and D. Dultzin acknowledge support form grants IN111610-3 and  IN107313  PAPIIT, DGAPA UNAM.

\clearpage

\begin{deluxetable}{lccccccccccc}
\tabletypesize{\scriptsize}
\tablecaption{Basic Properties of Sources and Log of Observations.\label{tab:obs}}
\tablewidth{0pt}
\tablehead{
\colhead{Object name} &\colhead{$\rm m_B$} & \colhead{$z$} & \colhead{$\Delta_z$} & \colhead{Line} & \colhead{$M_{\mathrm B}$} &
\colhead{Flux 6cm (mJy)} & \colhead{Date} & \colhead{DIT} & \colhead{N$_{\rm exp}$} & \colhead{Airmass} & \colhead{S/N} \\
\colhead{(1)}&\colhead{(2)}&\colhead{(3)}&\colhead{(4)}&\colhead{(5)}&\colhead{(6)}&
\colhead{(7)}& \colhead{(8)}&\colhead{(9)}&\colhead{(10)}&\colhead{(11)}&\colhead{(12)}
}
\startdata
J00103-0037	&	18.39	&	3.1546	&	0.0052	&	1	&	-25.68	&	0.402	&	2006-11-08	&	1139	&	3	&	1.13	&	60	\\
J00521-1108	&	18.70	&	3.2364	&	0.0018	&	2	&	-25.39	&	0.432	&	2007-01-01	&	1199	&	3	&	1.22	&	41	\\
J01225+1339	&	18.24	&	3.0511	&	0.0011	&	1	&	-25.80	&	*	&	2006-11-08	&	1259	&	2	&	1.34	&	92	\\
J02287+0002	&	18.20	&	2.7282	&	0.0097	&	1	&	-25.72	&	0.351	&	2006-12-16	&	1259	&	2	&	1.11	&	67	\\
J02390-0038	&	18.68	&	3.0675	&	0.0083	&	1	&	-25.36	&	0.429	&	2006-11-07	&	1199	&	3	&	1.47	&	57	\\
J03036-0023	&	17.65	&	3.2319	&	0.0016	&	1	&	-26.44	&	0.339	&	2006-12-16	&	1259	&	2	&	1.13	&	88	\\
J20497-0554	&	18.29	&	3.1979	&	0.0068	&	1	&	-25.79	&	*	&	2006-11-04	&	1259	&	2	&	1.61	&	54	\\
J23509-0052	&	18.67	&	3.0305	&	0.0041	&	1	&	-25.36	&	0.411	&	2006-11-07	&	1199	&	3	&	1.12	&	62	\\
\enddata
\tablenotetext{*}{Not in FIRST}
\end{deluxetable}

\begin{deluxetable}{lcccccccccccc}
\setlength{\tabcolsep}{3pt}
\tablecaption{Line Fluxes $^{a}$ \label{tab:fluxes}}
\tabletypesize{\tiny}
\rotate
\tablewidth{0pt}
\tablehead{
 & \multicolumn{2}{c}{\ciii}  & &&  &   \multicolumn{3}{c}{\civ}  & &  \multicolumn{3}{c}{\siiv+\oiv}  \\
\cline{2-3}
\cline{7-9}
\cline{11-13} 
\colhead{Object} & \colhead{$_{BC}$} & \colhead{$_{VBC}$} & \colhead{\siiii} & \colhead{\aliii}& \colhead{\siii} &  \colhead{$_{BC}$} & \colhead{$_{BLUE}$} & \colhead{$_{VBC}$} & \colhead{} & \colhead{$_{BC}$} & \colhead{$_{BLUE}$} & \colhead{$_{VBC}$} 
}
\startdata
J00103-0037	&	5.0	$\pm$	2.1	&	1.2	$\pm$	1.4	* &	2.9	$\pm$	1.0	&	2.0	$\pm$	0.8	&	1.1	$\pm$	0.8	:	&	14.3	$\pm$	8.3	&	6.3	$\pm$	1.8	&	8.7	$\pm$	2.4	&&	4.3	$\pm$	2.4	&	0.8	$\pm$	1.5	&	0.5	$\pm$	0.9	\\
J00521-1108	&	3.1	$\pm$	0.4	&	0.0	$\pm$	0.1	*&	2.8	$\pm$	0.7	&	1.6	$\pm$	0.9	&	1.1	$\pm$	0.9		&	10.9	$\pm$	2.4	&	1.4	$\pm$	1.3	&	8.3	$\pm$	2.6	&&		\ldots		&		\ldots		&		\ldots		\\
J01225+1339	&	10.7	$\pm$	1.4	&		\ldots		&	8.3	$\pm$	1.3	&	4.4	$\pm$	2.1	&	1.0	$\pm$	1.0		&	22.7	$\pm$	6.0	&	14.2	$\pm$	1.5	&		\ldots		&&	11.2	$\pm$	2.9	&	6.7	$\pm$	1.9	&		\ldots		\\
J02287+0002 $^{(1)}$	&	6.5	$\pm$	2.0	&		\ldots		&	6.0	$\pm$	2.2	&	2.5	$\pm$	1.1	&	0.6	$\pm$	0.7		&	9.9	$\pm$	3.6	&	7.9	$\pm$	2.7	&		\ldots		&&	7.2	$\pm$	3.4	&	2.1	$\pm$	1.3	&		\ldots		\\
J02287+0002 $^{(2)}$	&	7.7	$\pm$	2.0	&		\ldots		&	4.2	$\pm$	2.2	&	1.6	$\pm$	1.1	&	0.9	$\pm$	0.7		&	17.4	$\pm$	11.2	&	2.1	$\pm$	2.7	&		\ldots		&&	7.5	$\pm$	3.4	&	0.7	$\pm$	1.3	&		\ldots		\\
J02390-0038	&	4.8	$\pm$	1.3	&	0.6	$\pm$	1.0	*&	4.7	$\pm$	1.2	&	2.2	$\pm$	0.6	&	0.8	$\pm$	0.7		&	9.8	$\pm$	2.4	&	7.6	$\pm$	1.2	&	2.0	$\pm$	1.3	&&	3.2	$\pm$	0.7	&	1.8	$\pm$	0.8	&	0.2	$\pm$	0.3	\\
J03036-0023	&	13.2	$\pm$	1.1	&		\ldots		&	11.8	$\pm$	1.2	&	5.2	$\pm$	1.5	&	1.5	$\pm$	1.2	:	&	29.5	$\pm$	3.5	&	20.6	$\pm$	3.3	&		\ldots		&&	11.1	$\pm$	1.9	&	7.4	$\pm$	2.4	&		\ldots		\\
J20497-0554	&	8.0	$\pm$	1.1	&		\ldots		&	7.4	$\pm$	0.6	&	3.0	$\pm$	1.3	&	1.5	$\pm$	1.5	:	&	18.4	$\pm$	2.2	&	9.2	$\pm$	2.2	&		\ldots		&&	6.6	$\pm$	2.5	&	2.0	$\pm$	1.1	&		\ldots		\\
J23509-0052	&	5.2	$\pm$	1.2	&		\ldots		&	4.6	$\pm$	1.6	&	1.5	$\pm$	0.5	&	0.4	$\pm$	0.4		&	9.2	$\pm$	1.2	&	7.8	$\pm$	2.4	&		\ldots		&&	3.6	$\pm$	1.5	&	2.9	$\pm$	1.3	&		\ldots		\\
\enddata
\tablecomments{(a) Units are $10^{-14}$ ergs s$^{-1}$ cm$^{-2}$ \AA$^{-1}$.  (1) Considering $z_{OI}$. (2) Considering $z_{CIII]}$. (:) \siii\ approximated values due the line is affected by telluric absorptions (see Fig. \ref{fig:sample}). We do not measure \siiv\ for J00521-1108 because they have low S/N. (*) Consistent with no emission.}
\end{deluxetable}

\begin{deluxetable}{lcccc}
\tablecaption{Weak lines around \civ. $^{a}$ \label{tab:weak}}
\tabletypesize{\scriptsize}
\tablewidth{0pt}
\tablehead{
\colhead{Object} & \colhead{\niv} & \colhead{\siiiuv} &  \multicolumn{2}{c} {\heiiuv}\\
\cline{4-5}
&  &  &  \colhead{$_{BC}$} & \colhead{$_{BLUE}$} 
}
\startdata
J00103-0037	&	2.4	$\pm$	1.9	&	1.1	$\pm$	0.8	&	1.3	$\pm$	0.4	&	2.0	$\pm$	1.8	\\
J00521-1108	&	0.1	$\pm$	0.2	&	1.1	$\pm$	1.1	&	1.6	$\pm$	1.2	&	0.1	$\pm$	0.2	\\
J01225+1339	&		\ldots		&	1.0	$\pm$	1.8	&	3.2	$\pm$	3.4	&	4.3	$\pm$	1.8	\\
J02287+0002 $^{(1)}$	&		\ldots		&		\ldots		&	0.2	$\pm$	0.3	&	0.2	$\pm$	0.2	\\
J02287+0002 $^{(2)}$	&		\ldots		&		\ldots		&	0.6	$\pm$	0.3	&	0.4	$\pm$	0.2	\\
J02390-0038	&	0.6	$\pm$	0.7	&	0.3	$\pm$	1.0	&	0.0	$\pm$	0.4	&	1.8	$\pm$	0.9	\\
J03036-0023	&		\ldots		&	1.5	$\pm$	1.2	&	0.5	$\pm$	0.5	&	9.2	$\pm$	3.4	\\
J20497-0554	&	0.3	$\pm$	0.6	&	1.5	$\pm$	1.5	&	1.8	$\pm$	0.8	&	4.6	$\pm$	2.7	\\
J23509-0052	&		\ldots		&	0.4	$\pm$	0.4	&	0.2	$\pm$	0.9	&	2.5	$\pm$	1.6	\\
\enddata
\tablecomments{(a) Units are $10^{-14}$ ergs s$^{-1}$ cm$^{-2}$ \AA$^{-1}$. (1) Considering $z_{OI}$. (2) Considering $z_{CIII]}$. We do not show \heiiuv$_{VBC}$ because is very weak, when is considered.}
\end{deluxetable}

\begin{deluxetable}{lccccccccc}
\rotate
\tablecaption{Equivalent Widths.\label{tab:ew} }
\tabletypesize{\tiny}
\tablewidth{0pt}
\tablehead{
\colhead{Object} & \colhead{\ciii$_{BC}$} & \colhead{\ciii$_{Tot}$} & \colhead{\siiii} & \colhead{\aliii}& \colhead{\siii} &  \colhead{\civ$_{BC}$} &\colhead{\civ$_{Tot}$} &  \colhead{(\siiv+\oiv)$_{BC}$} &\colhead{(\siiv+\oiv)$_{Tot}$} 
}
\startdata
J00103-0037	&	13.7	$\pm$	5.8	&	17.1	$\pm$	7.2	&	7.9	$\pm$	3.0	&	5.3	$\pm$	2.8	&	2.9	$\pm$	2.5	:	&	29.0	$\pm$	16.1	&	59.6	$\pm$	18.0	&	7.4	$\pm$	4.3	&	9.7	$\pm$	6.7	\\
J00521-1108	&	8.2	$\pm$	1.5	&	8.2	$\pm$	2.4	&	7.3	$\pm$	2.5	&	4.0	$\pm$	2.6	&	2.6	$\pm$	2.1		&	21.0	$\pm$	12.4	&	40.1	$\pm$	14.9	&		\ldots		&		\ldots		\\
J01225+1339	&	15.3	$\pm$	3.5	&		\ldots		&	11.6	$\pm$	3.1	&	6.0	$\pm$	3.6	&	1.4	$\pm$	1.6		&	25.5	$\pm$	9.1	&	41.6	$\pm$	9.7	&	11.1	$\pm$	3.9	&	17.7	$\pm$	4.7	\\
J02287+0002 $^{(1)}$	&	19.3	$\pm$	6.7	&		\ldots		&	17.8	$\pm$	5.7	&	7.4	$\pm$	3.8	&	1.8	$\pm$	2.2		&	26.6	$\pm$	6.7	&	48.0	$\pm$	20.5	&	18.8	$\pm$	6.0	&	24.3	$\pm$	6.7	\\
J02287+0002 $^{(2)}$	&	22.8	$\pm$	6.7	&		\ldots		&	12.3	$\pm$	5.7	&	4.7	$\pm$	3.8	&	2.6	$\pm$	2.2		&	48.2	$\pm$	9.6	&	54.0	$\pm$	24.6	&	18.8	$\pm$	6.0	&	20.7	$\pm$	6.7	\\
J02390-0038	&	14.5	$\pm$	4.1	&	16.5	$\pm$	5.3	&	13.9	$\pm$	3.7	&	6.4	$\pm$	2.2	&	2.1	$\pm$	1.7		&	21.3	$\pm$	5.6	&	42.2	$\pm$	7.3	&	6.0	$\pm$	1.5	&	9.7	$\pm$	3.7	\\
J03036-0023	&	12.4	$\pm$	1.8	&		\ldots		&	10.9	$\pm$	1.8	&	4.6	$\pm$	1.6	&	1.3	$\pm$	1.1	:	&	19.7	$\pm$	3.7	&	33.3	$\pm$	4.8	&	6.4	$\pm$	1.5	&	10.7	$\pm$	2.9	\\
J20497-0554	&	15.5	$\pm$	3.5	&		\ldots		&	14.1	$\pm$	2.3	&	5.6	$\pm$	2.9	&	2.7	$\pm$	2.5	:	&	25.4	$\pm$	5.6	&	38.1	$\pm$	7.1	&	8.0	$\pm$	3.7	&	10.3	$\pm$	4.0	\\
J23509-0052	&	15.8	$\pm$	4.0	&		\ldots		&	13.7	$\pm$	5.3	&	4.4	$\pm$	1.8	&	1.1	$\pm$	0.9		&	22.0	$\pm$	4.2	&	40.5	$\pm$	10.6	&	7.7	$\pm$	3.7	&	14.0	$\pm$	5.1	\\
\enddata
\tablecomments{(1) Considering $z_{OI}$. (2) Considering $z_{CIII]}$. (:) \siii\ approximated values due the line is affected by telluric absorptions.}
\end{deluxetable}

\clearpage

\begin{deluxetable}{lcccccccccccccc}
\rotate
\setlength{\tabcolsep}{1.5pt}
\tablecaption{Hydrogen Density and Ionization Parameter.\label{tab:neu} }
\tabletypesize{\scriptsize}
\tablewidth{0pt}
\tablehead{
\colhead{Object} & 
\multicolumn{3}{c}{Log \nh} & &
\multicolumn{3}{c}{Log $U$} & &
\multicolumn{3}{c}{Log \nhu} \\
\cline{2-4} \cline{6-8} \cline{10-12}
\colhead{ } & 
\colhead{$1Z_\odot$}& \colhead{$5Z_\odot$} & \colhead{$5Z_\odot$ SiAl} &\colhead{ } &
\colhead{$1Z_\odot$}& \colhead{$5Z_\odot$} & \colhead{$5Z_\odot$ SiAl} &\colhead{ } &
\colhead{$1Z_\odot$}& \colhead{$5Z_\odot$} & \colhead{$5Z_\odot$ SiAl}
}
\startdata
J00103-0037	&{\bf 	12.50	$\pm$	0.17	}&	\ldots	&	\ldots	&&{\bf 	-2.79	$\pm$	0.19	}&	\ldots	&	\ldots	&&{\bf 	9.71	$\pm$	0.22	}&	\ldots	&	\ldots	\\
J00521-1108	&{\bf 	12.40	$\pm$	0.26	}&	\ldots	&	\ldots	&&{\bf 	-2.80	$\pm$	0.15	}&	\ldots	&	\ldots	&&{\bf 	9.60	$\pm$	0.26	}&	\ldots	&	\ldots	\\
J01225+1339	&	12.43	$\pm$	0.22	&{\bf	11.99	$\pm$	0.31	}&	11.58	$\pm$	0.28	&&	-2.93	$\pm$	0.09	&{\bf 	-2.44	$\pm$	0.39	}&	-2.04	$\pm$	0.11	&&	9.51	$\pm$	0.21	&{\bf 	9.55	$\pm$	0.43	}&	9.54	$\pm$	0.27	\\
J02287+0002 $^{(1)}$	&	12.32	$\pm$	0.15	&	11.88	$\pm$	0.23	&{\bf	11.33	$\pm$	0.38	}&&	-2.96	$\pm$	0.24	&	-2.55	$\pm$	0.40	&{\bf	-2.13	$\pm$	0.22	}&&	9.36	$\pm$	0.24	&	9.33	$\pm$	0.40	&{\bf	9.21	$\pm$	0.38	}\\
J02287+0002 $^{(2)}$	&{\bf 	12.10	$\pm$	0.32	}&	\ldots	&	\ldots	&&{\bf 	-2.57	$\pm$	0.28	}&	-\ldots	&	\ldots	&&{\bf 	9.53	$\pm$	0.36	}&	\ldots	&	\ldots	\\
J02390-0038	&{\bf 	12.47	$\pm$	0.12	}&	\ldots	&	\ldots	&&{\bf 	-3.10	$\pm$	0.05	}& 	\ldots	&	\ldots	&&{\bf 	9.38	$\pm$	0.11	}&	\ldots	&	\ldots	\\
J03036-0023	&{\bf 	12.34	$\pm$	0.14	}&	\ldots	&	\ldots	&&{\bf 	-2.96	$\pm$	0.06	}&	\ldots	&	\ldots	&&{\bf 	9.39	$\pm$	0.14	}&	\ldots	&	\ldots	\\
J20497-0554	&{\bf 	12.28	$\pm$	0.25	}&	\ldots	&	\ldots	&&{\bf 	-2.93	$\pm$	0.12	}&	\ldots	&	\ldots	&&{\bf 	9.35	$\pm$	0.25	}&	\ldots	&	\ldots	\\
J23509-0052	&{\bf 	12.15	$\pm$	0.24	}&	\ldots	&	\ldots	&&{\bf 	-2.91	$\pm$	0.09	}&	\ldots	&	\ldots	&&{\bf 	9.24	$\pm$	0.24	}&	\ldots	&	\ldots	\\
\enddata
\tablecomments{(1) Considering $z_{OI}$. (2) Considering $z_{CIII]}$. We show in bold numbers the ones that we consider the best.}
\end{deluxetable}

\begin{deluxetable}{lcccccccccccccccc}
\tabletypesize{\scriptsize}
\rotate
\setlength{\tabcolsep}{1.8pt}
\tablecaption{The Size of the Broad Line Region and the Black Hole Masses.\label{tab:r_m}}
\tablewidth{0pt}
\tablehead{
\colhead{Object} & \colhead{$d_C$} & \colhead{f(1700\AA)$^{\mathrm{a}}$}& \colhead{f(1350\AA)$^{\mathrm{a}}$} & \colhead{FWHM$_{BC}$} &\colhead{Pop.} &  \multicolumn{2}{c} {Log($r_{BLR}$) [cm]} & & \multicolumn{4}{c}{Log($M_{BH}$) [$M_{\odot}$]$^{\mathrm{b}}$}\\
\cline{7-8}
\cline{10-13}
\colhead{}  & 
\colhead{$[Mpc]$} & 
\colhead{$X10^{-15}$} & 
\colhead{$X10^{-15}$} & 
\colhead{[km s$^{-1}$]}&
\colhead{} & 
\colhead{$\phi$*} &
\colhead{low dens*} &
\colhead{} &
\colhead{$\phi$*} &
\colhead{low dens*} &
\colhead{VP06$^{\mathrm{b}}$}&
\colhead{S12$^{\mathrm{c}}$}\\
\colhead{(1)}&\colhead{(2)}&\colhead{(3)}&\colhead{(4)}&\colhead{(5)}&\colhead{(6)}&\colhead{(7)}&\colhead{(8)}&&\colhead{(9)}&\colhead{(10)}&\colhead{(11)}&\colhead{(12)}
}
\startdata
J00103-0037	&	6510	&	4.8	$\pm$	1.0	&	6.8	$\pm$	1.4	&	4500	$\pm$	1200	&	B	&	18.10	$\pm$	0.12	&	17.90	$\pm$	0.14	&&	9.29	$\pm$	0.26	&	9.08	$\pm$	0.27	&	9.11	&	9.17	\\
J00521-1108	&	6586	&	6.1	$\pm$	1.5	&	8.8	$\pm$	2.1	&	5300	$\pm$	1600	&	B	&	18.21	$\pm$	0.14	&	17.94	$\pm$	0.25	&&	9.54	$\pm$	0.30	&	9.27	$\pm$	0.36	&	9.32	&	9.24	\\
J01225+1339	&	6415	&	8.1	$\pm$	1.6	&	10.9	$\pm$	2.2	&	4400	$\pm$	1000	&	A$^{\dagger}$	&	18.29	$\pm$	0.15	&	18.08	$\pm$	0.25	&&	9.45	$\pm$	0.25	&	9.24	$\pm$	0.32	&	9.19	&	9.25	\\
J02287+0002 $^{(1)}$	&	6091	&	7.4	$\pm$	2.7	&	8.4	$\pm$	3.0	&	3900	$\pm$	1200	&	A	&	18.42	$\pm$	0.21	&	18.22	$\pm$	0.21	&&	9.48	$\pm$	0.34	&	9.28	$\pm$	0.34	&	9.00	&	9.17	\\
J02287+0002 $^{(2)}$	&	6091	&	7.4	$\pm$	2.7	&	8.4	$\pm$	3.0	&	4100	$\pm$	1200	&	A$^{\dagger}$	&	18.26	$\pm$	0.20	&	18.19	$\pm$	0.21	&&	9.36	$\pm$	0.32	&	9.29	$\pm$	0.33	&	9.05	&	9.17	\\
J02390-0038	&	6448	&	6.9	$\pm$	2.1	&	9.9	$\pm$	3.0	&	4600	$\pm$	1000	&	A$^{\dagger}$	&	18.34	$\pm$	0.16	&	18.10	$\pm$	0.16	&&	9.55	$\pm$	0.25	&	9.30	$\pm$	0.25	&	9.21	&	9.24	\\
J03036-0023	&	6582	&	20.5	$\pm$	5.7	&	30.0	$\pm$	8.4	&	3700	$\pm$	600	&	A	&	18.58	$\pm$	0.10	&	18.39	$\pm$	0.15	&&	9.60	$\pm$	0.17	&	9.40	$\pm$	0.21	&	9.29	&	9.45	\\
J20497-0554	&	6552	&	6.6	$\pm$	1.3	&	9.5	$\pm$	1.9	&	3800	$\pm$	600	&	A	&	18.35	$\pm$	0.13	&	18.17	$\pm$	0.22	&&	9.39	$\pm$	0.19	&	9.21	$\pm$	0.26	&	9.04	&	9.22	\\
J23509-0052	&	6398	&	4.8	$\pm$	1.0	&	6.1	$\pm$	1.2	&	3600	$\pm$	800	&	A	&	18.33	$\pm$	0.19	&	18.22	$\pm$	0.23	&&	9.32	$\pm$	0.27	&	9.21	$\pm$	0.30	&	8.88	&	9.11	\\
\enddata
\tablecomments{(a) Units of the flux at 1350 and 1700\AA\, are in ergs s$^{-1}$ cm$^{-2}$ \AA$^{-1}$.  We show the comparison between our computations and those using:  (b) the \citet{vestergaardpeterson06} method; and (c) the \citet{shenliu12} work. They report an uncertainty of 0.66 and 0.54 dex at 2 sigma confidence respectively. ($^\dagger$) According to the FWHM it is classified as Pop. B, but has other spectral characteristics of pop. A objects. See Section \ref{sec:component_analysis}. (*) The listed  values computed assuming an average SED of \citet{laoretal97a} and \citet{mathewsferland87} with a scatter $\pm$ 0.17 dex. We used the bold numbers of Table \ref{tab:neu} to compute \rbp. (1) Considering $z_{OI}$. (2) Considering $z_{CIII]}$}
\end{deluxetable}

\begin{figure}
\epsscale{0.6}
\includegraphics[scale=0.35]{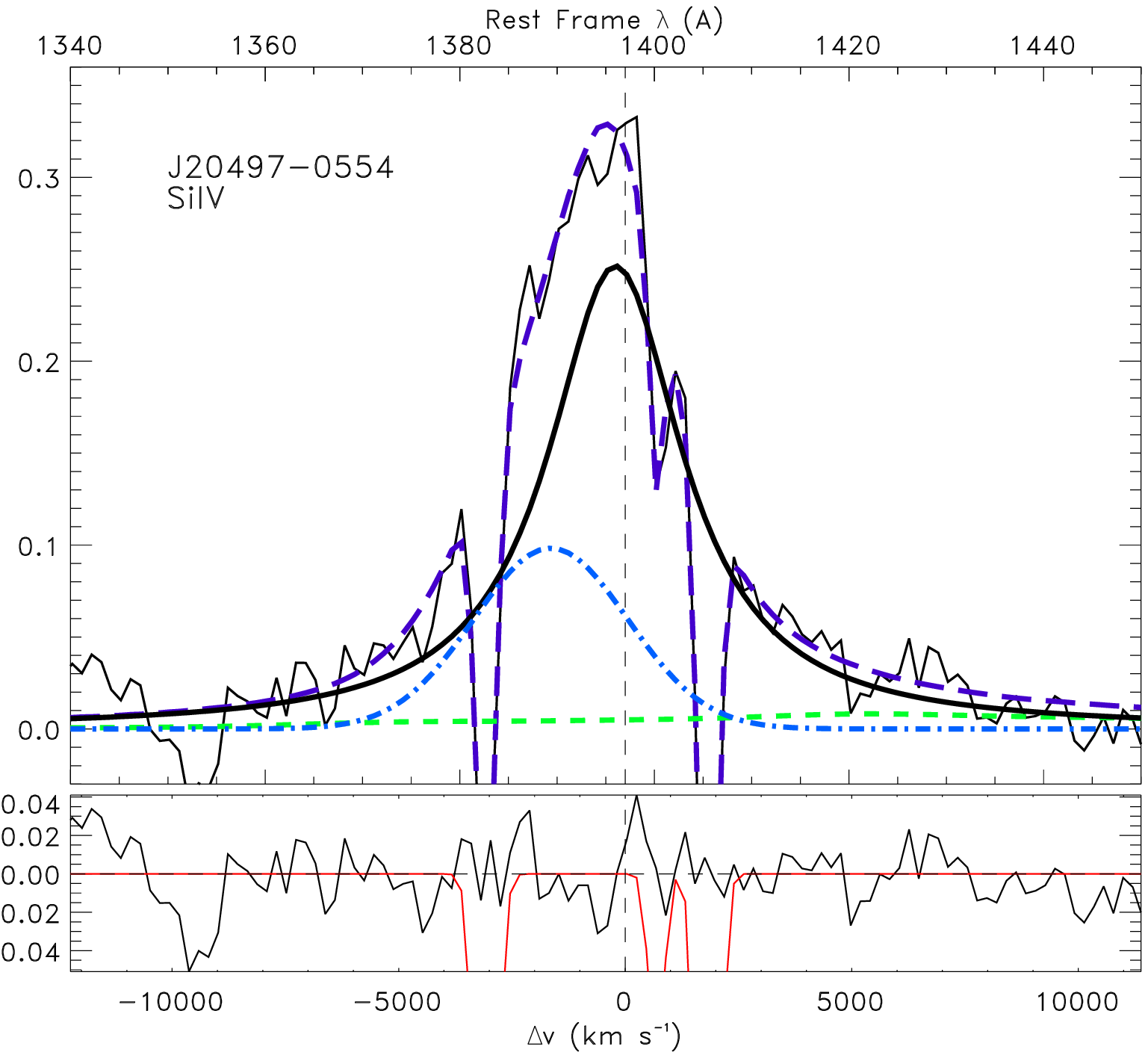}\includegraphics[scale=0.35]{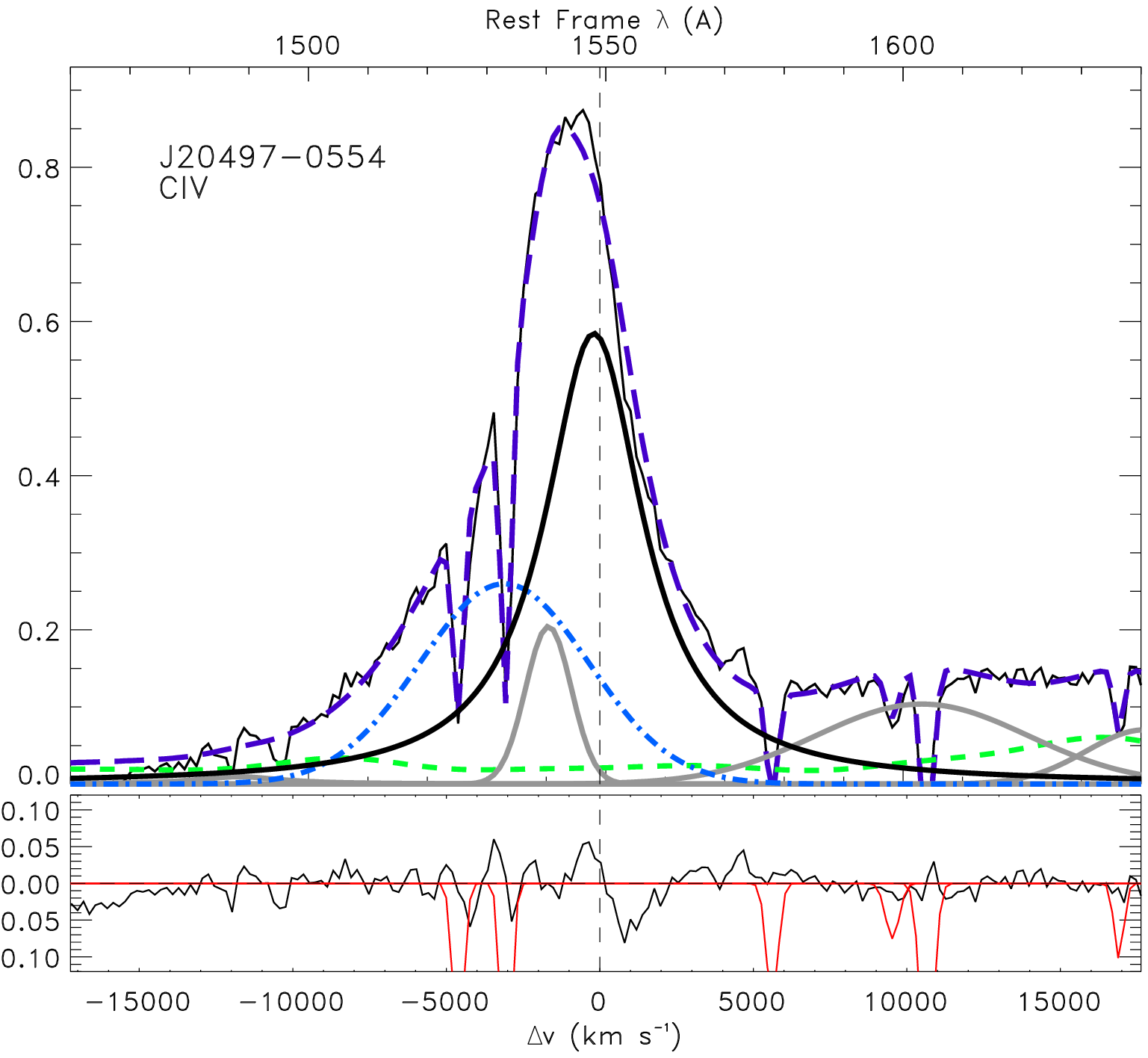}\includegraphics[scale=0.35]{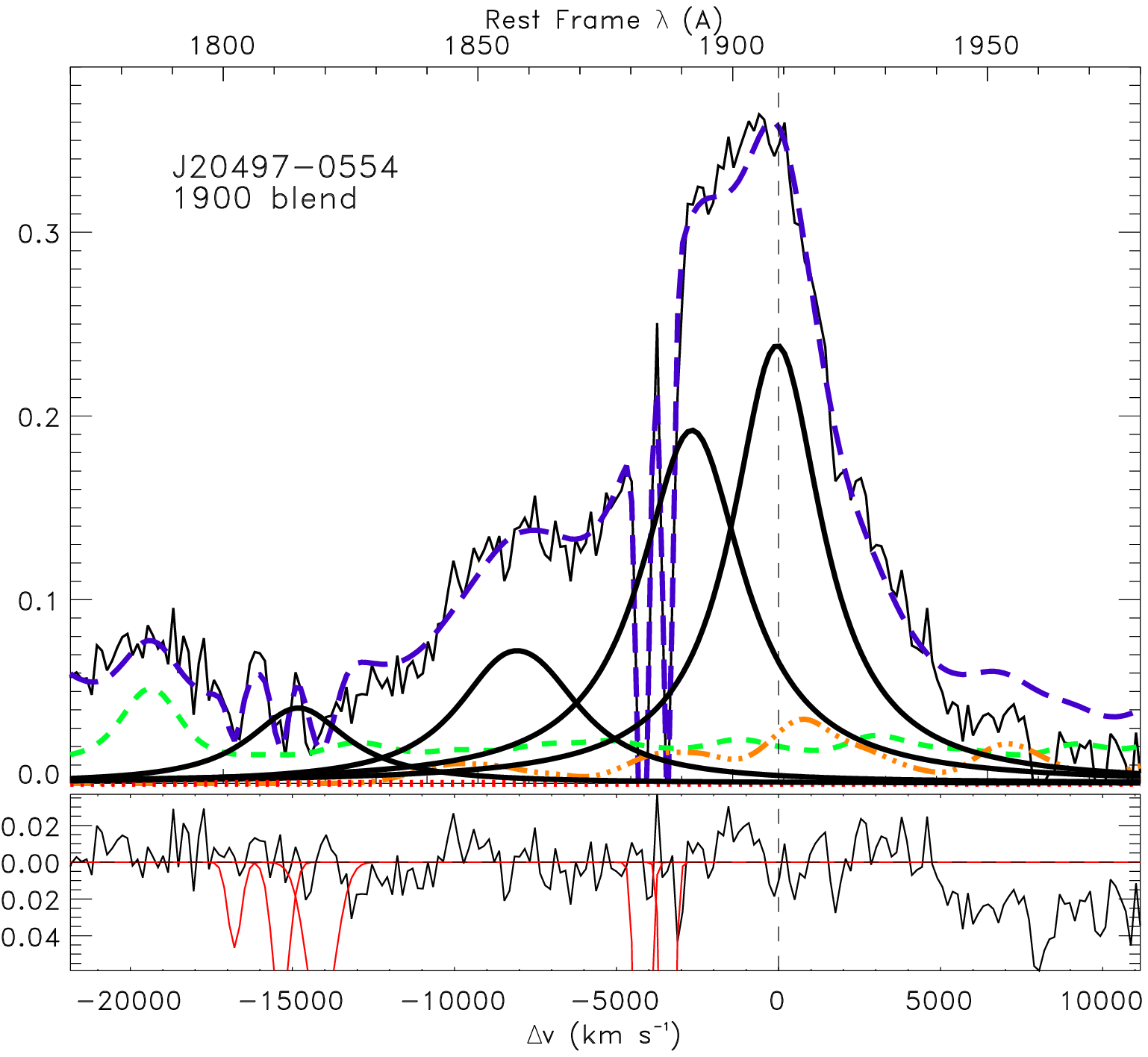}\\
\includegraphics[scale=0.35]{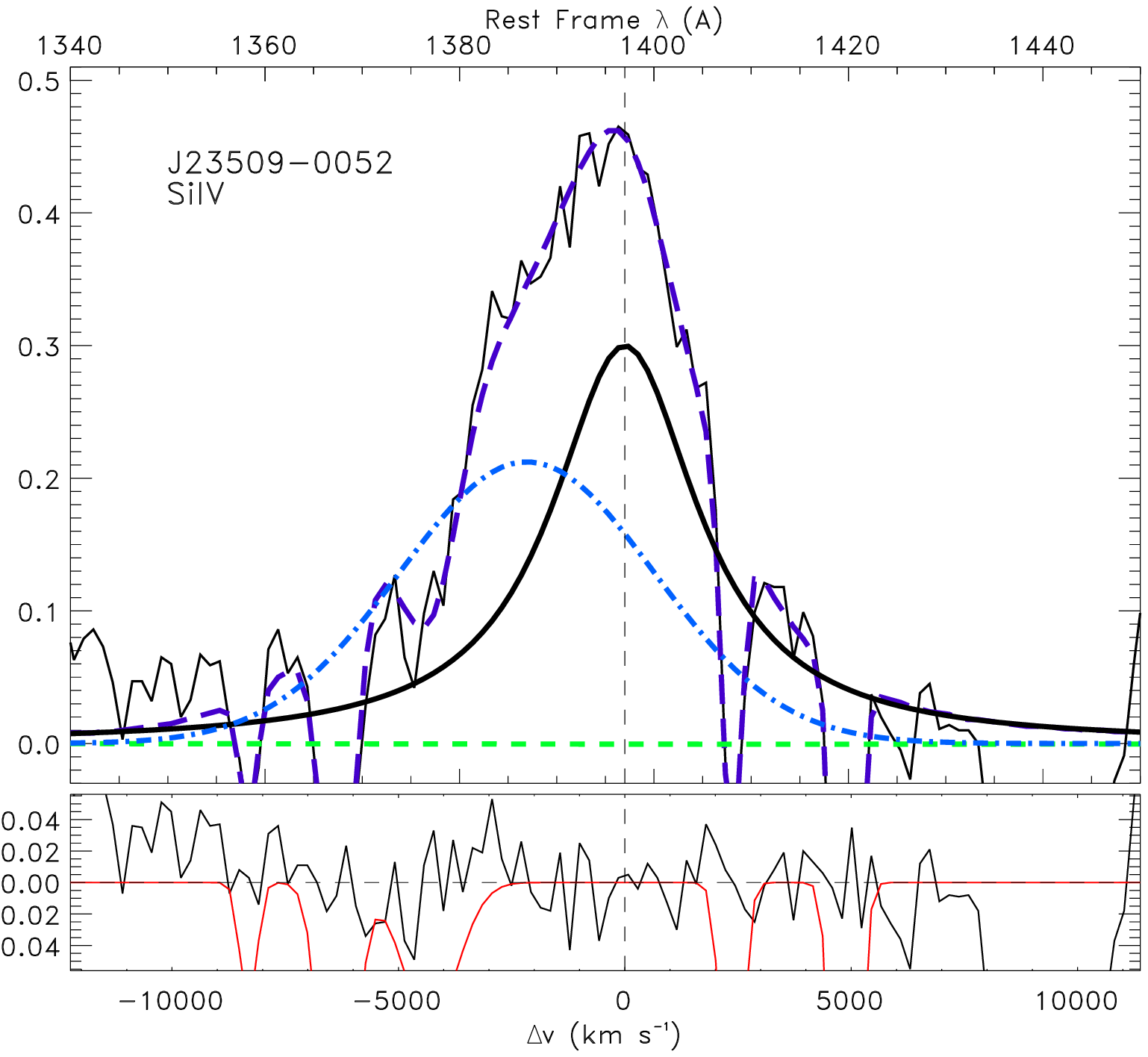}\includegraphics[scale=0.35]{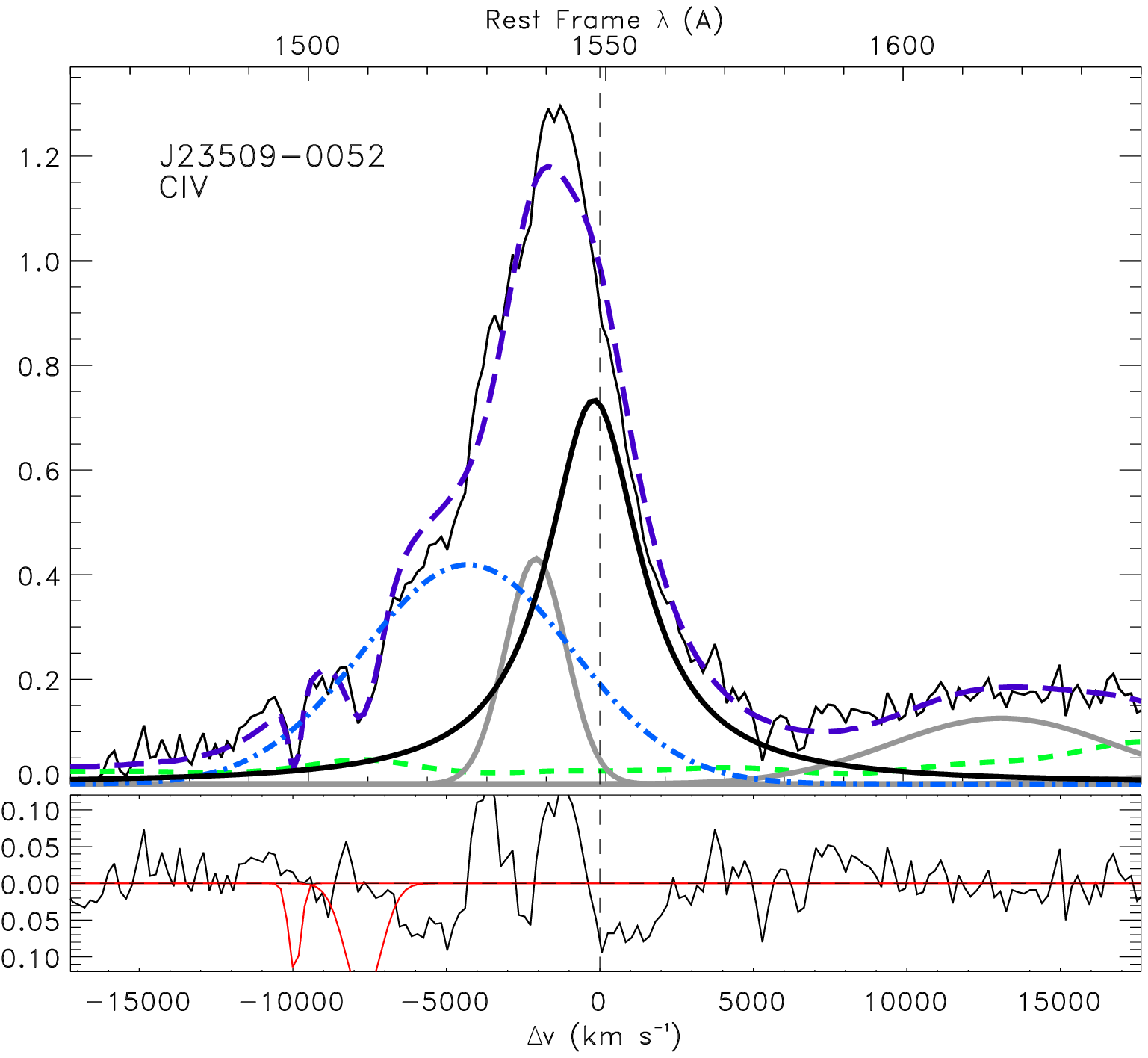}\includegraphics[scale=0.35]{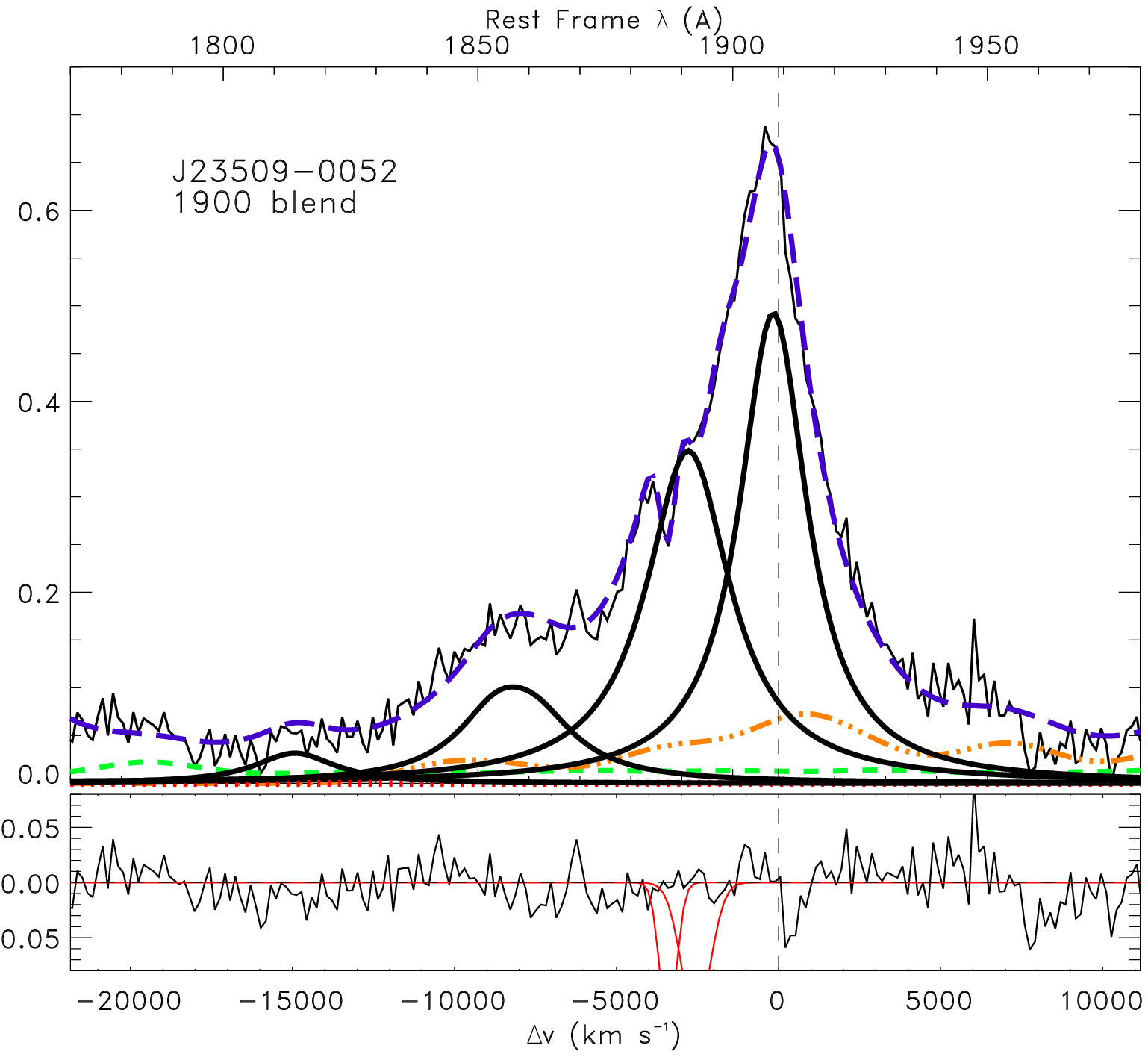}\\
{\footnotesize Fig. 2 -- Cont.\par}
\end{figure}

\begin{figure}
\includegraphics[scale=0.35]{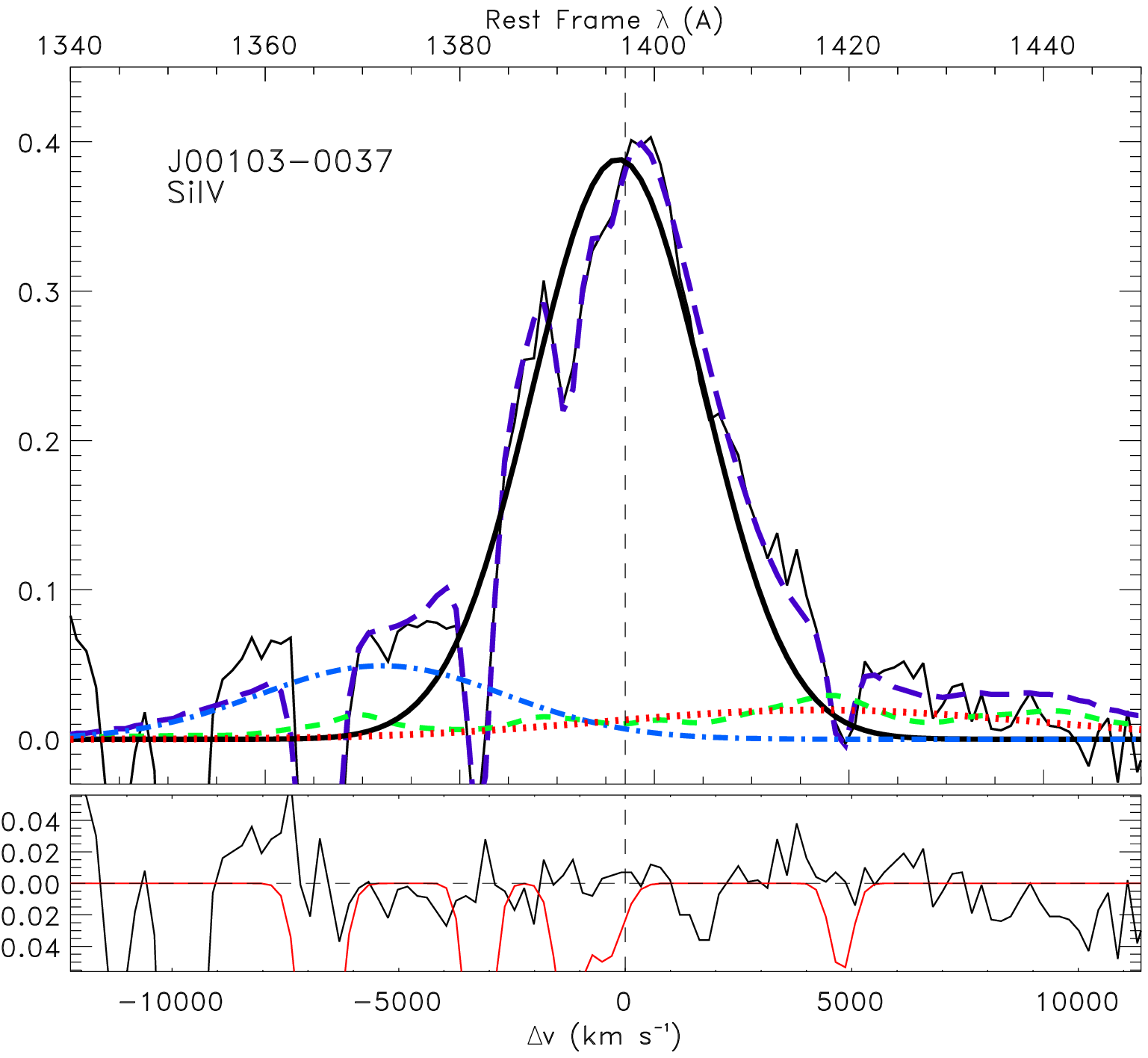}\includegraphics[scale=0.35]{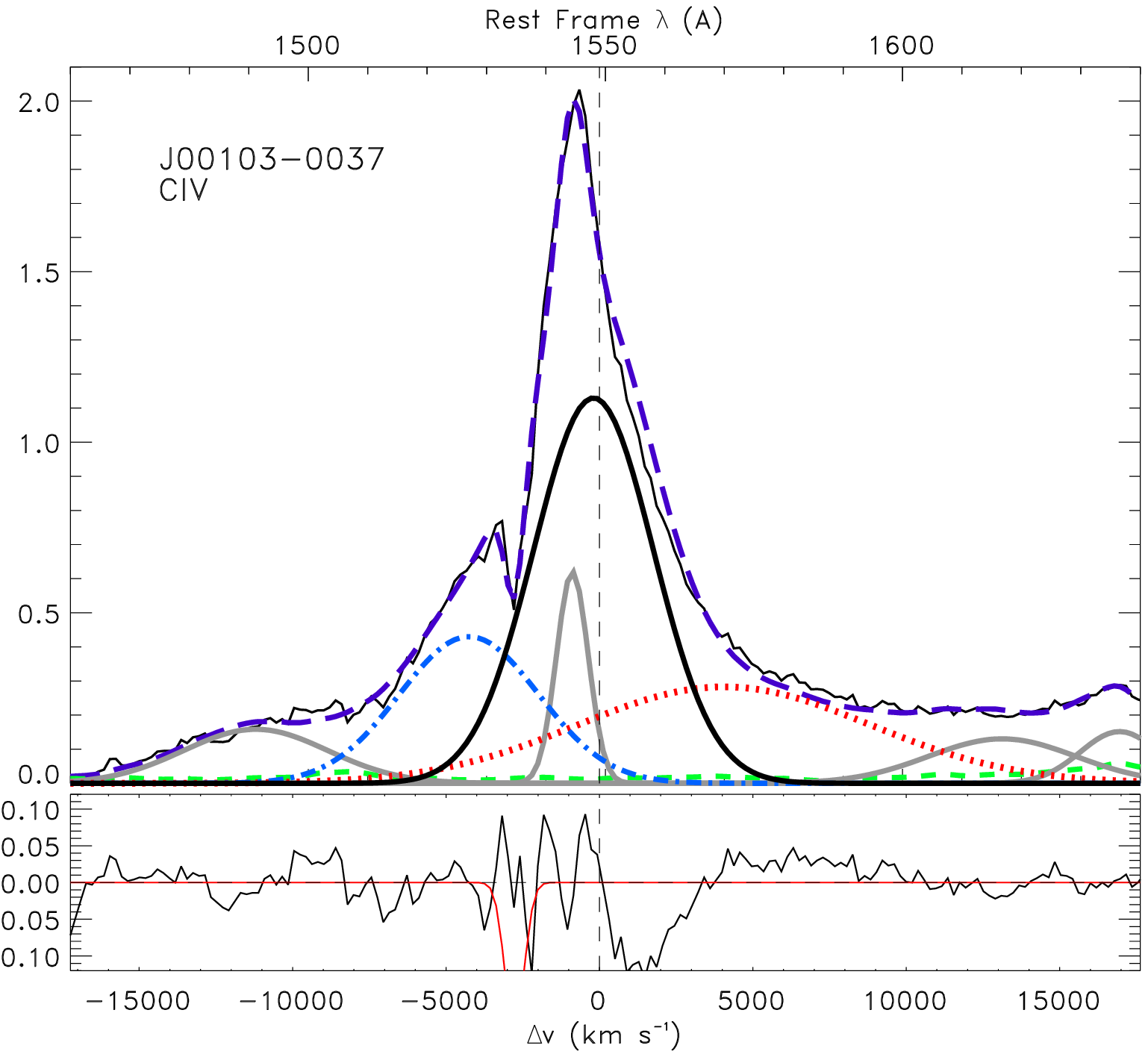}\includegraphics[scale=0.35]{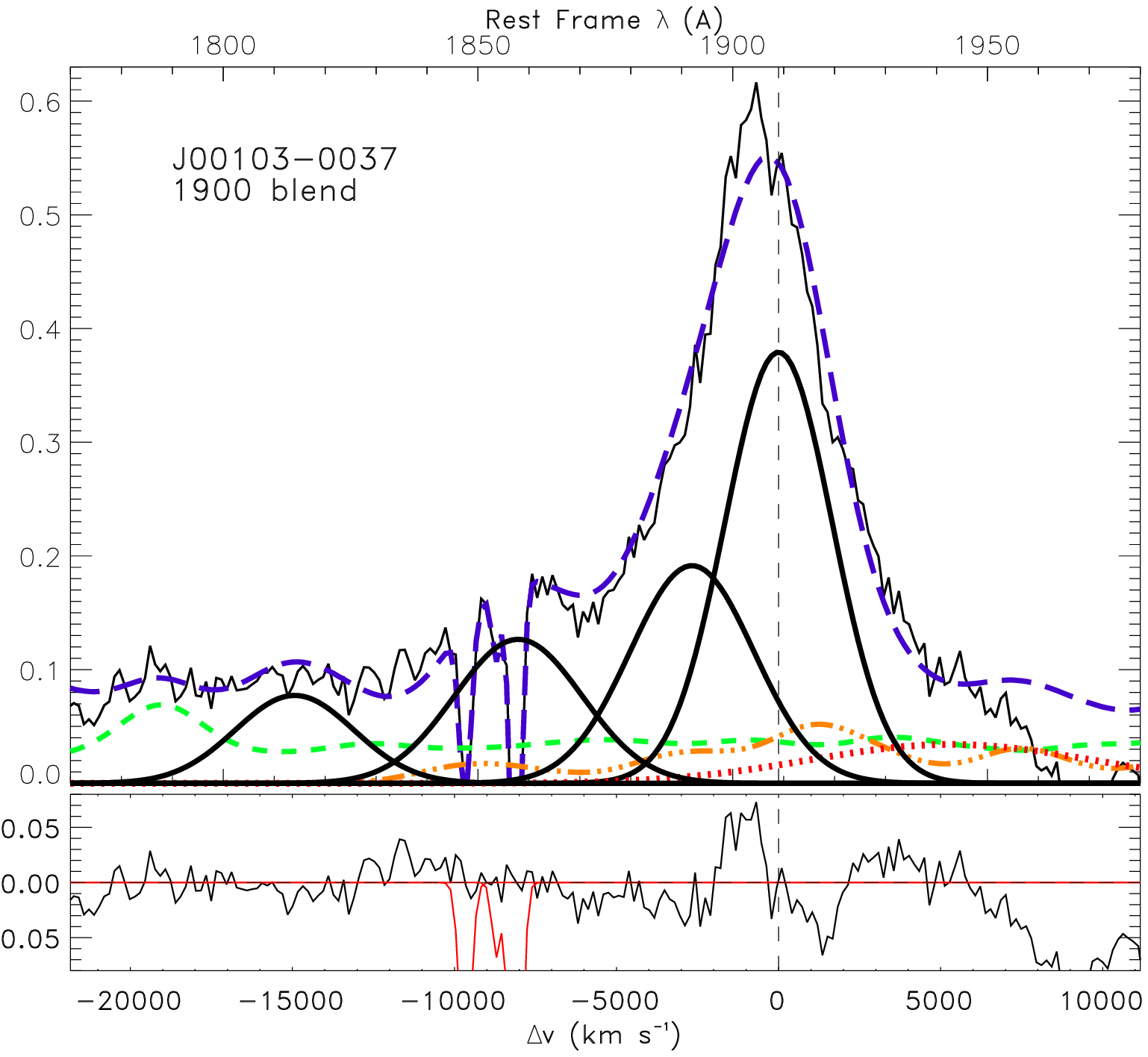}\\
\includegraphics[scale=0.35]{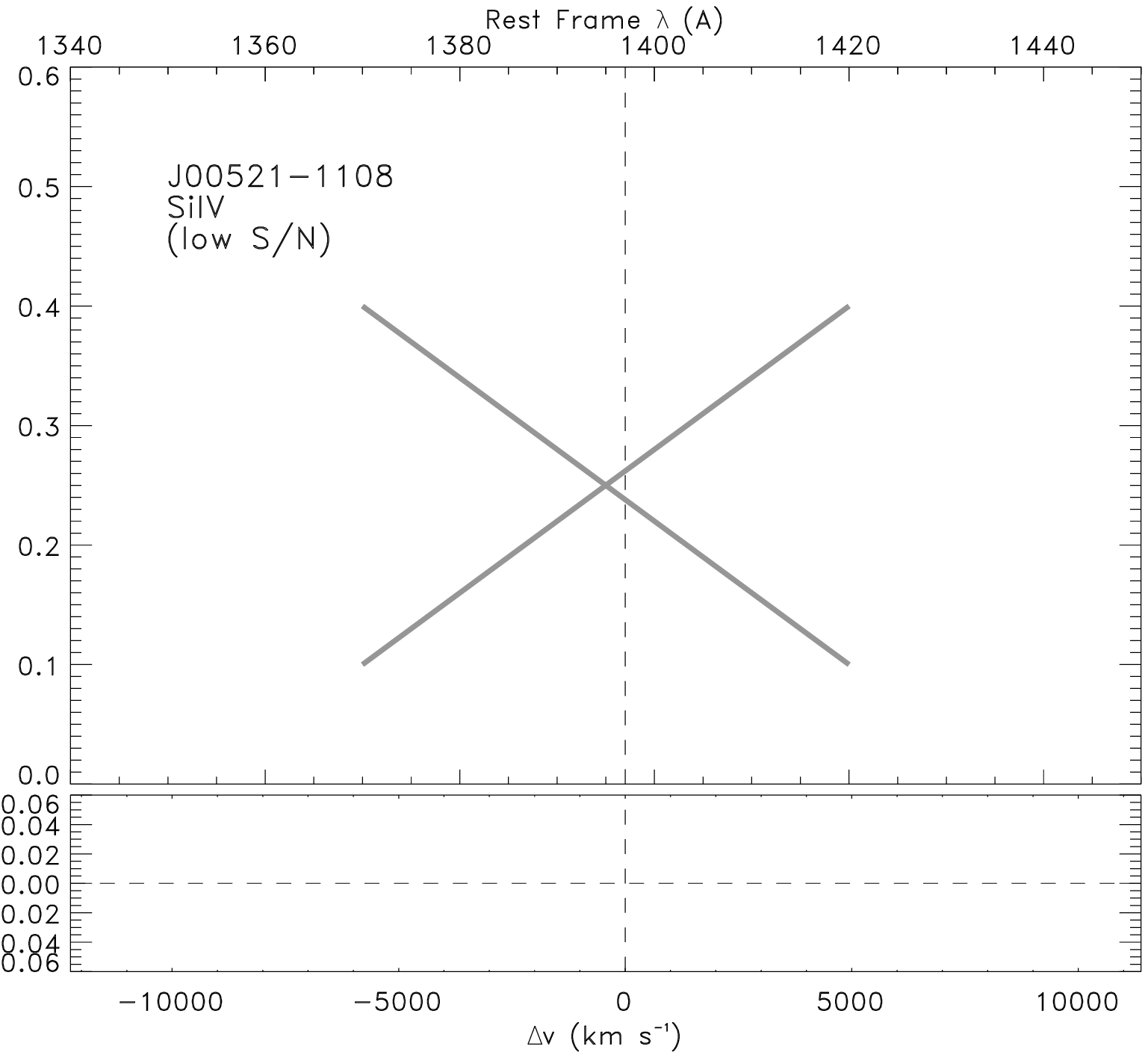}\includegraphics[scale=0.35]{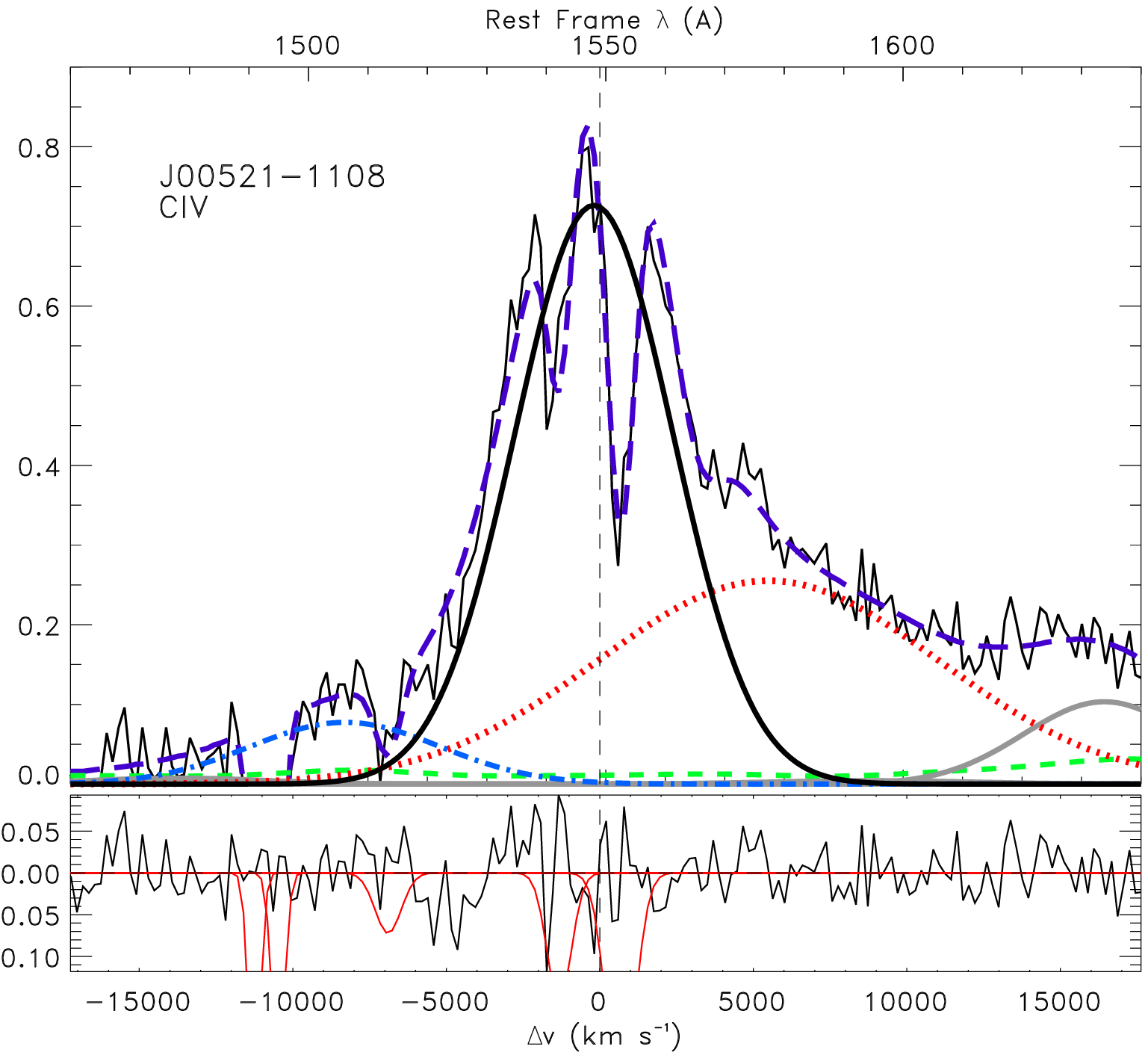}\includegraphics[scale=0.35]{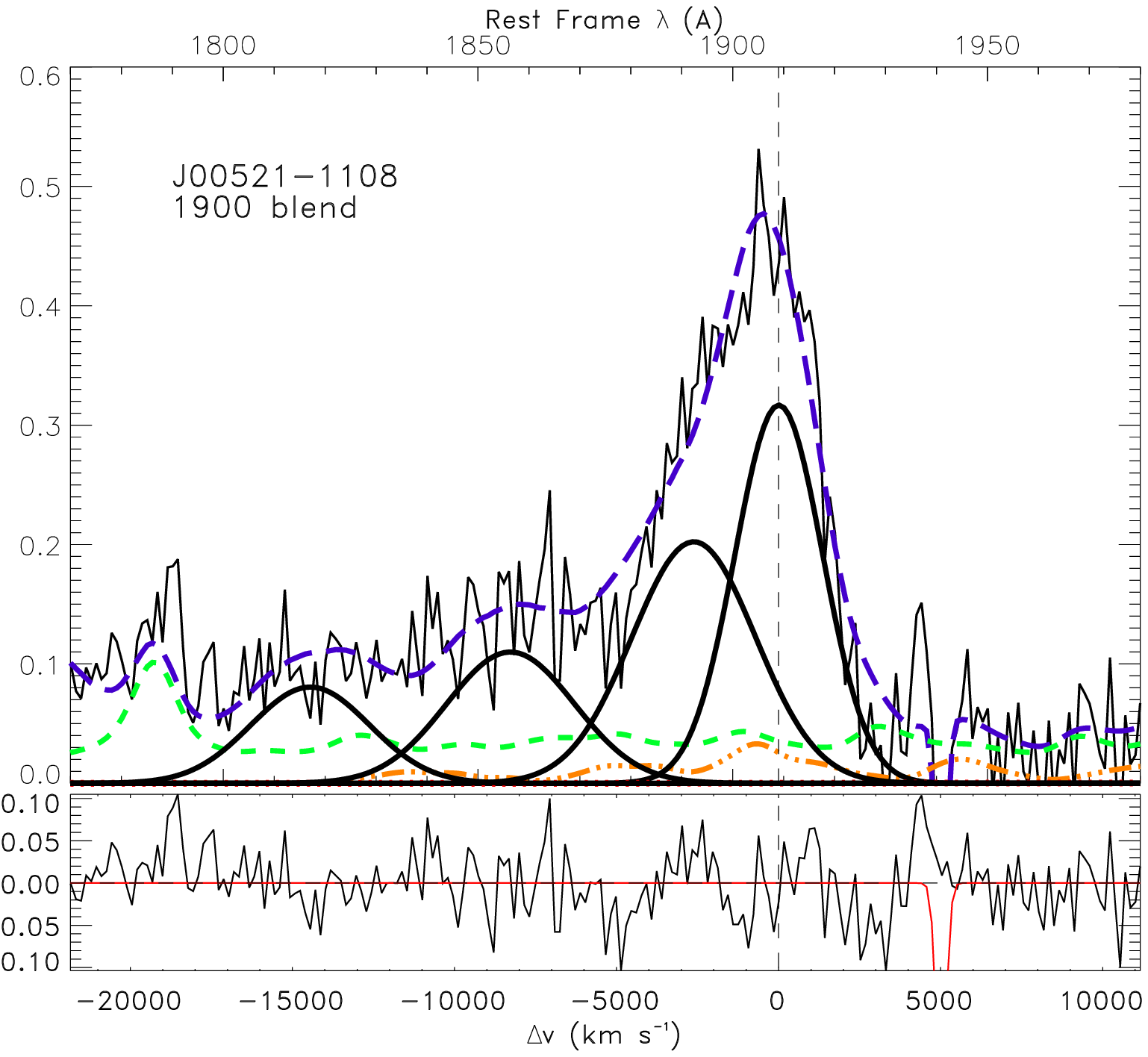}\\
\caption{Fits for Pop. B objects: up J00103-0037 and low J00521-1108. Units and meaning of symbols are the same of Fig. \ref{fig:fitsA}. \label{fig:fitsB}}
\end{figure}

\clearpage

\begin{figure}
\includegraphics[scale=0.35]{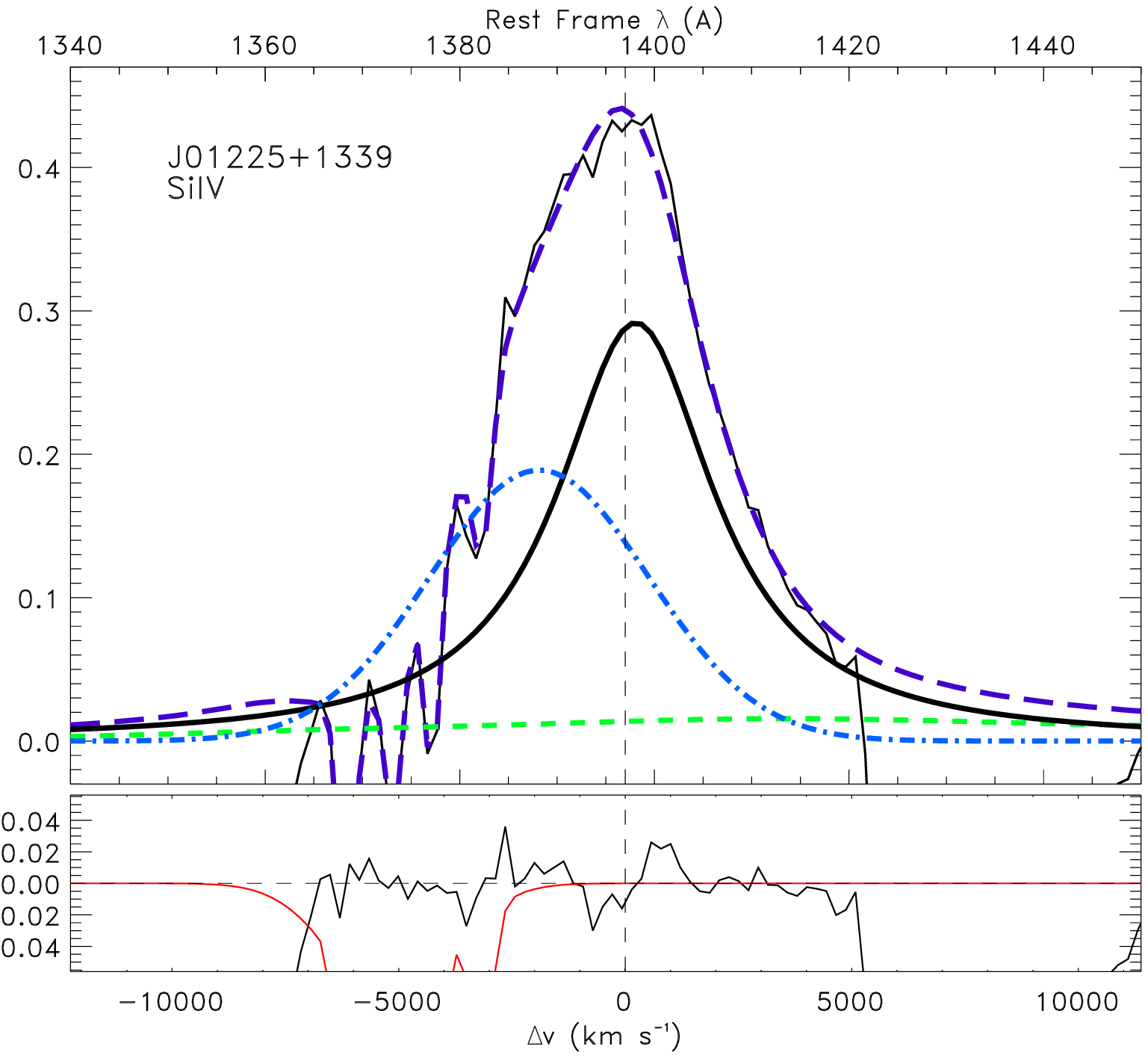}\includegraphics[scale=0.35]{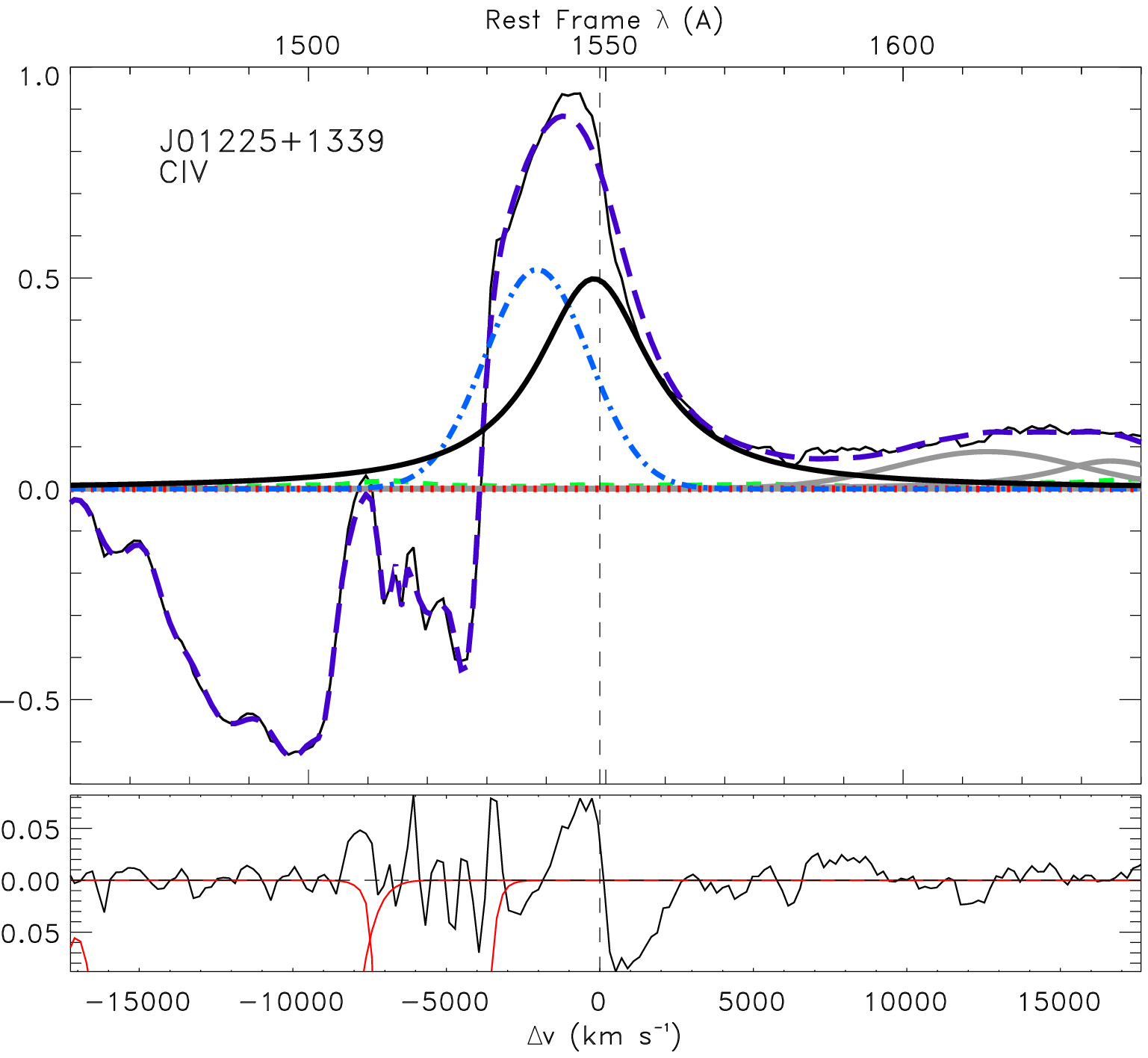}\includegraphics[scale=0.35]{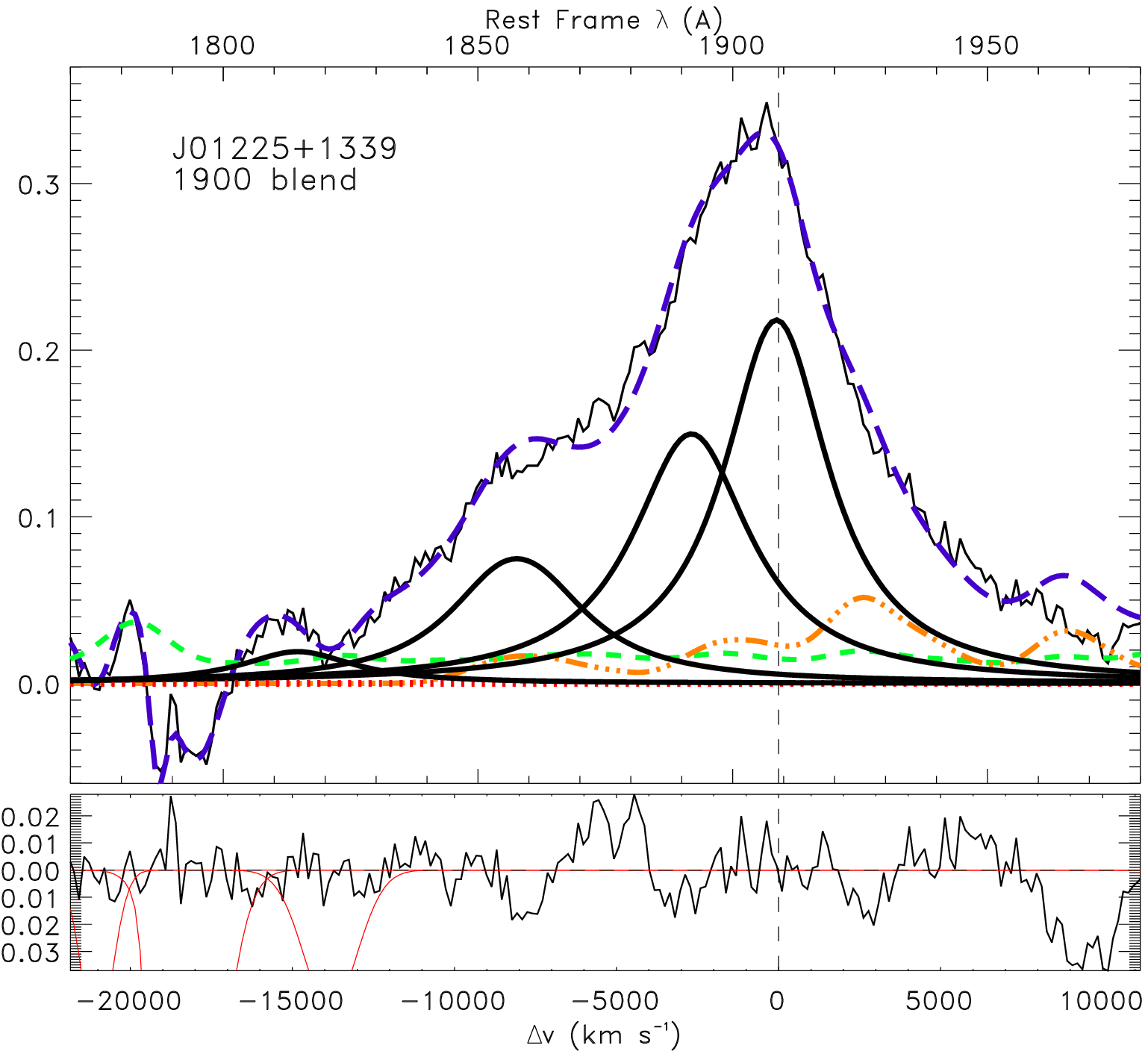}\\
\includegraphics[scale=0.35]{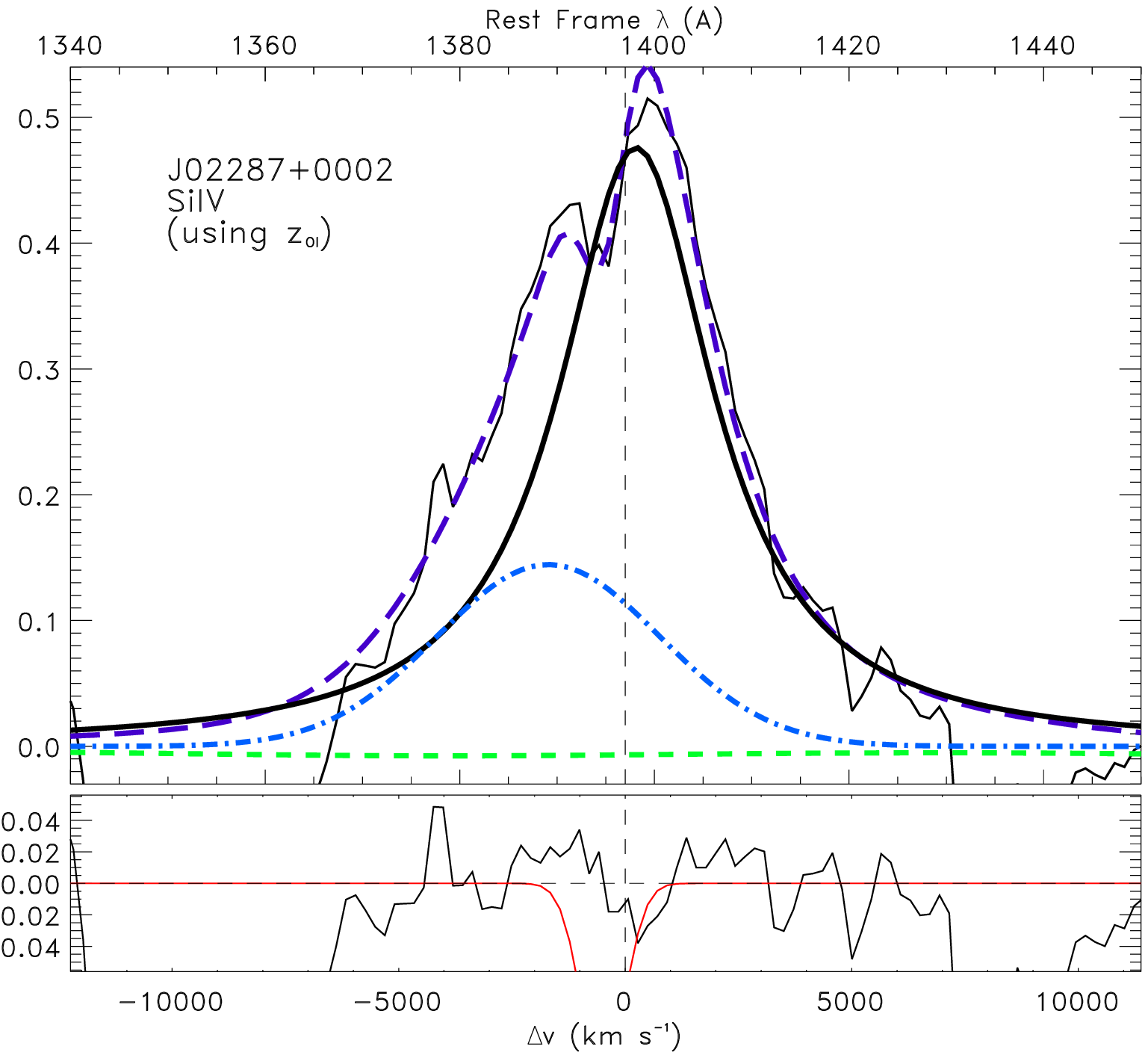}\includegraphics[scale=0.35]{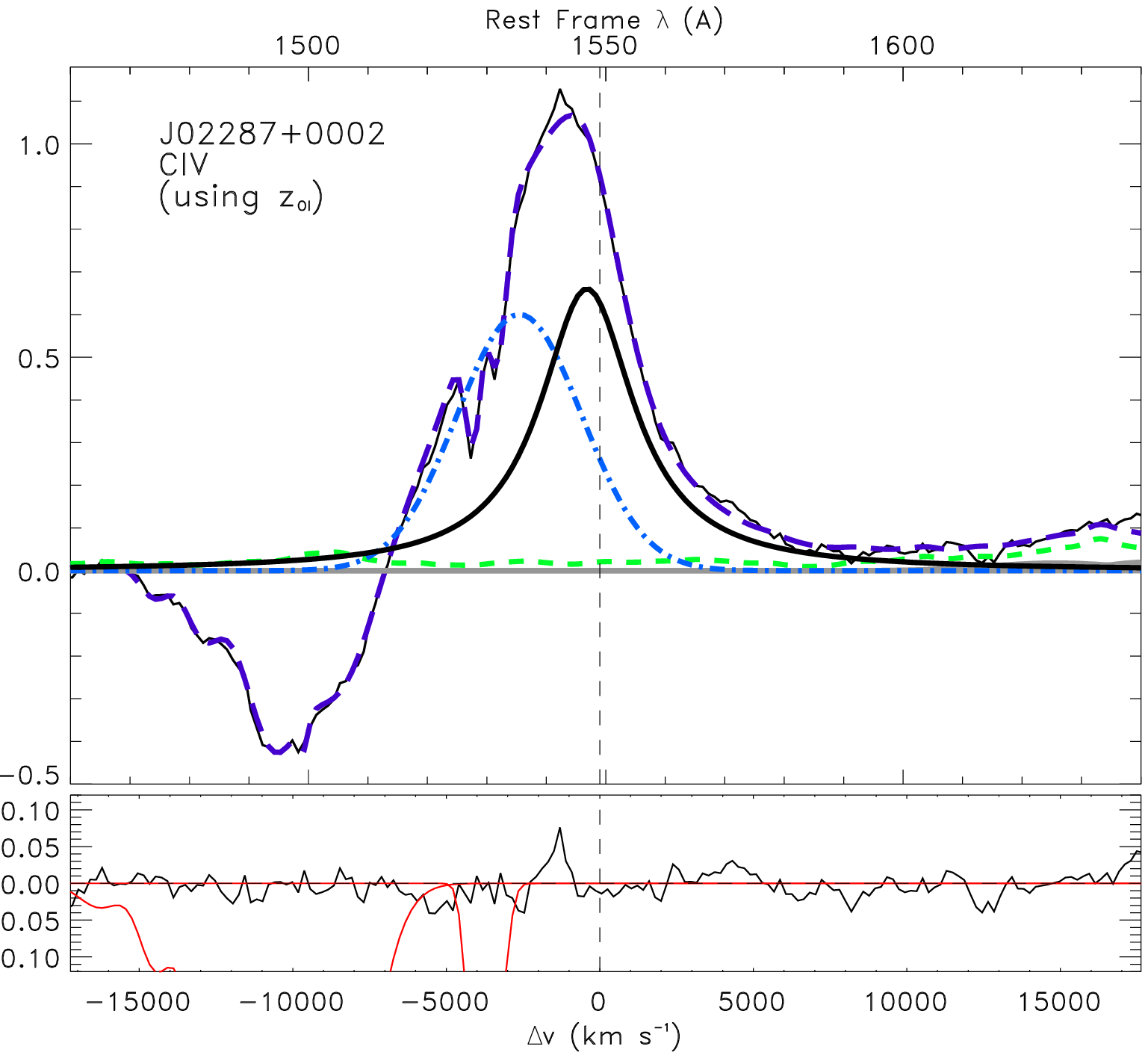}\includegraphics[scale=0.35]{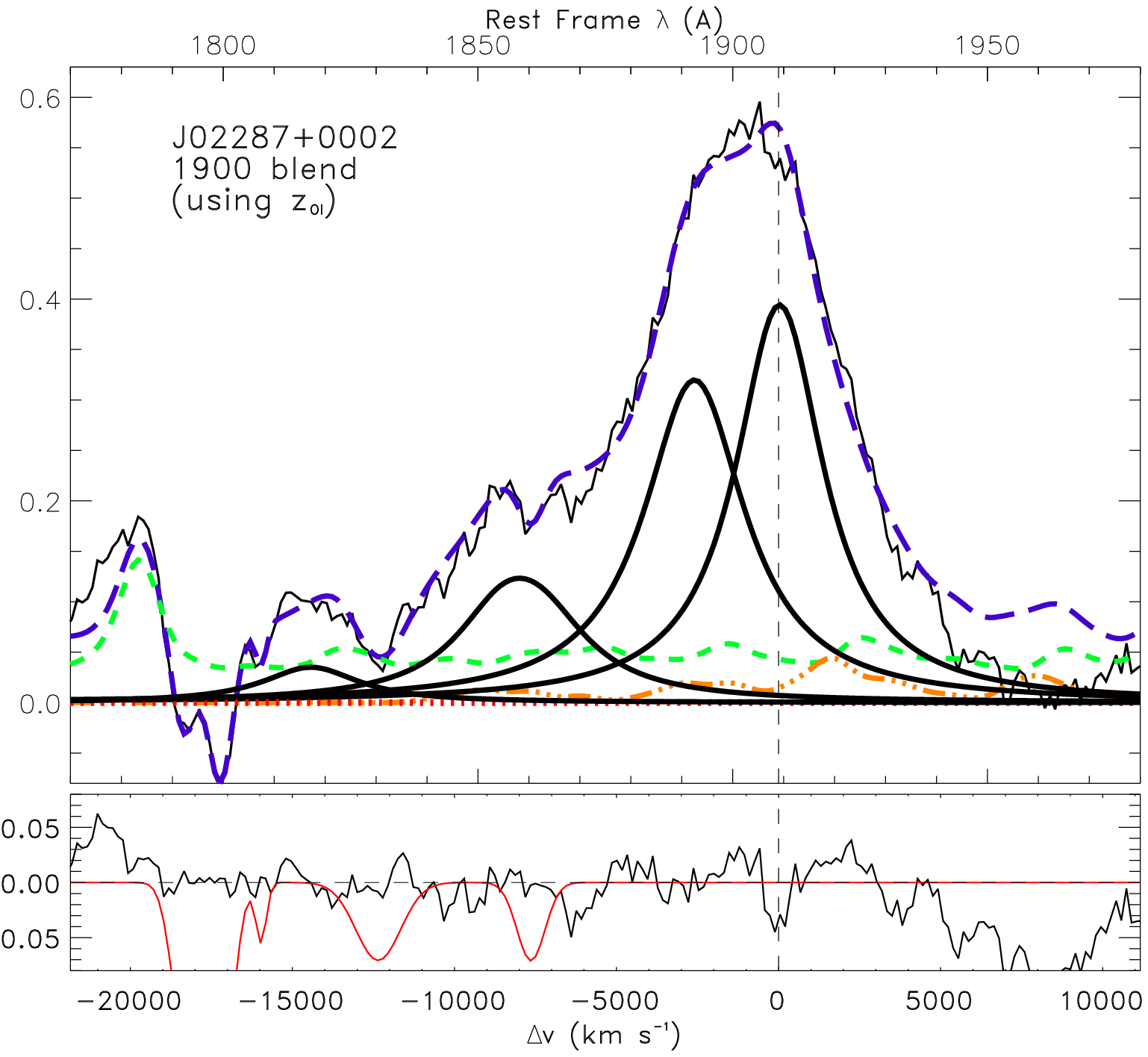}\\
\includegraphics[scale=0.35]{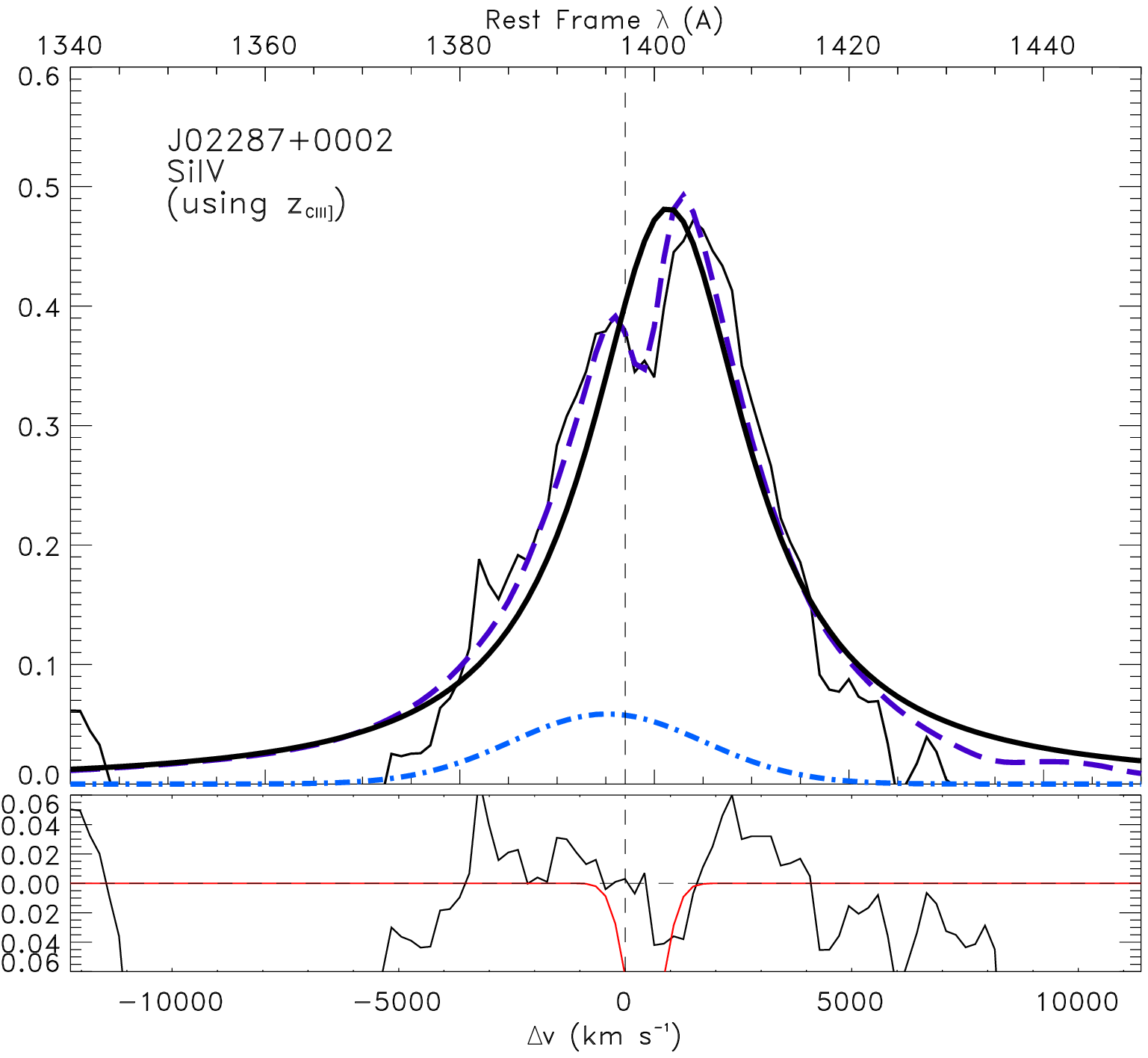}\includegraphics[scale=0.35]{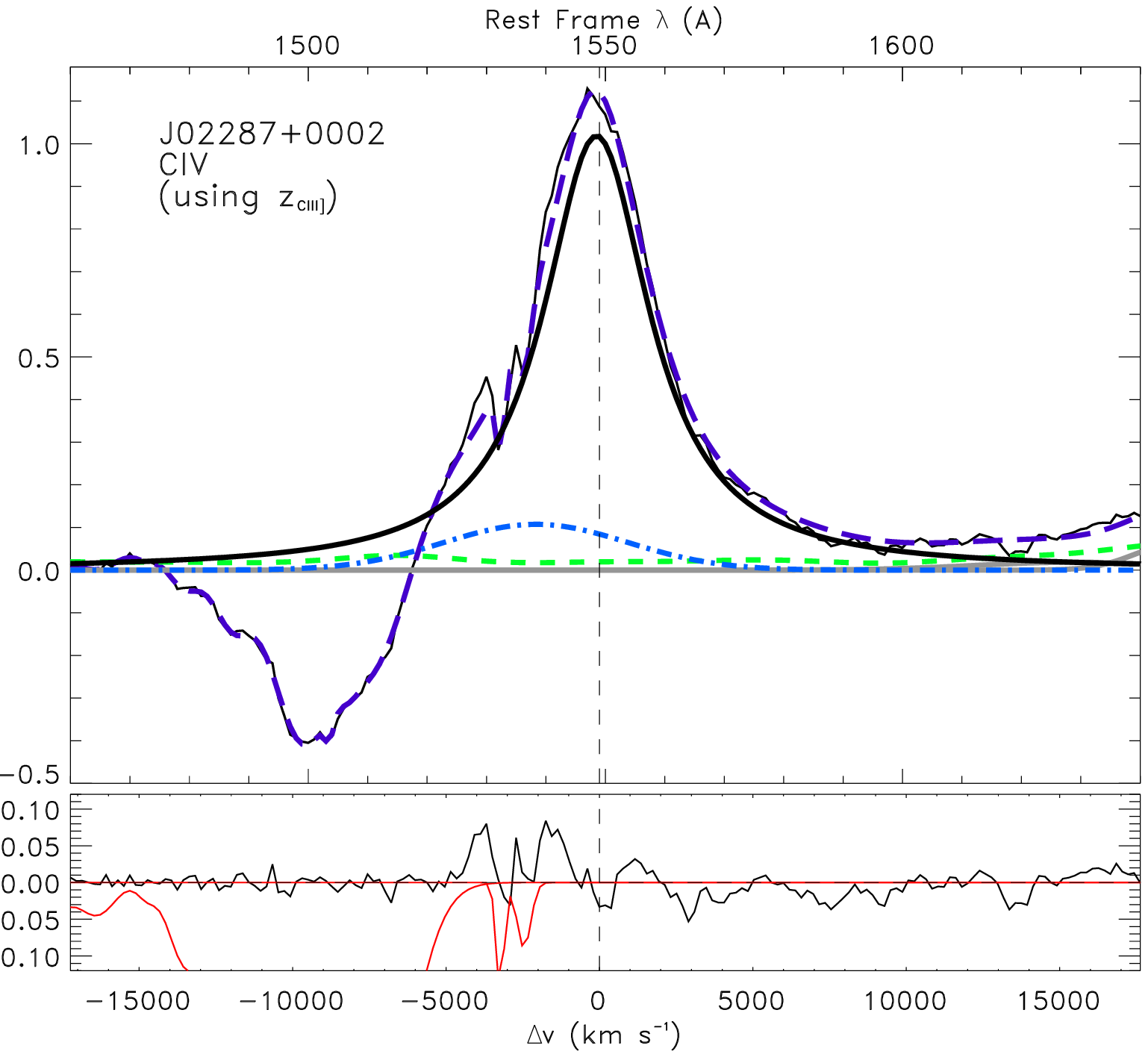}\includegraphics[scale=0.35]{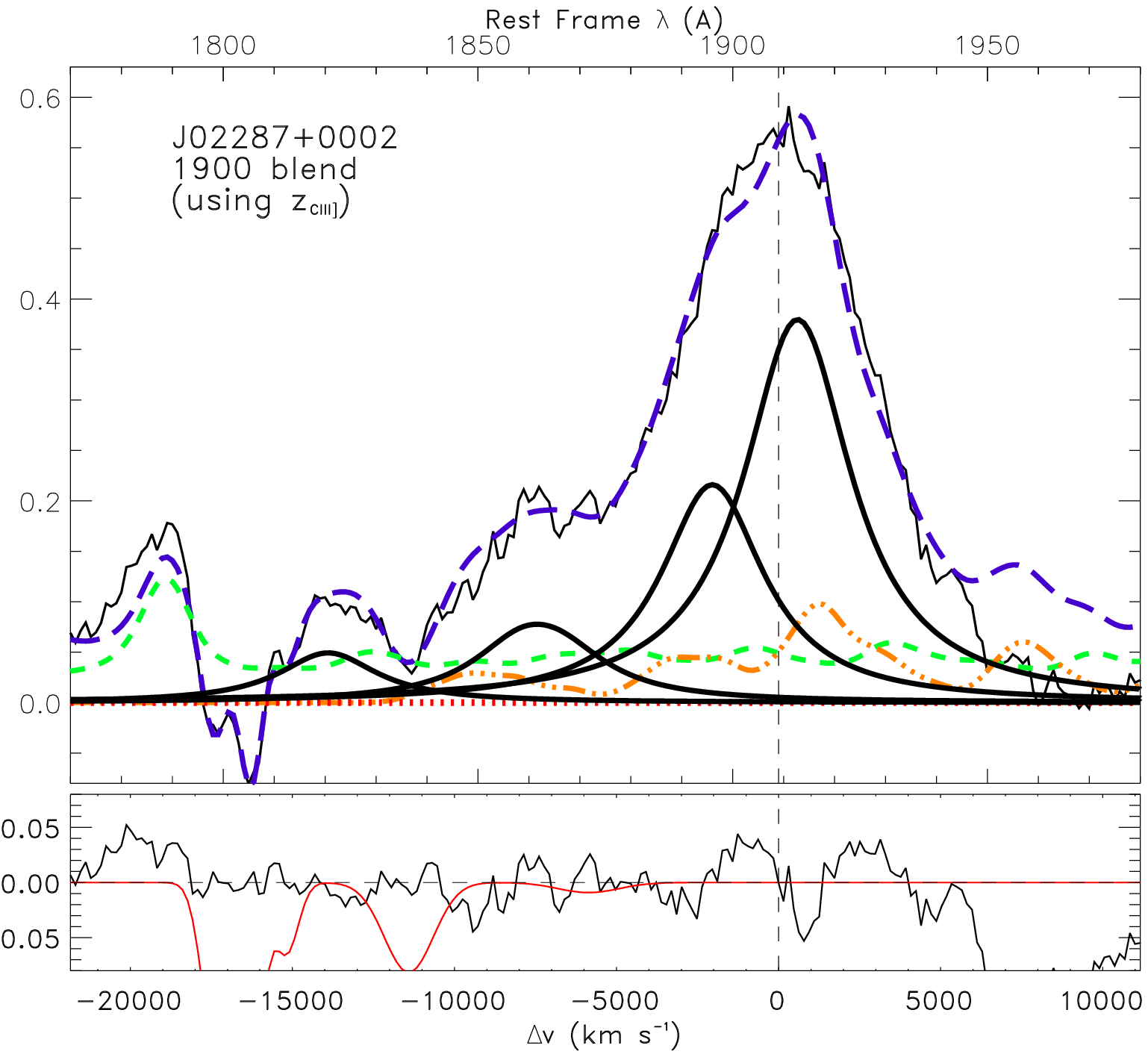}\\
\caption{Fits for BAL quasars: up J01225+1339, middle J02287+0002 using $z_{\mathrm O{\sc i}\lambda1304}$ rest frame, low J02287+0002 using $z_{\mathrm C{\sc iii}\lambda1909}$ rest frame. Note in J02287+0002 the line displacement  with the consequently line intensity changes, specially in \ciii, \siiii\ and \civ\ broad and blue-shifted components. Units and symbols are the same as in Fig. \ref{fig:fitsA}.   \label{fig:fits_bal}}
\end{figure}

\begin{figure}
\epsscale{1.1}
\includegraphics[scale=0.28, angle=0]{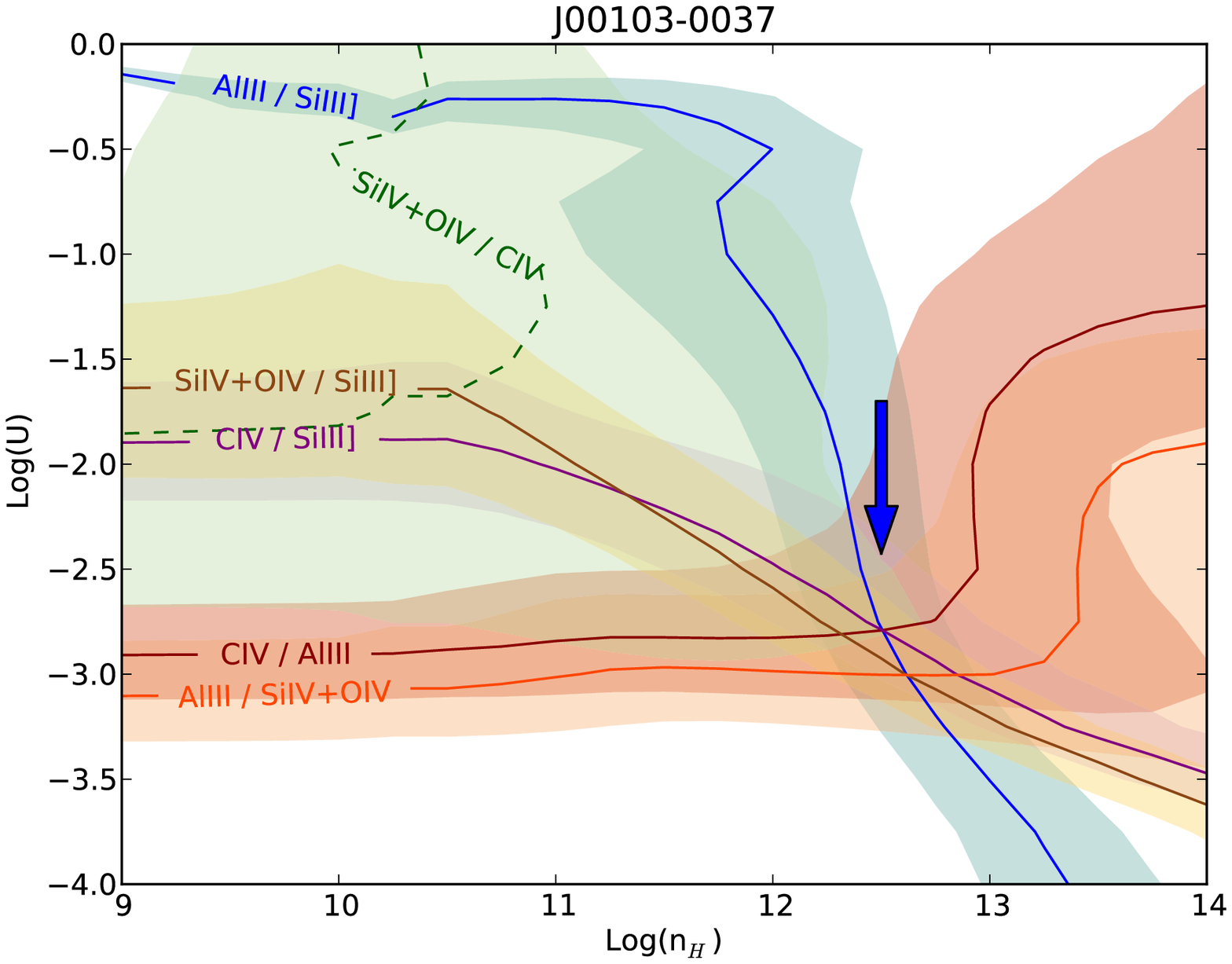}\includegraphics[scale=0.28, angle=0]{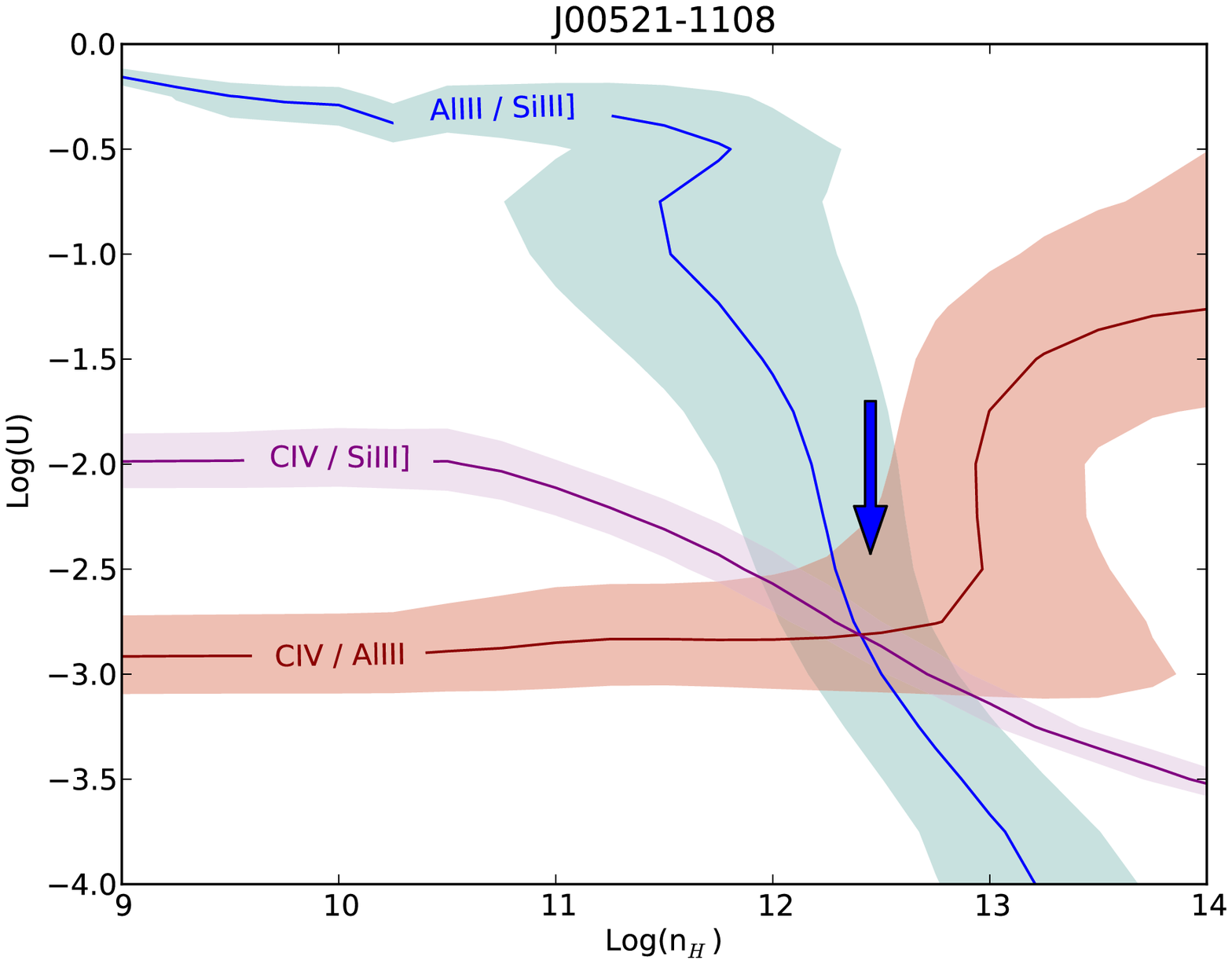}\includegraphics[scale=0.28, angle=0]{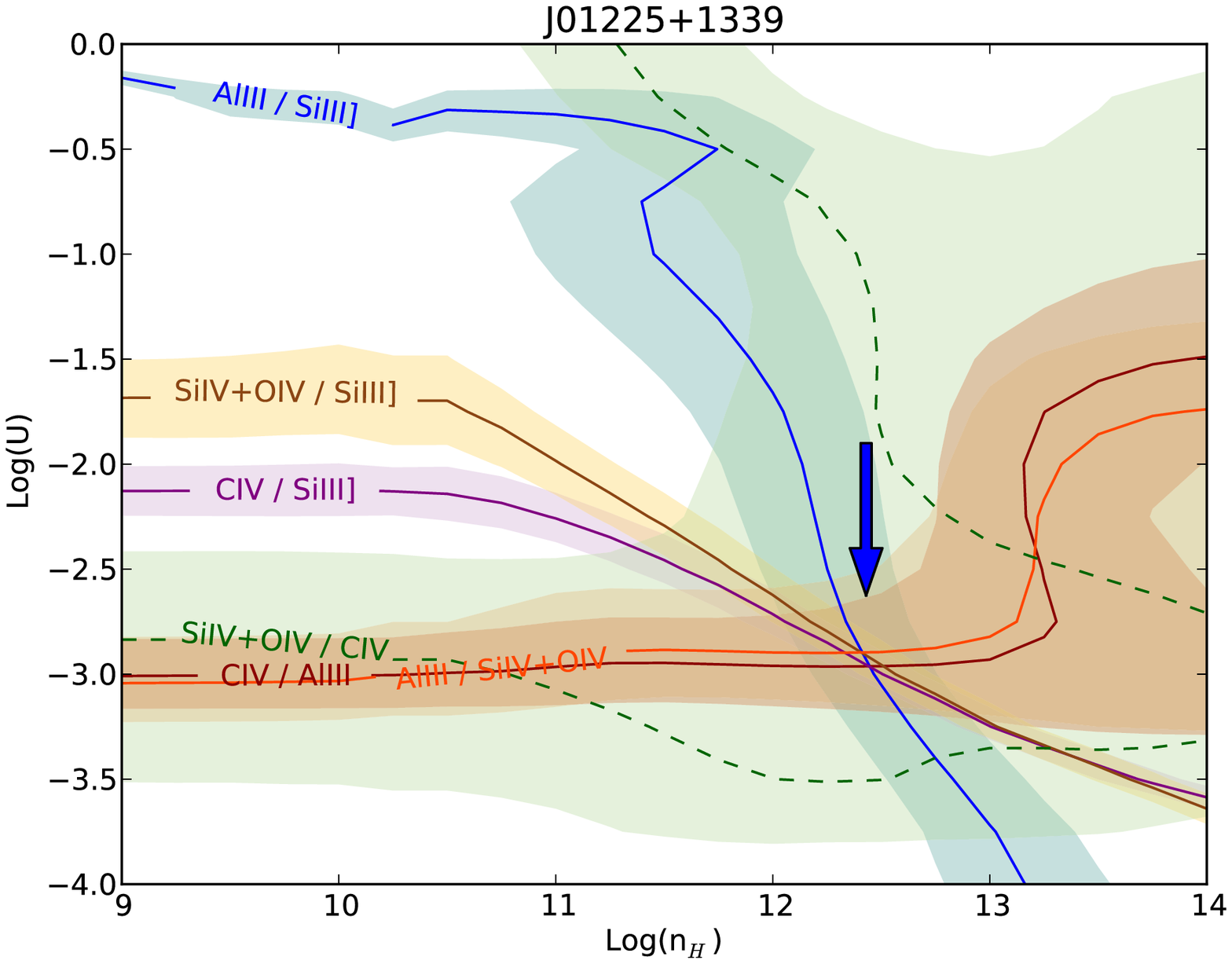}\\
\includegraphics[scale=0.28, angle=0]{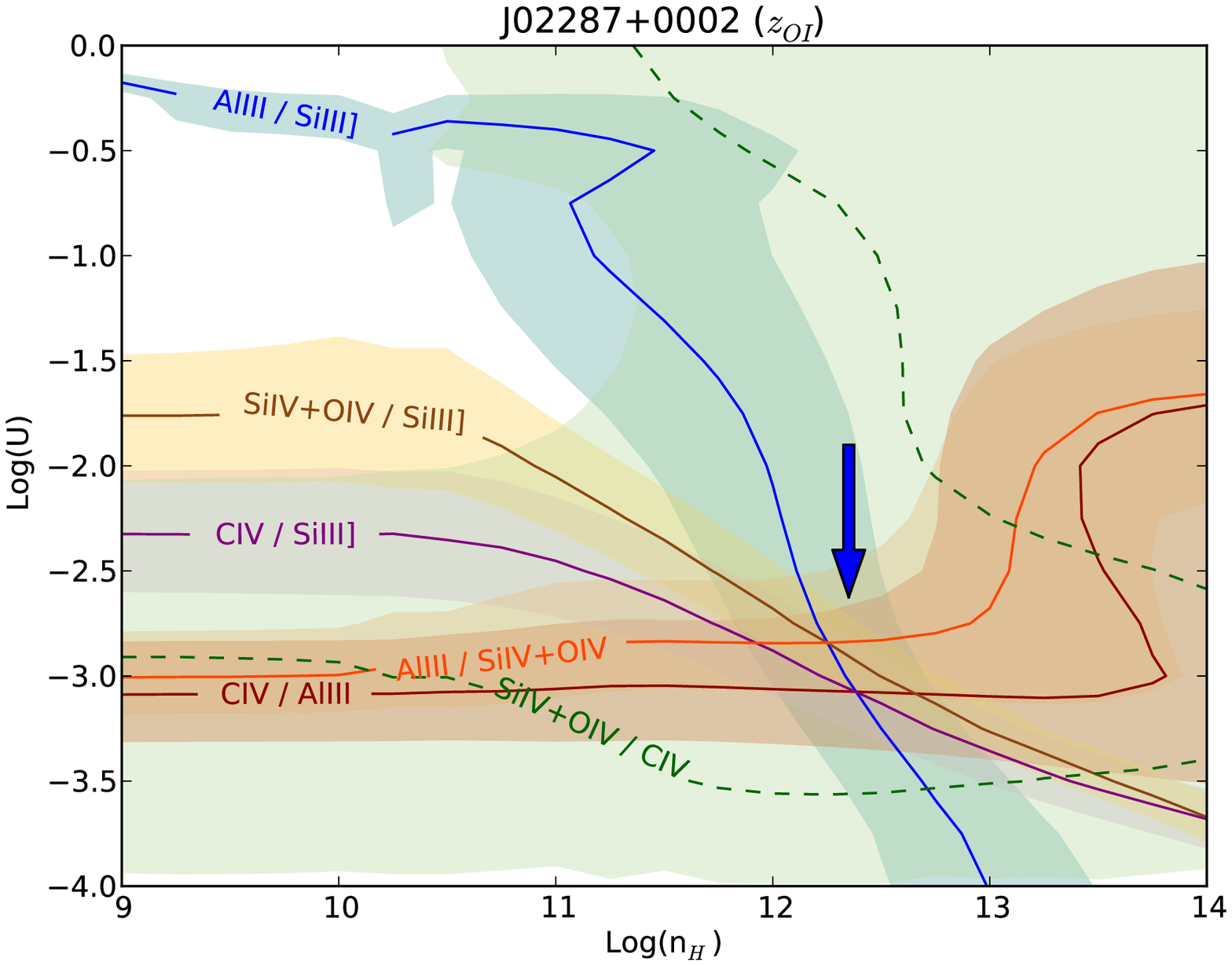}\includegraphics[scale=0.28, angle=0]{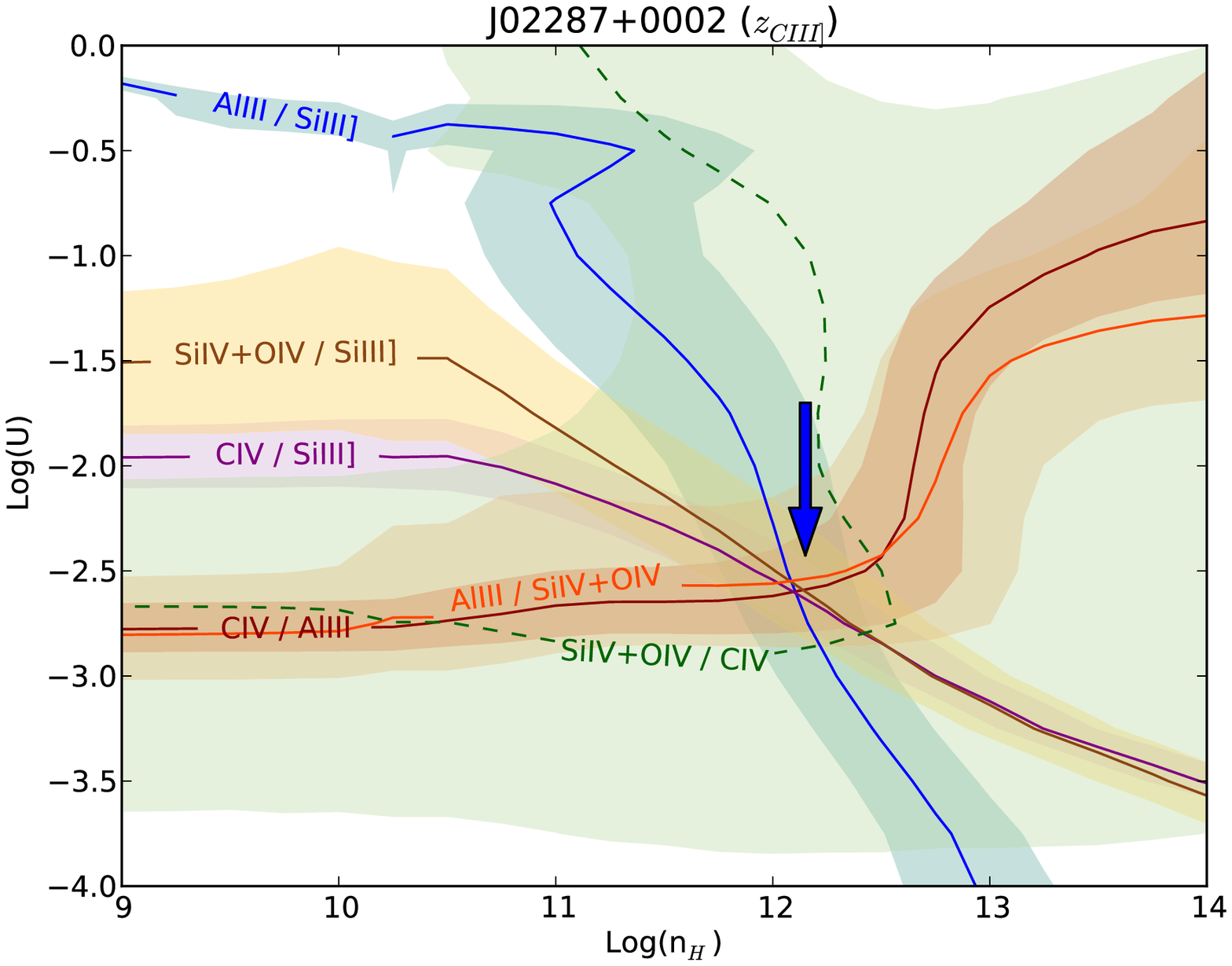}\includegraphics[scale=0.28, angle=0]{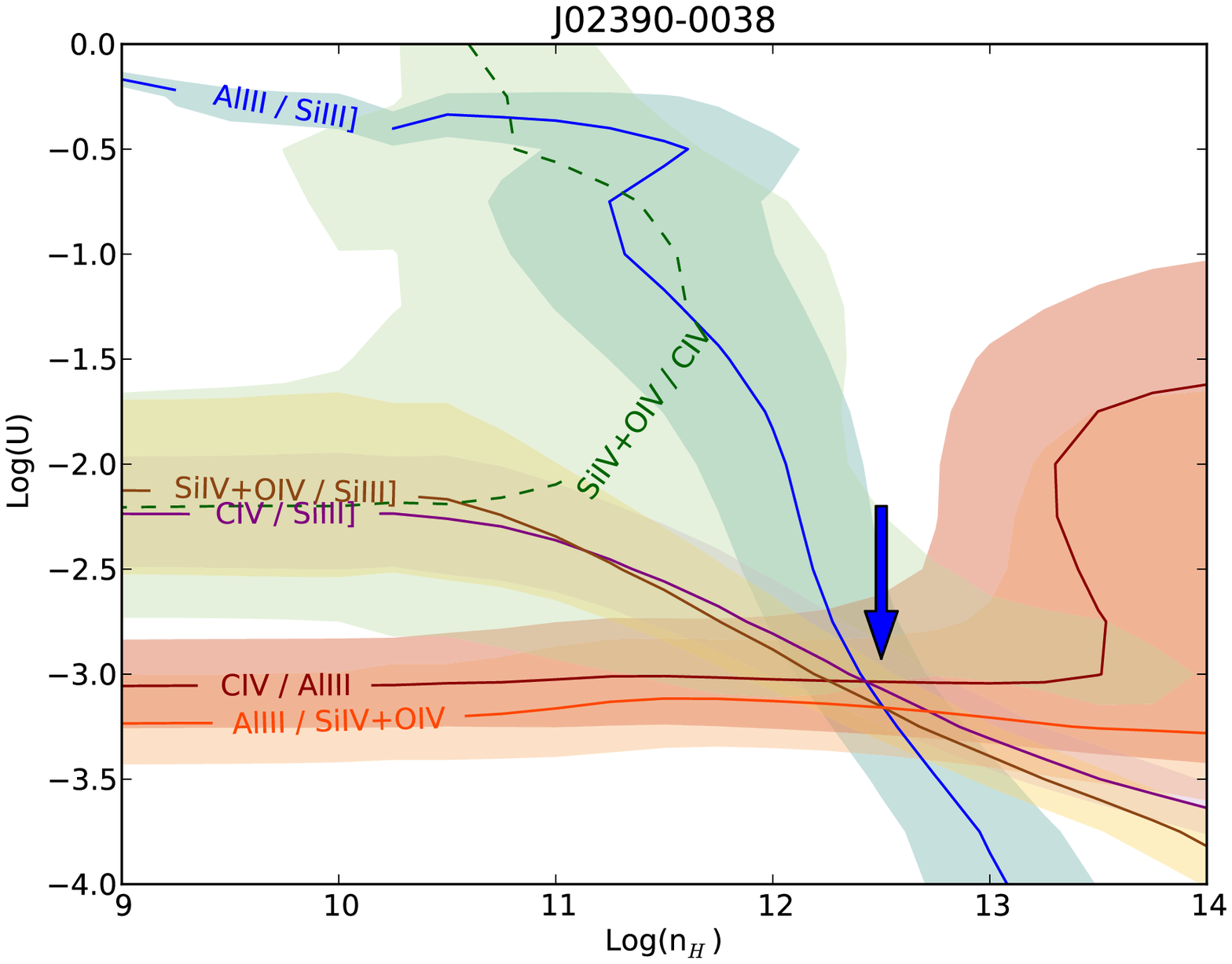}\\
\includegraphics[scale=0.28, angle=0]{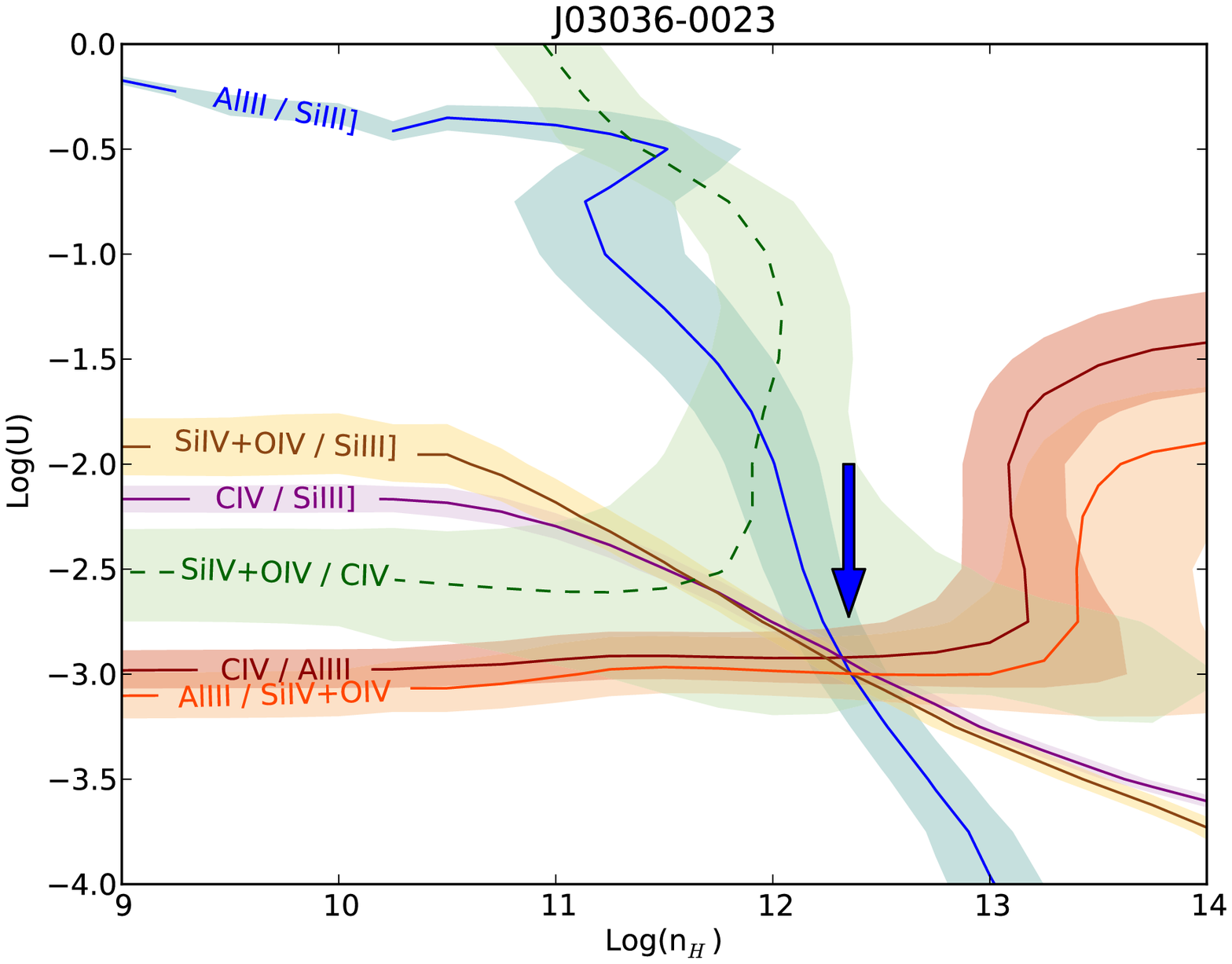}\includegraphics[scale=0.28, angle=0]{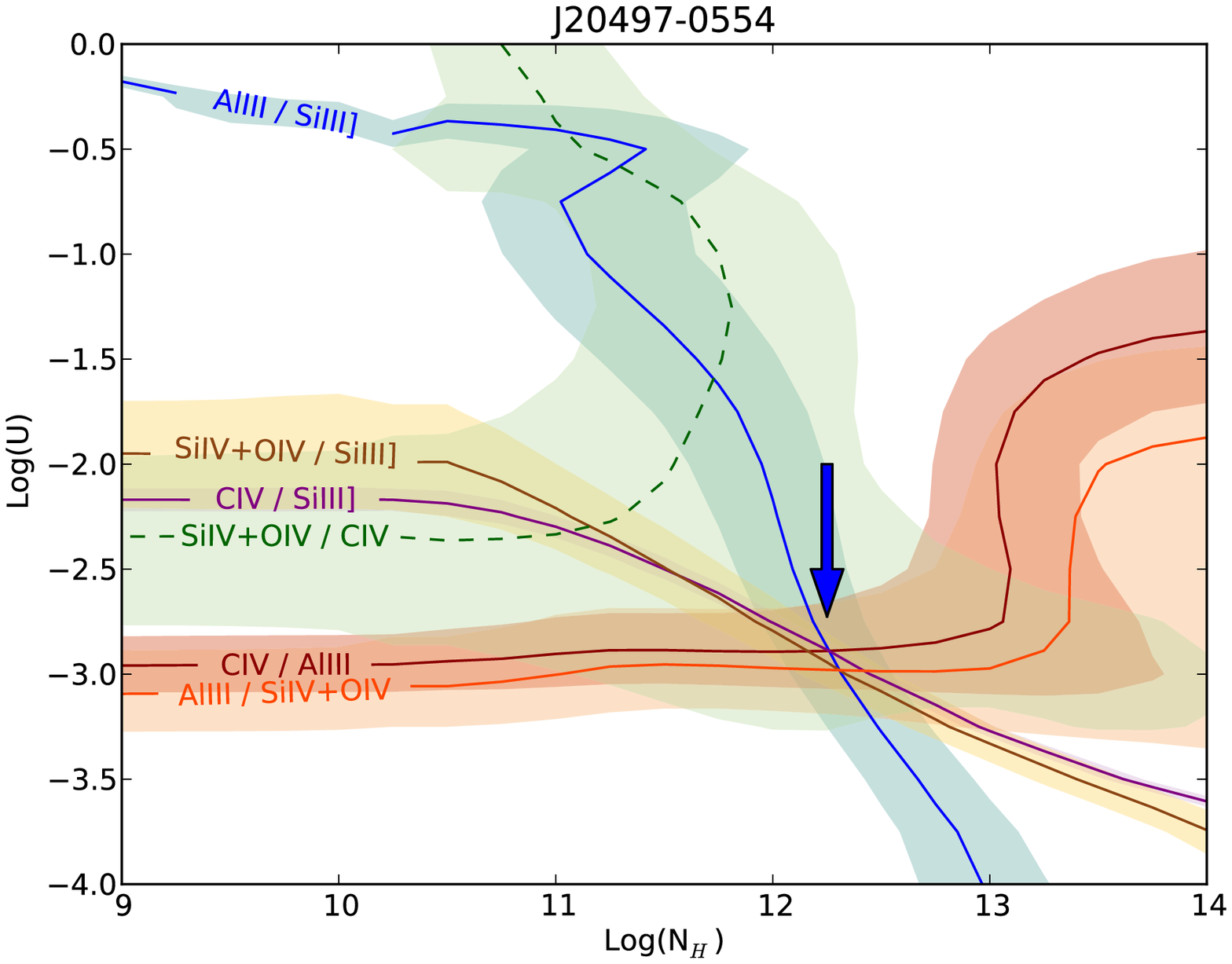}\includegraphics[scale=0.28, angle=0]{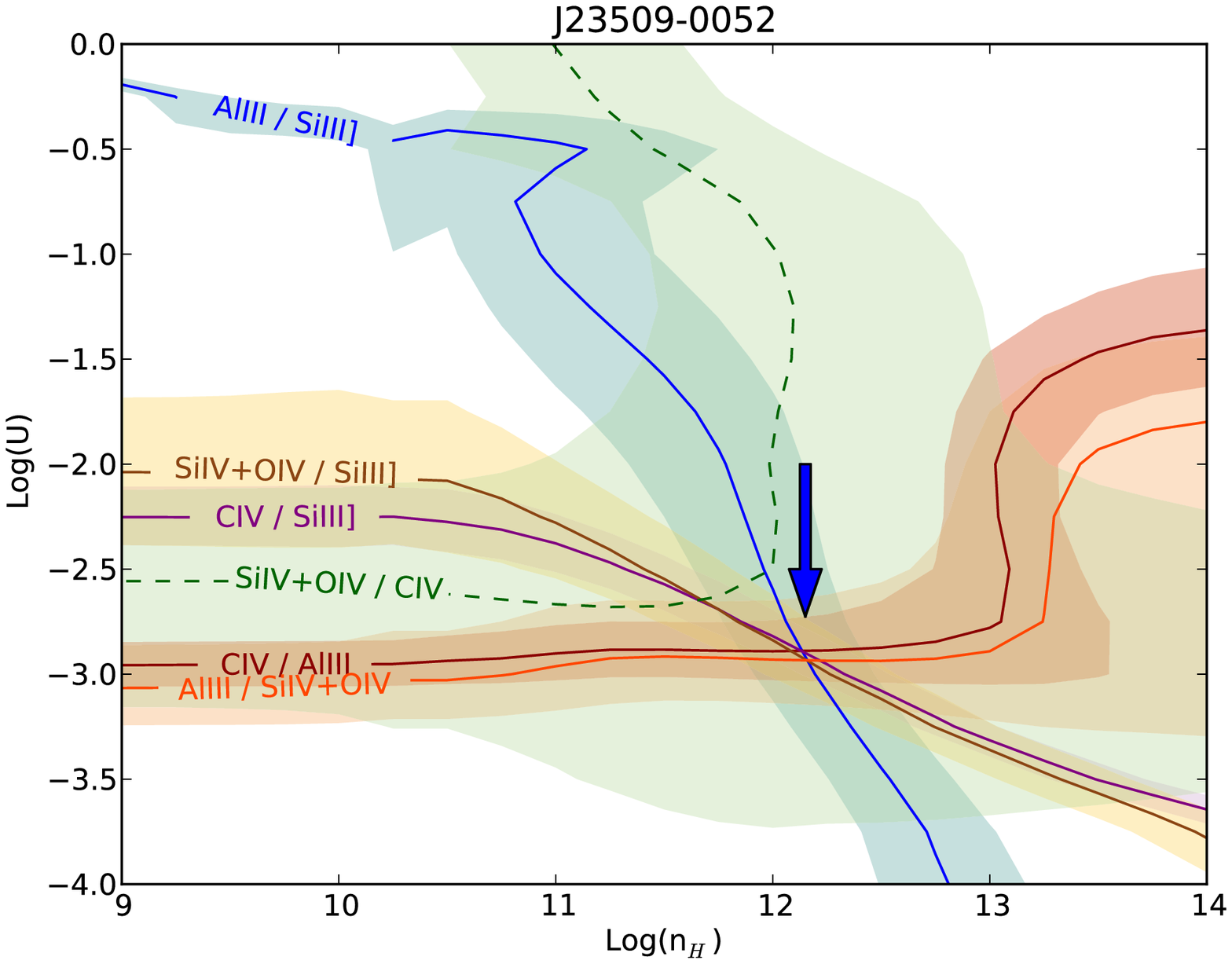}\\
 \caption{Contour plots for our sample. We show the two cases for the BAL quasar J02287+0002.  Abscissa is electron density in cm$^{-3}$, ordinate is the ionization parameter, both in logarithm scale. The point where the isocontours cross (marked with an arrow) determines the values of Log\nh\ and LogU. \siiv+\oiv/\civ\ is in dashed line because we do not use it for constrain \nhu, but to determine the metallicity (see Section \ref{metals}). The shaded area are the error bands at 2$\sigma$ confidence.
 \label{fig:neu}}
\end{figure}

\begin{figure}[h]
 \begin{center}
\includegraphics[scale=0.28, angle=0]{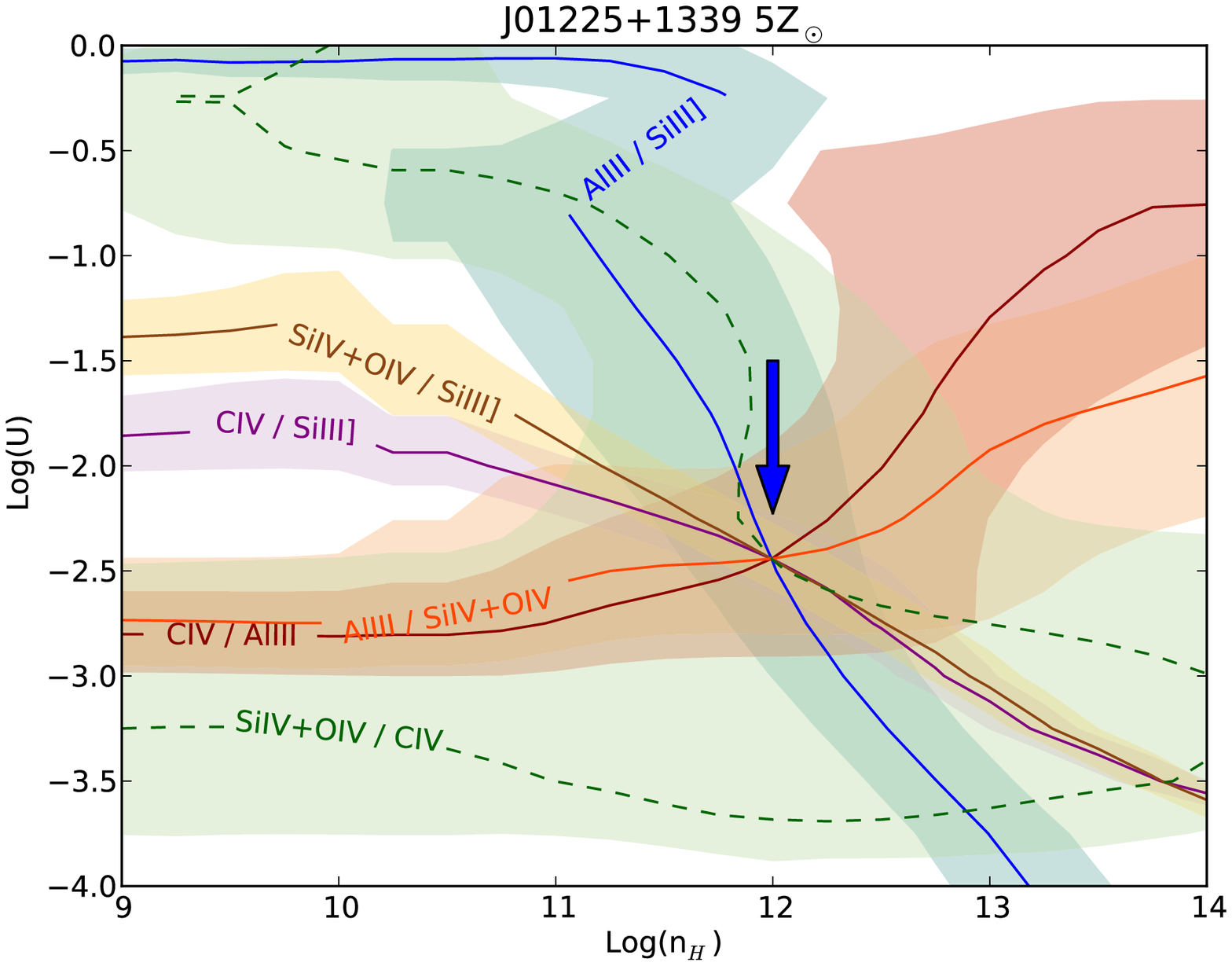}\includegraphics[scale=0.28, angle=0]{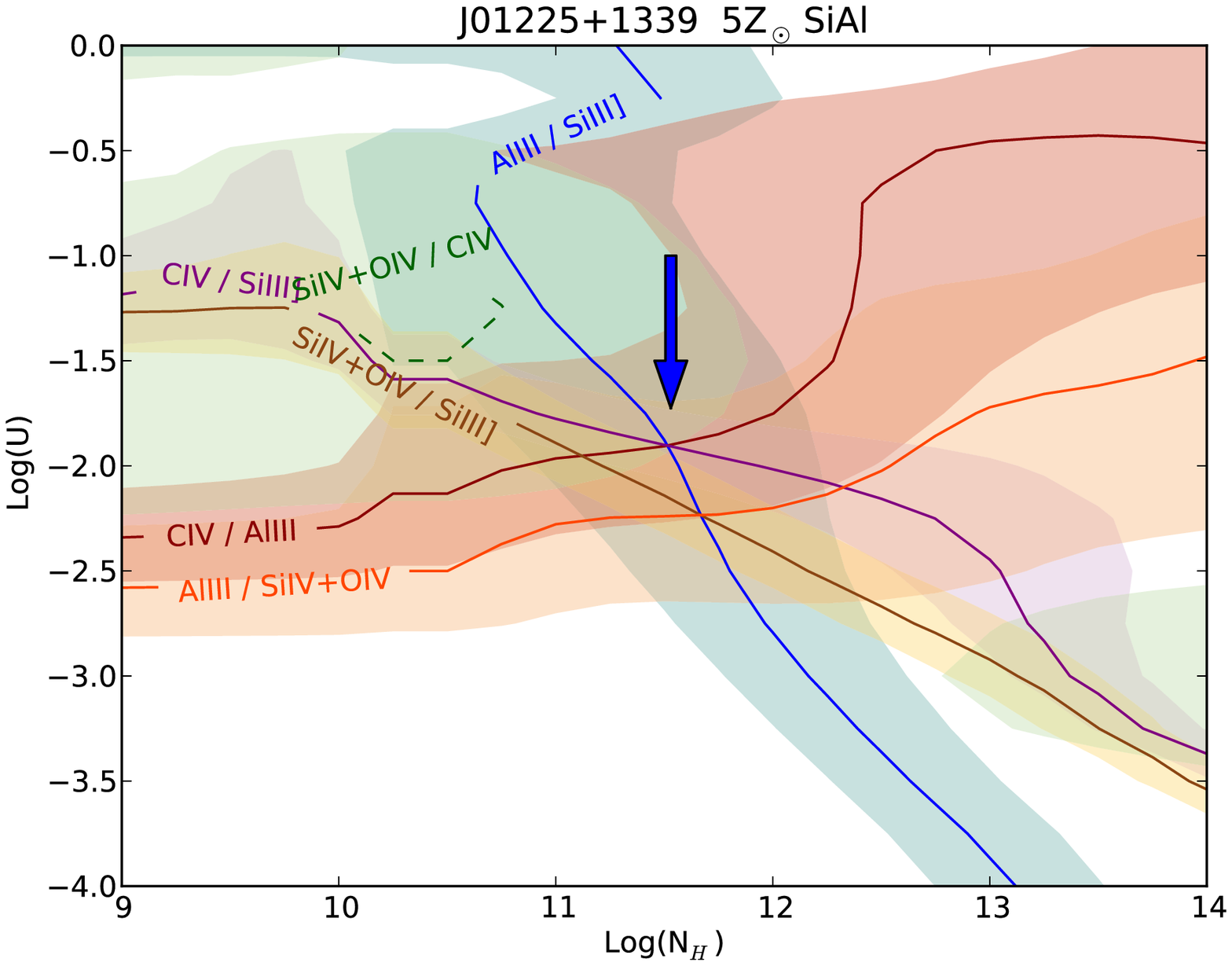}\\
\includegraphics[scale=0.28, angle=0]{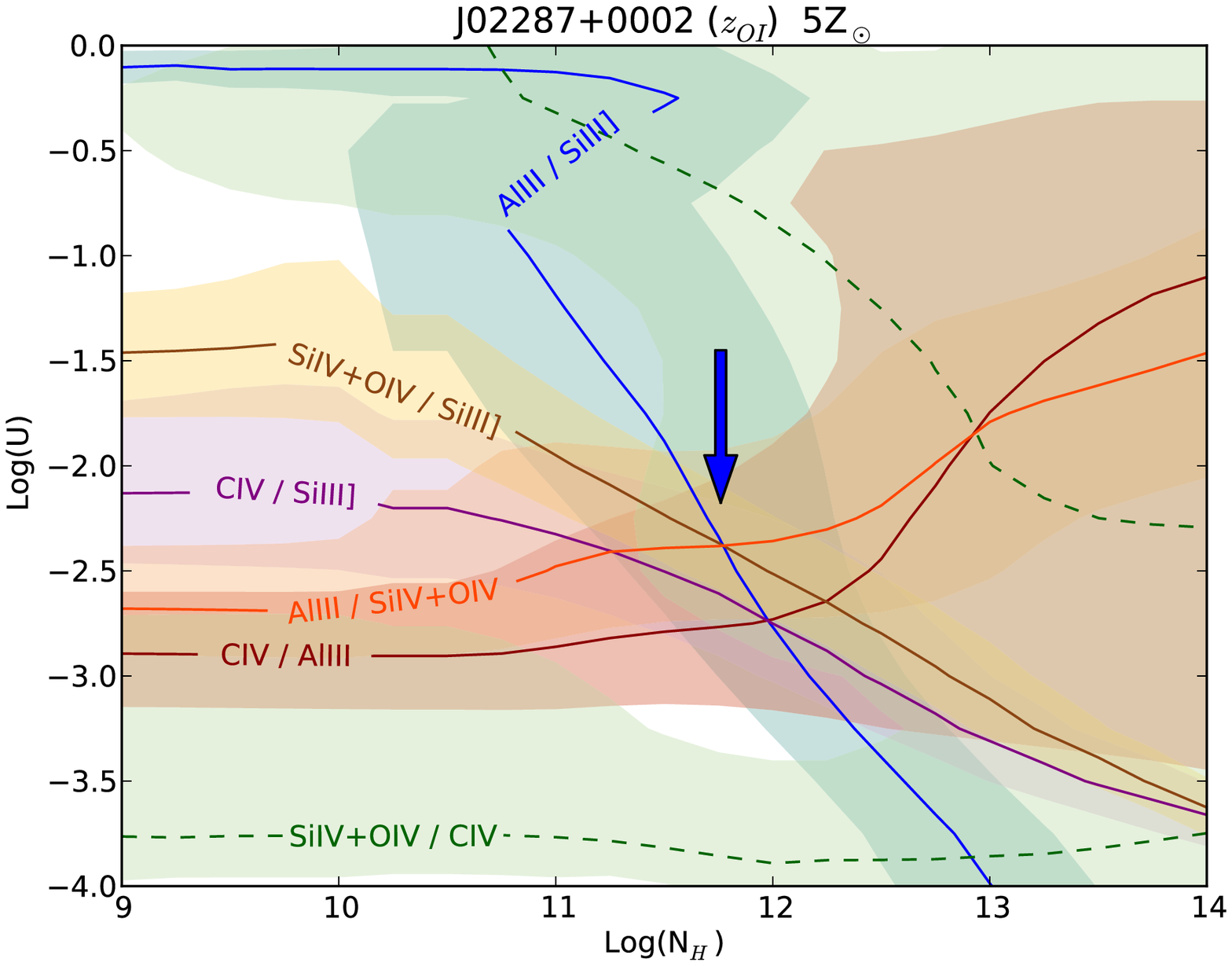}\includegraphics[scale=0.28, angle=0]{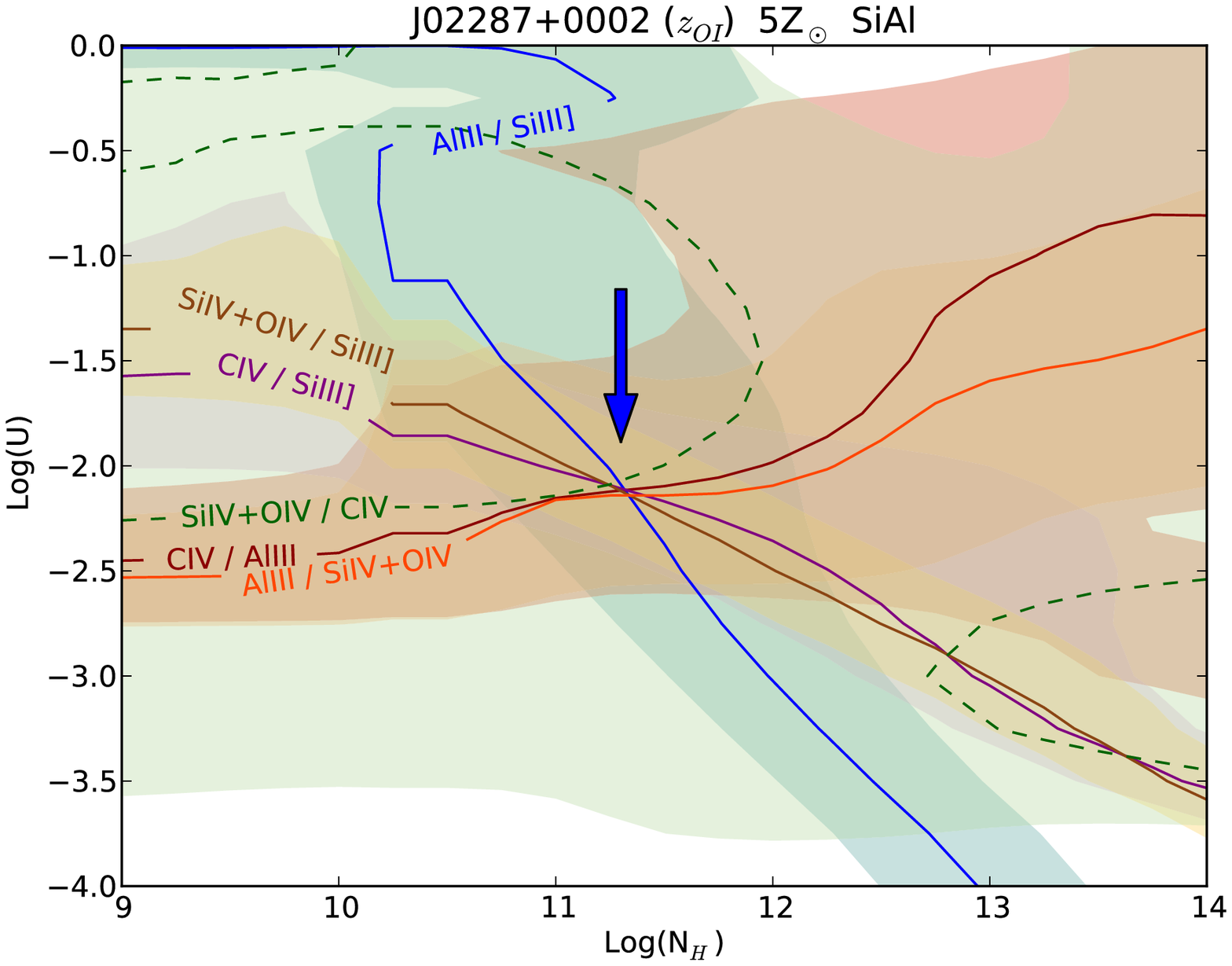}\\
\caption{Left: Contour plots from the array of simulations computed for $Z = 5Z_{\odot}$.  Right: Same of left panels  considering the case of over abundance of Si and Al. Upper panels are for J01225+1339. Lower panels are for J02287+0002 considering $z_{OI}$. Coordinates and symbols are as for Fig. \ref{fig:neu}. The  intersection point improves in certain cases, but in others is the same.\label{fig:z5} }
\end{center}
\end{figure}

\begin{figure}
\epsscale{0.45}
\plotone{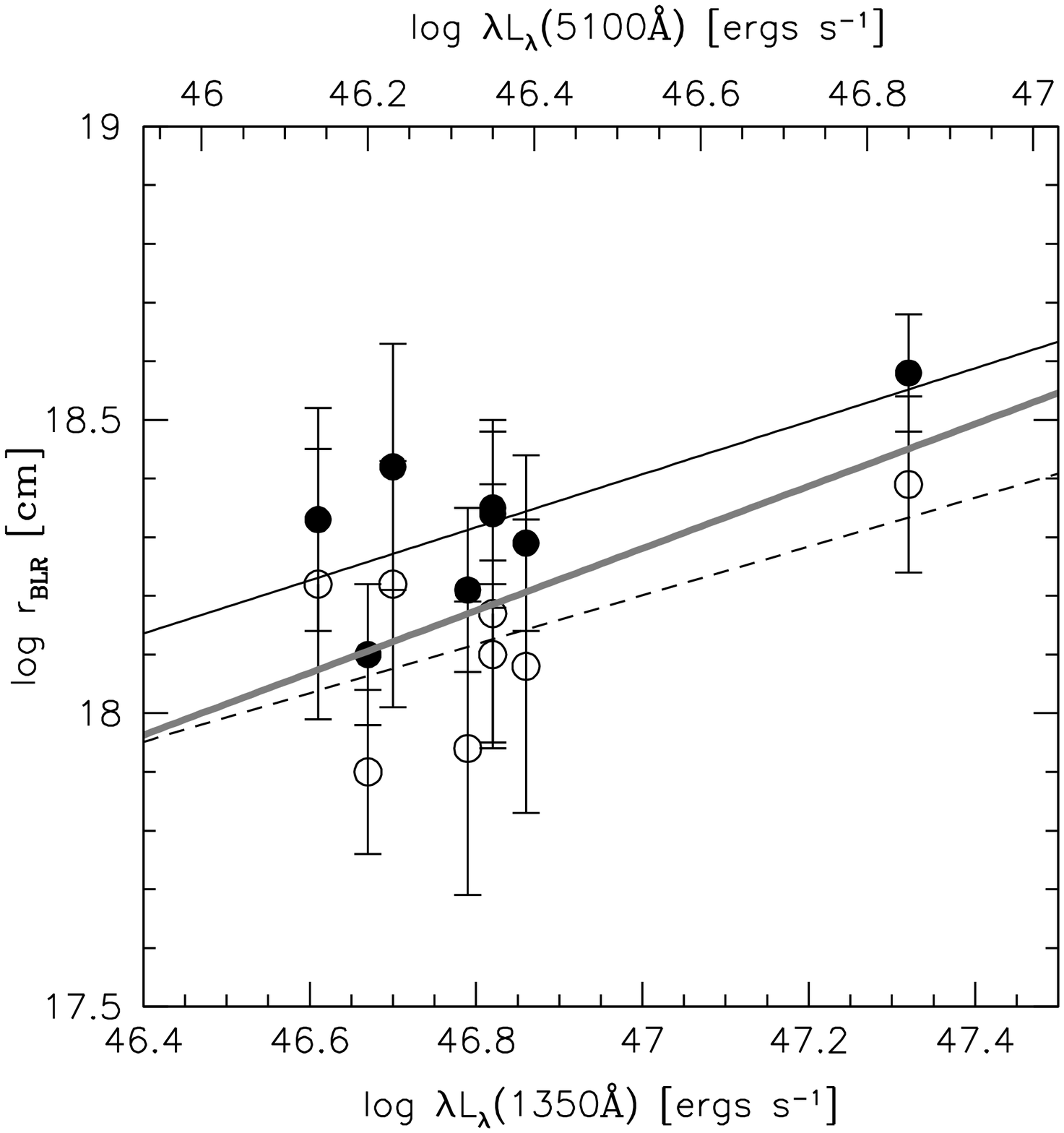}\\
\plotone{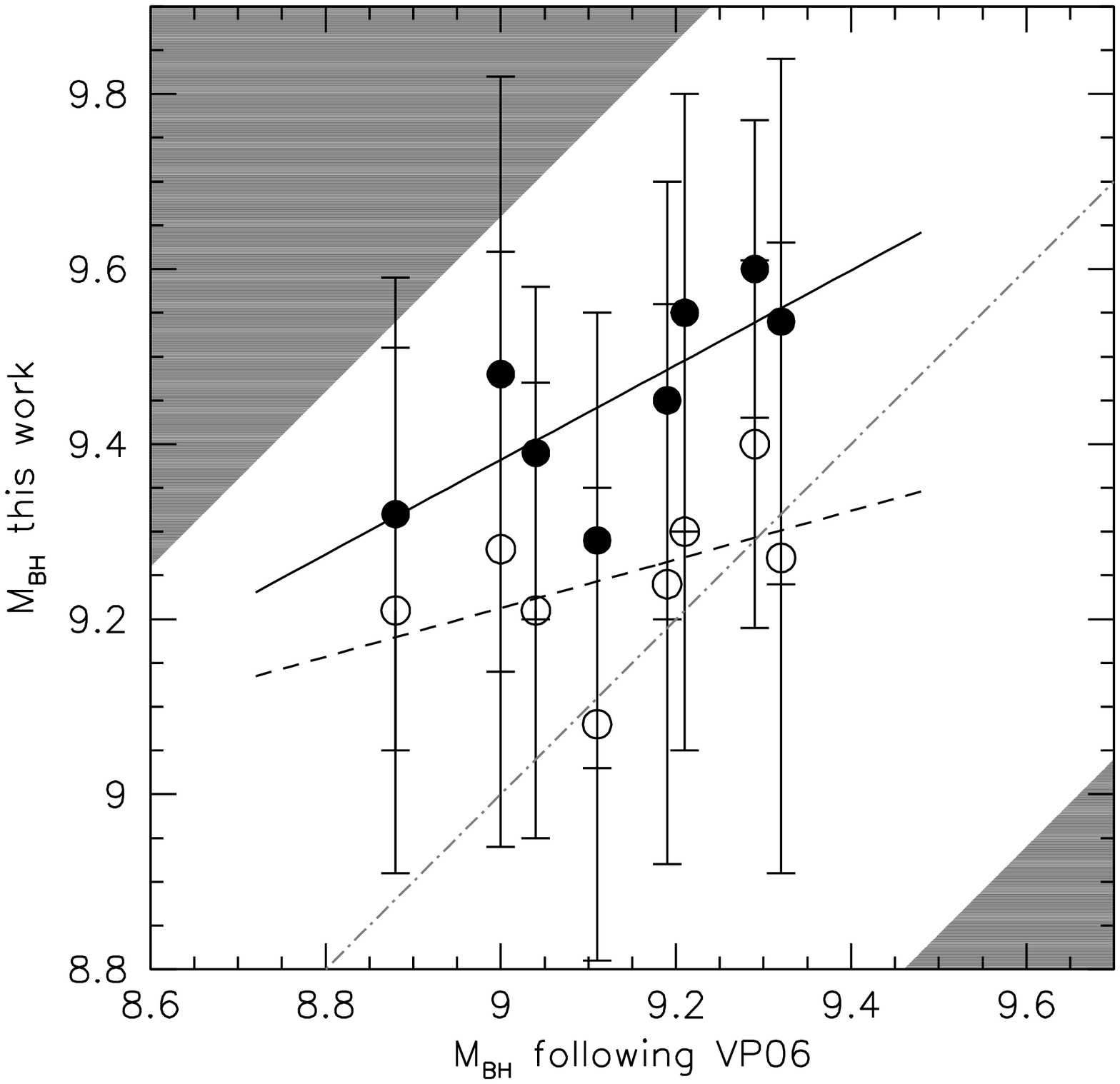}
\plotone{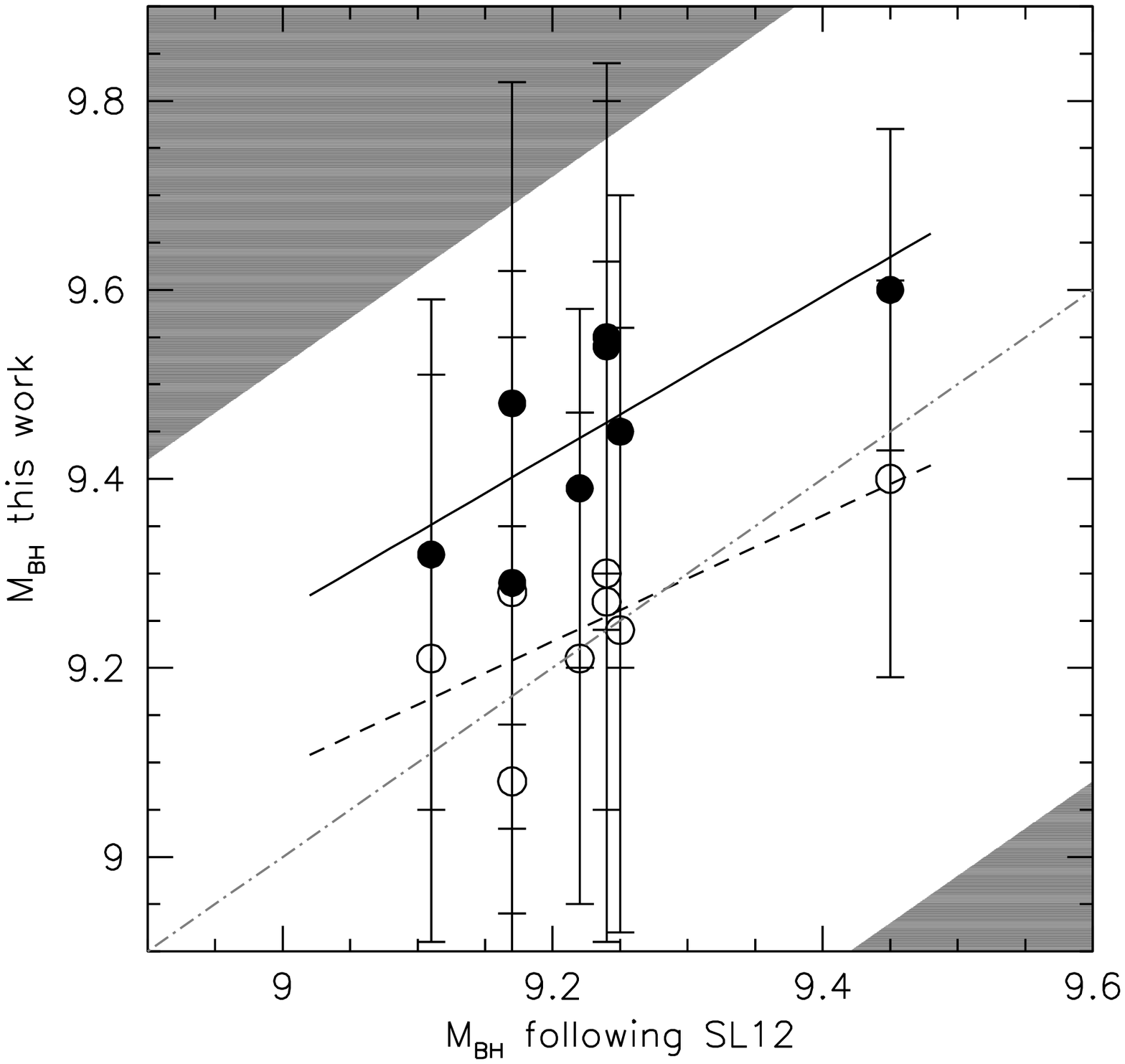}
\caption{
Upper panel: \rb\ estimates following the photoionization method with (open circles) and without (filled circles) correction for BLR low density/stratification, as a function of $\lambda L_\lambda$\ measured at 1350\AA. The thick grey line is the \rb - $\lambda L_\lambda$\ 5100\AA\ correlation as derived from \citet{bentzetal13}. The continuous and dashed lines are unweighted least square fits to uncorrected and corrected data, respectively. { $\log [ \lambda L_\lambda (1350)/\lambda L_\lambda (5100)] \approx$ 0.47, in accordance with typical SEDs of quasars (Elvis et al. 1994; Richards et al. 2006).} 
Middle panel: \mbh\ comparison for the high-$z$ sample.  The shaded bands  limit the $2\sigma$ confidence level spread expected on the basis of the \citet{vestergaardpeterson06} relation. The grey dot dashed line is the equality line. Filled symbols refer to uncorrected intensity ratios; open symbols are for intensity ratios corrected because of low-density emission.    
Bottom panel: same of middle panel, with \mbh\   computed from the \citet{shenliu12} relation. \label{fig:MBHcomparison}}
\end{figure}

\end{document}